\documentclass[12pt]{article}
\usepackage{a4p}
\usepackage{epsfig}
\usepackage{array}
\usepackage{cite}
\usepackage{pennames}
\usepackage{rotating}
\newcommand{\PPEnum} {CERN-EP/2003-053}
\newcommand{\Date}      {25th August 2003}
\newcommand{\inmath}[1] {\ifmmode#1\else$#1$\fi}
\newcommand{\definmath}[2] {\def#1{\ifmmode#2\else$#2$\fi}}
\newcommand{\alphas} {\alpha_{\mathrm{s}}}
\newcommand{\alphaem} {\mbox{$\alpha_{\mathrm{em}}$}}

\newcommand{\PZ}   {\mbox{$\mathrm{Z}$}}  % Z without a zero preferred
\definmath{\PWpm} {\mathrm{W}^{\pm}}      % W+-
\definmath{\Plp} {\ell^{+}}        % l+
\definmath{\Plm} {\ell^{-}}        % l-
\definmath{\Plpm}   {\ell^{\pm}}         % l+-
\definmath{\Pgtp} {\tau^{+}}        % tau+
\definmath{\Pgtm} {\tau^{-}}        % tau-
\definmath{\Pgtpm}   {\tau^{\pm}}         % tau+-
\definmath{\Pgn}  {\nu}          % neutrino
\definmath{\Pagn} {\overline{\nu}}     % anti-neutrino
\definmath{\Pf}      {\mathrm{f}}
\definmath{\Paf}  {\overline{\mathrm{f}}}
\definmath{\Pq}      {\mathrm{q}}
\definmath{\Paq}  {\overline{\mathrm{q}}}
\definmath{\Pu}      {\mathrm{u}}
\definmath{\Pau}  {\overline{\mathrm{u}}}
\definmath{\Pd}      {\mathrm{d}}
\definmath{\Pad}  {\overline{\mathrm{d}}}
\definmath{\Ps}      {\mathrm{s}}
\definmath{\Pas}  {\overline{\mathrm{s}}}
\definmath{\Pc}      {\mathrm{c}}
\definmath{\Pac}  {\overline{\mathrm{c}}}
\definmath{\Pb}      {\mathrm{b}}
\definmath{\Pab}  {\overline{\mathrm{b}}}
\definmath{\Pt}      {\mathrm{t}}
\definmath{\Pat}  {\overline{\mathrm{t}}}
\definmath{\Pap}  {\overline{\mathrm{p}}}
\definmath{\Pan}  {\overline{\mathrm{n}}}
\definmath{\PaD}  {\overline{\mathrm{D}}}
\definmath{\PaDz} {\overline{\mathrm{D}}^{0}}
\definmath{\PaB}  {\overline{\mathrm{B}}}
\definmath{\PaBz} {\overline{\mathrm{B}}^{0}}
\definmath{\PsDpm}   {\mathrm{D}^{\pm}_{\mathrm{s}}}  % Ds+-
\definmath{\PcgLpm}  {\Lambda^{\pm}_{\mathrm{c}}}  % Lambda_c+-
\definmath{\PD} {\mathrm{D}}     % D
\definmath{\PDst} {\mathrm{D}^{*}}     % D*
\definmath{\PgLz} {\Lambda^{0}}        % Lambda0

\newcommand{\zp}{\mbox{$\mathrm{Z'}$}}
\def\tm{\theta_{\rm M}}
\def\te6{\theta_{\rm E6}}
\def\tw{\theta_{\rm W}}
\def\alr{\alpha_{\rm LR}}
\newcommand{\massof}[1] {m_{\smash{#1}\mathstrut}}
\newcommand{\mtop}   {\massof{\mathrm{top}}}
\newcommand{\mHiggs} {\massof{\mathrm{Higgs}}}

\newcommand{\mPZ} {\massof{\mathrm{Z}}}
\newcommand{\mzp} {\massof{\mathrm{Z'}}}

\newcommand{\GammaZ} {\Gamma_{\mathrm{Z}}}

\newcommand{\AFB}    {A_{\mathrm{fb}}}
\newcommand{\AFBSM}  {A_{\mathrm{fb}}^{\mathrm{SM}}}
\newcommand{\jtoth}  {j^{\rm tot}_{\rm had}}

\newcommand{\thacol}{\theta_{\mathrm{acol}}}

\def\gZ{$\gamma - \mbox{Z}$}
\newcommand{\epem}   {\Pep\Pem}
\newcommand{\gamgam} {\Pgg\Pgg}
\newcommand{\mumu}   {\Pgmp\Pgmm}
\newcommand{\tautau} {\Pgtp\Pgtm}
\newcommand{\ffbar}  {\Pf\Paf}
\newcommand{\fpfp}   {\mathrm{f'}\overline{\mathrm{f'}}}

\newcommand{\qqbar}  {\Pq\Paq}
\newcommand{\uubar}  {\Pu\Pau}
\newcommand{\ddbar}  {\Pd\Pad}

\newcommand{\ud}     {\uubar + \ddbar}

\newcommand{\WW}{\ensuremath{\mathrm{W}^+\mathrm{W}^-}}
\newcommand{\eetoWW}    {\epem\to\WW}
\newcommand{\eetogg}    {\epem\to\gamgam}
\newcommand{\eetoee}    {\epem\to\epem}

\newcommand{\eetomumu}     {\epem\to\mumu}
\newcommand{\eetotautau}   {\epem\to\tautau}

\newcommand{\eetoqq}    {\epem\to\qqbar}

\newcommand{\dsdcc}        {{\rm d}\sigma/{\rm d}\!\cos\theta}
\newcommand{\dsdabscc}     {{\rm d}\sigma/{\rm d}|\cos\theta|}
\def\lapproxeq {\mbox{{\lower .7ex\hbox{$\;\stackrel{\textstyle
                  <}{\sim}\;$}}}}
\def\gapproxeq  {\mbox{{\lower .7ex\hbox{$\;\stackrel{\textstyle
                  >}{\sim}\;$}}}}
\newcommand{\roots} {\sqrt{s}}

\newcommand{\Evis}   {\mbox{$E_{\mathrm{vis}}$}}
\newcommand{\Rvis}   {\mbox{$R_{\mathrm{vis}}$}}
\newcommand{\Rbal}   {\mbox{$R_{\mathrm{bal}}$}}
\newcommand{\Eclus}  {E_{\mathrm{clus}}}

\newcommand {\ct}      {\mbox{$\cos \theta$}}
\newcommand {\absct}   {\mbox{$|\cos \theta |$}}
\newcommand {\ctem}    {\mbox{$\cos \theta_{{\rm e}^-}$}}
\newcommand {\absctem} {\mbox{$|\cos \theta_{{\rm e}^-} |$}}

\newcommand {\absctep} {\mbox{$|\cos \theta_{{\rm e}^+} |$}}
\newcommand {\absctepem} {\mbox{$|\cos \theta_{{\rm e}^\pm} |$}}

\definmath{\GeV}  {\mathrm{GeV}}
\definmath{\GeVc} {\mathrm{GeV}\!/c}
\definmath{\GeVcc}   {\mathrm{GeV}\!/c^2}
\definmath{\MeV}  {\mathrm{MeV}}
\definmath{\MeVc} {\mathrm{MeV}\!/c}
\definmath{\MeVcc}   {\mathrm{MeV}\!/c^2}
\definmath{\MVm}  {\mathrm{MV}\!/\mathrm{m}}
\definmath{\keV}  {\mathrm{keV}}
\definmath{\keVcm}   {\mathrm{keV}\!/\mathrm{cm}}
\definmath{\kV}      {\mathrm{kV}}
\definmath{\km}      {\mathrm{km}}
\definmath{\meter}   {\mathrm{m}}
\definmath{\cm}      {\mathrm{cm}}
\definmath{\mm}      {\mathrm{mm}}
\definmath{\micron}  {\mu\mathrm{m}}
\definmath{\nm}      {\mathrm{nm}}
\definmath{\kg}      {\mathrm{kg}}
\definmath{\gram} {\mathrm{g}}
\definmath{\second}  {\mathrm{s}}
\definmath{\microsec}   {\mu\mathrm{s}}
\definmath{\degree}  {^\circ}
\definmath{\degC} {^\circ\mathrm{C}}
\definmath{\ohm}  {\Omega}
\definmath{\Mohm} {\mathrm{M}\Omega}
\definmath{\rad}  {\mathrm{rad}}
\definmath{\mrad} {\mathrm{mrad}}
\definmath{\nb}      {\mathrm{nb}}
\newcommand{\eqref}[1]  {(\ref{#1})}
\newcommand{\PhysLett}  {Phys.~Lett.}
\newcommand{\PRL} {Phys.~Rev.\ Lett.}

\newcommand{\PhysRev}   {Phys.~Rev.}
\newcommand{\NPhys}  {Nucl.~Phys.}
\newcommand{\NIM} {Nucl.~Instr.\ and Meth.}
\newcommand{\ZPhys}  {Z.~Phys.}

\newcommand{\CPC} {Comp. Phys. Comm.}
\newcommand{\EPJ} {Eur.~Phys.~J.} 
\newcommand{\LEPone}    {LEP\,1}

\newcommand{\LEPtwo}    {LEP\,2}
\newcommand{\ALEPHColl}   {ALEPH Collab.}
\newcommand{\DELPHIColl}  {DELPHI Collab.}

\newcommand{\LthreeColl}  {L3 Collab.}
\newcommand{\OPALColl}    {OPAL Collab.}
\newcommand{\ALIBABA}{\mbox{A{\sc libaba}}}
\newcommand{\ARIADNE}{\mbox{A{\sc riadne}}}
\newcommand{\BHLUMI}{\mbox{B{\sc hlumi}}}
\newcommand{\BHWIDE}{\mbox{B{\sc hwide}}}
\newcommand{\HERWIG}{\mbox{H{\sc erwig}}}
\newcommand{\KANDY}{\mbox{K{\sc and}Y}}
\newcommand{\KK}{\mbox{{\cal KK}2f}}
\newcommand{\KORALW}{\mbox{K{\sc oral}W}}
\newcommand{\KORALZ}{\mbox{K{\sc oral}Z}}
\newcommand{\PYTHIA}{\mbox{P{\sc ythia}}}
\newcommand{\RADCOR}{\mbox{R{\sc adcor}}}
\newcommand{\PHOJET}{\mbox{P{\sc hojet}}}
\newcommand{\SMATASY}{\mbox{S{\sc matasy}}}
\newcommand{\TWOGEN}{\mbox{T{\sc wogen}}}
\newcommand{\TEEGG}{\mbox{T{\sc eegg}}}
\newcommand{\ZEFIT}{\mbox{Z{\sc efit}}}
\newcommand{\ZFITTER}{\mbox{Z{\sc fitter}}}

\newcolumntype{L} {>{$}l<{$}}
\newcolumntype{C} {>{$}c<{$}}
\newcolumntype{R} {>{$}r<{$}}

\newcommand{\phz} {\phantom{0}}
   \newcommand{\lept}
{\ell^+\ell^-}   

\newcommand{\epsz}  {\varepsilon_0}   \newcommand{\lamm}   {\Lambda_-}
\newcommand{\lamp} {\Lambda_+} 

\begin{document}
%%%%%%%%%%%%%%%%%%%%%%%%%%%%%%%%%%%%%%%%%%%%%%%%%%%%%%%%%%%%%%%%%%%%%%%%
%
%  Title Page
%
\begin{titlepage}
%     Header
%
\begin{center}
    \Large EUROPEAN ORGANIZATION FOR NUCLEAR RESEARCH
\end{center}
\bigskip 
\begin{flushright}
    \large \PPEnum \\  \Date \\ 
%    \large \PNnum \\  \Date \\ 
\end{flushright}
%
%     Main title
%
\begin{center}
    \huge\bf\boldmath Tests of the Standard Model and Constraints
    on New Physics from Measurements of Fermion-Pair Production at 
    189--209~\GeV\ at LEP 
\end{center}
\vspace{0.5cm} 
%
%     Author names
%
\begin{center}
    \LARGE   The  OPAL   Collaboration   \\  
\vspace{0.5cm} 
%
%     Abstract
%
\begin{abstract}%=======================================================
Cross-sections and angular distributions for hadronic and lepton-pair final
states in \epem\ collisions at centre-of-mass energies between 189~GeV
and 209~GeV, measured with the OPAL detector at LEP, are presented and 
compared with the predictions of the Standard Model. The measurements are 
used to determine the electromagnetic coupling constant \alphaem\ at \LEPtwo\ 
energies. In addition, the results are used together with OPAL measurements 
at 91--183~GeV within the S-matrix formalism to determine the $\gamma$--\PZ\ 
interference term and to make an almost model-independent measurement of the 
Z mass. Limits on extensions to the Standard Model described by effective 
four-fermion contact interactions or the addition of a heavy \zp\ boson
are also presented.
\end{abstract}%=========================================================
\end{center}
\vspace{0.5cm}
%\begin{center}
%{\bf \large FINAL DRAFT -- DO NOT QUOTE}
%\end{center}
%\begin{center}
%{\bf This note describes preliminary OPAL results}
%\end{center}
\begin{center}
%\vspace{0.5cm}  
%    \large {\bf Authors:} \\ 
%    \normalsize \LEPtwo\ Standard Model Group: Giovanni Abbiendi, 
%        Chris Ainsley, Giorgios Anagnostou, Tatsuo Kawamoto,
%        Michael Kobel, Kirsten Sachs, Pat Ward, David Ward, Pete Watkins \\
%    \large  {\bf Editorial Board:} \\  
%    \normalsize   Dick Kellogg, Achim Stahl, Andreas Mutter \\
%\vspace{0.5cm}
%{\bf Comments to {\tt Patricia.Ward@cern.ch} before 0900 Wednesday 13th August}
\end{center}
\begin{center}
{\large Submitted to European Journal of Physics C}
\end{center}
%
%  End of title page
%
\end{titlepage}
\begin{center}{\Large        The OPAL Collaboration
}\end{center}\bigskip
\begin{center}{
%begin authorlist PLEASE DO NOT DELETE THIS COMMENT
G.\thinspace Abbiendi$^{  2}$,
C.\thinspace Ainsley$^{  5}$,
P.F.\thinspace {\AA}kesson$^{  3,  y}$,
G.\thinspace Alexander$^{ 22}$,
J.\thinspace Allison$^{ 16}$,
P.\thinspace Amaral$^{  9}$, 
G.\thinspace Anagnostou$^{  1}$,
K.J.\thinspace Anderson$^{  9}$,
S.\thinspace Arcelli$^{  2}$,
S.\thinspace Asai$^{ 23}$
% ...
%end authorlist PLEASE DO NOT DELETE THIS COMMENT
}\end{center}\bigskip
\bigskip
%begin institutes
$^{  1}$School of Physics and Astronomy, University of Birmingham,
Birmingham B15 2TT, UK
\newline
$^{  2}$Dipartimento di Fisica dell' Universit\`a di Bologna and INFN,
I-40126 Bologna, Italy
\newline
$^{  3}$Physikalisches Institut, Universit\"at Bonn,
D-53115 Bonn, Germany
\newline
% ...
$^{  5}$Cavendish Laboratory, Cambridge CB3 0HE, UK
\newline
% ...
$^{  9}$Enrico Fermi Institute and Department of Physics,
University of Chicago, Chicago IL 60637, USA
\newline
% ...
$^{ 16}$Department of Physics, Schuster Laboratory, The University,
Manchester M13 9PL, UK
\newline
% ...
$^{ 22}$Department of Physics and Astronomy, Tel Aviv University,
Tel Aviv 69978, Israel
\newline
$^{ 23}$International Centre for Elementary Particle Physics and
Department of Physics, University of Tokyo, Tokyo 113-0033, and
Kobe University, Kobe 657-8501, Japan
\newline
% ...
$^{  y}$ now at CERN
\newline
% ...
%end notes

%=======================================================================
%       Main Text
%=======================================================================

%%-----------------------------------------------------------------------
%\section{Introduction}                                 \label{sec:intro}      
%%-----------------------------------------------------------------------
%\input intro.tex
%-----------------------------------------------------------------------
\section{Introduction}           \label{sec:intro}
%-----------------------------------------------------------------------
Measurements of fermion-pair production in \epem\ collisions at high
energies provide a sensitive test of Standard Model predictions, and
allow limits to be set on many possible new physics 
processes~\cite{bib:OPAL-SM172,bib:OPAL-SM183,bib:OPAL-SM189,bib:ADL-SM}.
In this paper we present measurements of cross-sections and angular
distributions for hadronic and lepton-pair final states at centre-of-mass 
energies $\sqrt{s}$ between 189~GeV and 209~GeV; forward-backward asymmetries 
for the leptonic states are also given. The data were collected by the OPAL 
detector at LEP in 1998, 1999 and 2000. 

In the Standard Model, fermion-pair production proceeds via $s$-channel 
photon and $\PZ$ diagrams, except for the $\epem$ final state where 
$t$-channel diagrams dominate. A general feature of $\epem$ collision data
at these energies is radiative return to the $\PZ$. If one or more 
initial-state radiation photons are emitted which reduce the effective 
centre-of-mass energy of
the subsequent $\epem$ collision, $\sqrt{s'}$, to the region of the $\PZ$
resonance, the cross-section is greatly enhanced. A separation can be made
between these radiative events and non-radiative events for which
$\sqrt{s'}\simeq\sqrt{s}$. While the properties of radiative events are
similar to those measured in $\PZ$ decays at \LEPone, modified only by
the boost due to recoil against hard initial-state radiation, non-radiative
events have different properties, reflecting the increased
relative importance of photon-exchange processes above the $\PZ$
resonance. At the centre-of-mass energies considered here, the contribution
of the photon-exchange diagram to the cross-section is about four times 
greater than that of the $\PZ$-exchange diagram for $\mumu$ and $\tautau$ 
final states with $\sqrt{s'}\simeq\sqrt{s}$.

The analyses presented here are similar to those already presented at lower 
energies~\cite{bib:OPAL-SM172,bib:OPAL-SM183,bib:OPAL-SM189}. We use 
identical techniques to measure $s'$ 
%,the square of the centre-of-mass energy of the \epem\ system after 
%initial-state radiation, 
and to separate non-radiative events, which have little initial-state
radiation, from radiative return to the Z peak. We define non-radiative 
events as those having $s'/s > 0.7225$, while inclusive measurements 
correspond to $s'/s > 0.01$. We correct our measurements of hadronic, \mumu\ 
and \tautau, but not \epem, events to remove the effect of interference 
between initial- and final-state radiation, as in our previous publications. 
The treatment of the four-fermion contribution to the two-fermion final 
states is similar to that at lower energies. The precise signal definition is
discussed in Section~\ref{sec:theory}. 

While the event selection for hadronic and \epem\ final states is
essentially unchanged from previous analyses, an improvement in the
rejection of cosmic ray events in the \mumu\ final state has led to a 
significant reduction in the uncertainty in the residual background. 
The event selection for \tautau\ events has been tightened to reduce the 
background in this channel, also reducing one of the larger systematic
uncertainties. 
In all channels, the higher luminosity and hence higher statistics now 
available have enabled a more thorough study of systematic effects. 
Combining data from three years has led to a significant reduction of the 
experimental systematic uncertainties compared with previous analyses;
for example, the systematic error on the non-radiative hadronic
cross-section has been reduced by $\sim$40\%. To take advantage of these
reduced systematic errors in fits to the Standard Model and searches for
new physics we present updated results for the high statistics data at 
189~GeV, together with new results at higher energies; the 189~GeV results 
supersede those presented in~\cite{bib:OPAL-SM189}.  

Measurements of fermion-pair production up to 189~GeV have shown very 
good agreement with Standard Model
expectations~\cite{bib:OPAL-SM172,bib:OPAL-SM183,bib:OPAL-SM189,bib:ADL-SM}. 
Here we repeat our measurement of the electromagnetic coupling constant 
\alphaem($\sqrt{s}$) including the higher energy data. In addition, we
combine results at energies above the \PZ\ peak (\LEPtwo) with those from data 
taken around the \PZ\ peak (\LEPone) to determine the mass of the \PZ\ boson 
and the size of the $\gamma$--\PZ\ interference contribution within the 
framework of the S-matrix formalism~\cite{bib:smatrix}. 
Including data at higher energies also allows us to extend the searches for 
new physics presented in~\cite{bib:OPAL-SM189}. In particular we obtain 
improved limits on the energy scale of a possible four-fermion contact 
interaction, and also present limits on the mass of a possible heavy
\zp\ boson.

The paper is organized as follows. In Section~\ref{sec:theory} we discuss
the signal definition, theoretical considerations and the corrections
made to the data to obtain measurements corresponding to this definition. 
The data and Monte Carlo samples used in the analysis are described in 
Section~\ref{sec:dataMC}, while Section~\ref{sec:meas} describes the data 
analysis and the cross-section and asymmetry measurements. 
In Section~\ref{sec:SM} we
compare our measurements to the predictions of the Standard Model
and use them to measure the energy dependence of $\alphaem$. The
S-matrix analysis is presented in Section~\ref{sec:smat} 
and the results of searches for new physics in Section~\ref{sec:new_phys}.

%%-----------------------------------------------------------------------
%\section{Signal Definition}                            \label{sec:theory}
%%-----------------------------------------------------------------------
%-----------------------------------------------------------------------
%\subsection{Initial-final State Photon Interference}   \label{sec:ifsr}
%-----------------------------------------------------------------------
%-----------------------------------------------------------------------
%\subsection{Four-fermion Effects}                      \label{sec:4f}
%-----------------------------------------------------------------------
%-----------------------------------------------------------------------
%\subsection{\bf \epem\ Final States}                   
%-----------------------------------------------------------------------
%\input theory.tex
%%-----------------------------------------------------------------------
\section{Signal Definition}                            \label{sec:theory}
%%-----------------------------------------------------------------------
To make precise tests of the Standard Model, the measurements of two-fermion
processes must be compared with theoretical predictions calculated by, for
example, the semi-analytical program \ZFITTER~\cite{bib:zfitter}. We 
therefore need a signal definition for which the theoretical predictions
can be made, and also which corresponds closely to the experimental 
measurements.  
The definition of the two-fermion signal used in this paper is the same as
in previous publications~\cite{bib:OPAL-SM172,bib:OPAL-SM183,bib:OPAL-SM189}.
For the $\epem$ final state it is described in Section~\ref{sec:theory_ee}
below. For hadronic, $\mumu$ and $\tautau$ final states it is as follows:
\begin{itemize}
\item{} $s'$ is defined as the square of the mass of the \PZ/$\gamma$
        propagator. A `non-radiative' sample of events is defined by
        $s'/s > 0.7225$, while inclusive measurements correspond to
        $s'/s > 0.01$.
\item{} Interference between initial- and final-state radiation makes
        the definition of $s'$ ambiguous. To remove this ambiguity,
        the predicted contribution of interference is subtracted from the
        measured cross-sections, as described in Section~\ref{sec:ifsr}.
\item{} Four-fermion final states with a secondary pair arising from 
        an initial-state photon in an $s$-channel diagram 
        (i.e.\ initial state non-singlet photon diagrams, 
        ISNS$_\gamma$~\cite{bib:LEPMCWS}) 
        are considered to be signal if the primary pair passes the $s'$ cut. 
        Four-fermion final states arising from final-state photon
        diagrams (FS$_\gamma$) are included in the signal.
        This corresponds to definition 1 from~\cite{bib:LEPMCWS} page 346.
        The procedure is discussed in detail in Section~\ref{sec:4f}.
\item{} Cross-section and asymmetry measurements are corrected to full
        $4\pi$ acceptance.   
\end{itemize}
%-----------------------------------------------------------------------
\subsection{Interference between Initial- and Final-state Photons}  
\label{sec:ifsr}
%-----------------------------------------------------------------------
The data include the effects of interference between initial- and 
final-state radiation, which needs to be subtracted from the measurements
to form an unambiguous signal definition for comparison with
theoretical predictions. We have investigated two methods of performing
this subtraction. The first method is identical to that used in previous
analyses, described fully in~\cite{bib:OPAL-SM172}. We define a differential 
`interference cross-section', d$^2\sigma_{\mathrm{IFSR}}$/d$m_{\ffbar}$ d\ct, 
as the difference between the differential cross-section including 
interference between initial- and final-state radiation and that excluding 
interference, 
as calculated by \ZFITTER\footnote{Cross-sections including interference were
calculated by setting the flag INTF=2, those excluding
interference by setting INTF=0.}. The differential interference 
cross-section  may be either positive or negative, depending on the values of 
the cosine of the angle $\theta$ between the fermion and the electron beam 
direction, and the invariant mass of  the fermion pair $m_{\ffbar}$.  
We then estimate the fraction of this cross-section accepted by our selection 
cuts by assuming that, as a function of $\ct$ and $m_{\ffbar}$, its selection 
efficiency
%\begin{equation}
%\epsilon_{\rm{IFSR}}(\ct,m_{\ffbar}) =
%                               \epsilon_{\rm{noint}}(\ct,m_{\ffbar}), 
%\label{eq:ifsreff}
%\end{equation} 
$\epsilon_{\rm{IFSR}}(\ct,m_{\ffbar})$ is equal to
$\epsilon_{\rm{noint}}(\ct,m_{\ffbar})$, 
where $\epsilon_{\mathrm{noint}}$ has been determined from Monte Carlo events 
which do not include interference. The selected interference cross-section 
is then subtracted from the measured cross-section before efficiency 
correction, as for other backgrounds. As the accepted cross-section is 
estimated as a function of $\ct$, the correction is easily applied to 
total cross-sections, angular distributions or asymmetry measurements. 

The second method investigated uses special samples of Monte Carlo events
generated with the \KK\ program~\cite{bib:KK2f}. These samples were
generated including initial-final-state photon interference with event weights
allowing them to be reweighted to exclude the effects of interference. 
Applying selection cuts to these samples allowed the accepted `interference
cross-section' to be determined. The accepted cross-sections were in
good agreement with those derived from the first method. In the case of
hadronic events, the average of the two methods was used to correct the
data and half the difference between them taken as the associated systematic
error. For \mumu\ and \tautau\ final states 
%insufficient Monte Carlo statistics were available to allow use of the 
%second method, and as in previous publications 
the first method was used. As at lower energies, the 
systematic error was assessed by repeating the estimate assuming that the 
efficiency in each bin of \ct\ and $m_{\ffbar}$ was 
increased by half its difference with respect to the efficiency at 
$m_{\ffbar}=\sqrt{s}$ at the same \ct.
%the average of the efficiency in that bin and the efficiency in the bin 
%including $m_{\ffbar}=\sqrt{s}$ at the same \ct. 

The corrections to the final measured cross-sections and asymmetry
measurements are given in Table~\ref{tab:ifsr}. 

%-----------------------------------------------------------------------
\subsection{Four-fermion Effects}                      \label{sec:4f}
%-----------------------------------------------------------------------
The treatment of the four-fermion contribution to the two-fermion final 
states is similar to that at lower energies. Secondary pairs arising from
initial-state photons in $s$-channel diagrams (ISNS$_\gamma$) are 
considered to be signal if the primary pair satisfies the $s'$ cut. 
Pairs arising from final-state photons (FS$_\gamma$) are always considered 
to be signal. The overall 
efficiency of event selection cuts $\epsilon$ is calculated as 
\begin{equation}
  \epsilon = \left( 1 -
    \frac{\sigma_{\ffbar\fpfp}}{\sigma_{\rm{tot}}} \right)\epsilon_{\ffbar}
  + \frac{\sigma_{\ffbar\fpfp}}{\sigma_{\rm{tot}}}\epsilon_{\ffbar\fpfp}
\end{equation}
where $\epsilon_{\ffbar}$ and $\epsilon_{\ffbar\fpfp}$ are the efficiencies 
derived from two-fermion and four-fermion signal Monte Carlo events
respectively, $\sigma_{\ffbar\fpfp}$ is the generated four-fermion signal 
cross-section, and $\sigma_{\mathrm{tot}}$ is the total cross-section
from \ZFITTER\ including pair emission. Using this definition of efficiency,
effects of cuts on soft pair emission in the four-fermion generator
are correctly summed with vertex corrections involving virtual pairs.
For these analyses, a change has been made to the method used to separate the 
signal contribution from the background contribution in the four-fermion
Monte Carlo events. At lower energies, separate samples of $s$-channel and 
$t$-channel four-fermion Monte Carlo events were generated, and the signal 
contribution was defined by kinematic cuts on the $s$-channel events,
designed to include pair production via an initial-state photon but to
exclude pair production via a Z boson. In this analysis,
Monte Carlo samples including all four-fermion diagrams, generated
with either the grc4f~\cite{bib:grc4f} program or with 
\KORALW~\cite{bib:koralw} with grc4f matrix elements, were used. 
Each event was given a weight to be signal (or background) calculated using 
the matrix elements of the appropriate diagrams. This method gives a 
definition of signal which is closer to that employed in the semi-analytic 
calculations with which we compare our results, and also avoids the necessity 
of generating special Monte Carlo samples.

The inclusion of the four-fermion part of the signal reduces efficiencies
by about 0.3\% for inclusive hadrons, 0.8\% for inclusive muons and
1\% for inclusive taus. For non-radiative events the effects are much
smaller, less than 0.02\% for hadrons and around 0.2\% for muons and taus. 

%-----------------------------------------------------------------------
\subsection{\boldmath \epem\ Final States}
\label{sec:theory_ee}                       
%-----------------------------------------------------------------------
The discussion above applies to hadronic, \mumu\ and \tautau\ final
states. Because of ambiguities arising from the $t$-channel contribution, 
the acceptance for the \epem\ final state is defined in terms of the angle 
$\theta$ of the electron or positron with respect to the electron beam 
direction and the acollinearity angle $\thacol$ between the electron and 
positron; a cut on $s'$ is not used. It is thus unnecessary to subtract
interference between initial- and final-state radiation to make an
unambiguous signal definition. Cross-sections and asymmetries for 
\epem\ are not corrected for interference between initial- and final-state 
radiation; they are compared to theoretical predictions which include 
interference. In principle the $t$-channel process with a second fermion pair 
arising from the conversion of a virtual photon emitted from an initial- or 
final-state electron should be included as signal, as well as the $s$-channel
diagrams. 
%Since this process is not included in any program we use for 
%comparison, we simply ignore such events: they are not treated as background 
%as this would underestimate the cross-section.
Diagrams with real pairs are included in four-fermion Monte Carlo generators, 
but diagrams with virtual pairs are not included, and neither real nor
virtual pairs are included in the program we use for comparison with the data.
It would be improper to treat real pairs alone either as signal or background.
We choose the best alternative, and simply ignore such events in 
both efficiency and background calculations.
If the efficiency for four-fermion events is similar to that for two-fermion
events, as is expected to be the case here, we are effectively comparing
data including the four-fermion contribution to theory without. This does
not introduce a large error because real and virtual pair contributions have 
opposite sign, and thus their effects tend to cancel in any total 
cross-section.

%%-----------------------------------------------------------------------
%\section{Data and Monte Carlo Simulations}             \label{sec:dataMC}
%%-----------------------------------------------------------------------
%-----------------------------------------------------------------------
%\subsection{Data}                                      \label{sec:data}
%-----------------------------------------------------------------------
%-----------------------------------------------------------------------
%\subsection{Monte Carlo Simulations}                  \label{sec:MC}
%-----------------------------------------------------------------------
%\input dataMC.tex
%%-----------------------------------------------------------------------
\section{Data and Monte Carlo Simulations}             \label{sec:dataMC}
%%-----------------------------------------------------------------------

%-----------------------------------------------------------------------
\subsection{Data}                                      \label{sec:data}
%-----------------------------------------------------------------------
The OPAL detector\footnote{OPAL uses a right-handed coordinate system in
which the $z$ axis is along the electron beam direction and the $x$
axis is horizontal. The polar angle $\theta$ is measured with respect
to the $z$ axis and the azimuthal angle $\phi$ with respect to the
$x$ axis.}, trigger and data acquisition system are fully described 
elsewhere~\cite{bib:OPAL-detector,bib:OPAL-SI,bib:OPAL-lumi,bib:OPAL-TR,
bib:OPAL-DAQ}. The high redundancy of the trigger system leads to
negligible trigger inefficiency for all channels discussed here.

\begin{figure}
\begin{center}
\epsfxsize=\textwidth
 %\epsfbox{lumi.eps}
\epsfbox{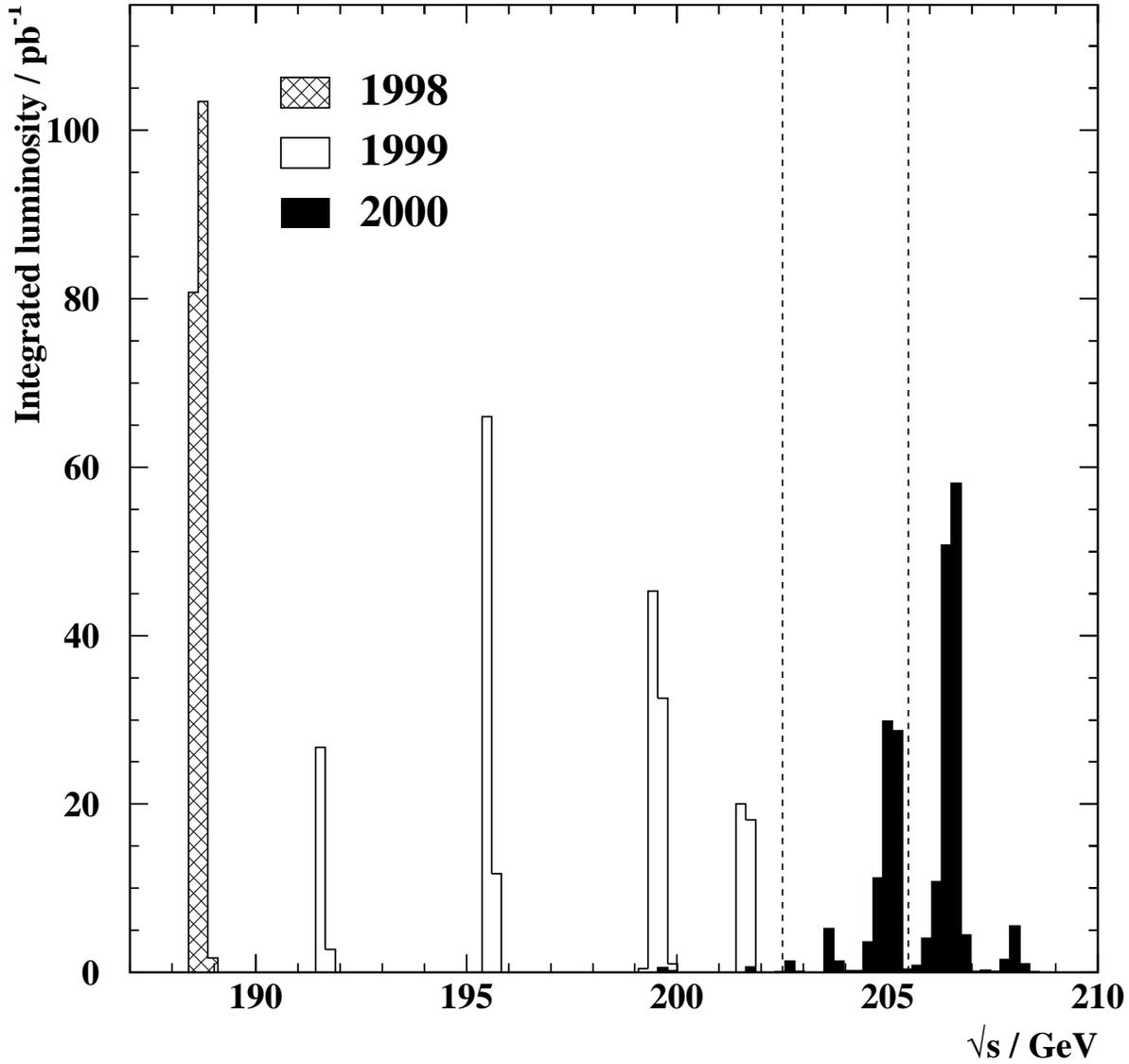}
\caption
{
  Integrated luminosity collected by OPAL, and used in these analyses, 
  during 1998, 1999 and 2000. The dashed lines indicate the division of 
  the 2000 data into the two centre-of-mass energy bins, 
  202.5~GeV$< \sqrt{s} < $205.5~GeV and $\sqrt{s} >$205.5~GeV.
  The precise amount of data used in each analysis varies slightly from 
  channel to channel.
}
\label{fig:lumi}
\end{center}
\end{figure}

The analyses presented in this paper use data recorded during 1998,
1999 and 2000. The 1998 data were recorded at a centre-of-mass energy
near 189~GeV. In 1999, data were taken at four different centre-of-mass 
energy points, close to 192~GeV, 196~GeV, 200~GeV and 202~GeV. 
In 2000, a small amount of data was taken at centre-of-mass energies near
200~GeV and 202~GeV; these data have been included with the 200~GeV and
202~GeV data sets taken in 1999.
The bulk of the 2000 data was taken at a range of centre-of-mass energies 
between 203~GeV and 209~GeV, as shown in Fig.\ref{fig:lumi}. Also, in 2000,
the beam energy was changed within a run by a series of 
`miniramps', resulting in a broad distribution of centre-of-mass energies
rather than a series of discrete values as in previous years. Data taken 
while the beam energy was changing and other data with a poor beam energy 
measurement have been removed from these analyses. These data amount to 
about 1.1\% of the total 2000 data set. For analysis the good data have been 
divided into two centre-of-mass energy ranges: 202.5~GeV -- 205.5~GeV and 
$>$ 205.5~GeV (henceforth referred to as 205~GeV and 207~GeV respectively); 
these are 
the energy ranges used for the combination of data from all LEP experiments.
The mean centre-of-mass energy and approximate total integrated luminosity 
collected at each energy point are shown in Table~\ref{tab:lumi}; the
actual amount of data varies slightly from channel to channel because of
differing requirements on data quality.

%-----------------------------------------------------------------------
\subsection{Monte Carlo Simulations}                  \label{sec:MC}
%-----------------------------------------------------------------------
The estimation of efficiencies and background processes makes extensive 
use of Monte Carlo simulations of many different final states. For studies
of $\epem\to\qqbar$ we used the \KK~\cite{bib:KK2f} program,
version 4.13. In \KK\ photon radiation is modelled using Coherent Exclusive 
Exponentiation (CEEX) and complete ${\cal O}(\alpha^{2})$ matrix elements for 
initial-state radiation are included. Hadronization was performed according 
to the \PYTHIA6.150~\cite{bib:pythia} string model. Samples hadronized with 
the \HERWIG6.2~\cite{bib:herwig} cluster model or 
\ARIADNE4.11~\cite{bib:ariadne}
colour dipole model were used for systematic studies. In all cases input 
parameters have been optimized by a study of global event shape variables 
and particle production rates in \PZ\ decay data~\cite{bib:OPAL-tune}.
Final-state radiation from quarks was simulated as part of the hadronization
process, and not by the \KK\ program.
For $\eetoee$ we used the \BHWIDE1.04~\cite{bib:bhwide} Monte Carlo  program, 
and for $\eetomumu$ and $\eetotautau$ \KK\ was used with 
\KORALZ4.0~\cite{bib:koralz} for comparison.

Four-fermion events were modelled with the grc4f~\cite{bib:grc4f} generator
or with the \KORALW~\cite{bib:koralw} program with grc4f matrix elements.
The latter has superior modelling of initial-state radiation in channels
without electrons. Final states containing quarks were hadronized using 
\PYTHIA, with \HERWIG\ and \ARIADNE\ used for systematic studies, as for the 
$\epem\to\qqbar$ events. Two-photon background 
processes with hadronic final states were simulated using \PYTHIA\ and
\PHOJET~\cite{bib:phojet} at low  $Q^2$. At high $Q^2$ the
\TWOGEN~\cite{bib:twogen} program (with the `perimiss'
 option~\cite{bib:OPAL-f2gam}), \HERWIG\ and \PHOJET\ were used.
In the following, the terms `tagged' and `untagged' are used to denote
the high- and low-$Q^2$ samples respectively.
The BDK generator~\cite{bib:bdk} was used to simulate two-photon
processes resulting in $\epem\mumu$ and $\epem\tautau$ final states,
while the Vermaseren generator~\cite{bib:vermaseren} was used
for the $\epem\epem$ final state. The
$\eetogg$ background in the \Pep\Pem\ final state was modelled with
the \RADCOR~\cite{bib:radcor} program, while the contribution from
$\epem\gamma$ where the photon and one of the charged particles are
inside the detector acceptance was modelled with
\TEEGG~\cite{bib:teegg}.  

Monte Carlo samples were generated at 189~GeV, at the four nominal energy 
values of the data collected in 1999, and at several energies spanning the 
range 204~GeV to 208~GeV for simulation of the data taken in 2000.
All samples were processed through the OPAL detector simulation 
program~\cite{bib:gopal} and reconstructed in the same way as for
real data. Efficiencies and backgrounds at the centre-of-mass energy values 
corresponding to the data were determined by fitting the energy dependence 
of these quantities.
  
For the measurement of the luminosity, the cross-section for 
small-angle Bhabha scattering was calculated using the Monte Carlo program 
\BHLUMI~\cite{bib:bhlumi}, using generated events processed through a 
program which parameterizes the response of the 
luminometer~\cite{bib:OPAL-lumi}.

%%-----------------------------------------------------------------------
\section{Cross-section and Asymmetry Measurements}     \label{sec:meas}
%%-----------------------------------------------------------------------
%-----------------------------------------------------------------------
%\subsection{Luminosity}                                \label{sec:lumi} 
%-----------------------------------------------------------------------
%\input lumi.tex
%-----------------------------------------------------------------------
\subsection{Luminosity}                                \label{sec:lumi} 
%-----------------------------------------------------------------------
The integrated luminosity was measured using small-angle Bhabha scattering 
events, $\epem\to\epem$, recorded in the silicon-tungsten 
luminometer~\cite{bib:OPAL-lumi}. The luminometer consisted of two finely
segmented silicon-tungsten calorimeters placed around the beam pipe, 
symmetrically on the left and right sides of the OPAL detector, 2.5~m from
the interaction point. Each calorimeter covered angles from the beam between 
25~mrad and 59~mrad. The luminosity determination closely followed the 
procedure used for the precise determination at \LEPone~\cite{bib:OPAL-lumi}. 
However, before \LEPtwo\ data-taking, tungsten shields designed to protect the 
tracking detectors from synchrotron radiation were installed. These introduced 
about 50~radiation lengths of material in front of the calorimeter
between 26~mrad and 33~mrad from the beam axis, thus reducing the useful 
acceptance of the detector.

\begin{figure}
\begin{center} 
\epsfxsize=\textwidth
 %\epsfbox{lumi00.eps}
\epsfbox{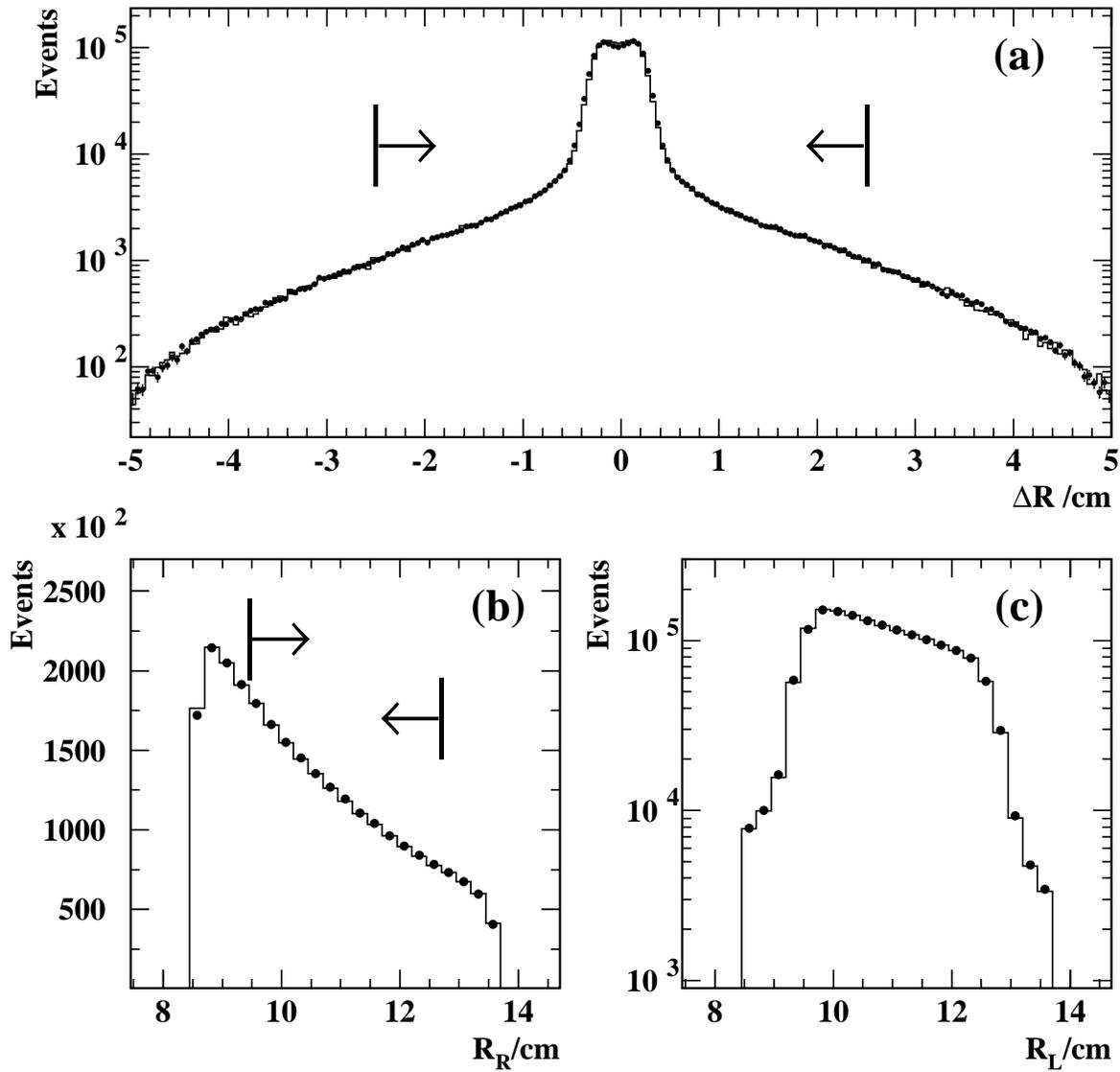}
\caption
{
  (a) The distribution of the difference in radial coordinate between
  the two clusters in Bhabha scattering events used for the silicon-tungsten
  luminosity measurement. Distributions of the radial coordinates of clusters 
  are shown for (b) the `narrow' side and (c) the `wide' side calorimeter.
  Distributions are shown after all cuts except the acollinearity cut in (a) 
  and the inner and outer radial acceptance cuts, on that side, in (b). 
  Points show the data taken in the year 2000, while the histograms show the 
  Monte Carlo expectation. The vertical bars show the positions of the cuts 
  which define the acceptance, with the arrows pointing into the accepted 
  region.
  %while the dashed lines in (c) indicate the `wide' side acceptance cuts.
}
\label{fig:swlumi}
\end{center}
\end{figure}

Bhabha scattering events were selected by requiring a high energy cluster in 
each side of the detector, using asymmetric acceptance cuts. The energy in 
each calorimeter had to be at least half the beam energy, and the total 
energy in the fiducial region of both calorimeters had to be at least three 
quarters of the centre-of-mass energy.
The two highest energy clusters were required to be back-to-back in $\phi$,
\mbox{$||\phi_{R} - \phi_{L}| - \pi| <$ 200~mrad}, 
where $\phi_{R}$ and $\phi_{L}$
are the azimuthal angles of the cluster in the right- and left-hand 
calorimeter respectively. They were also required to be collinear, by
placing a cut on the difference between the radial positions, 
$\Delta R\equiv R_{R} - R_{L}$, at $|\Delta R| <$ 2.5~cm, where $R_{R}$ and 
$R_{L}$ are the 
radial coordinates of the clusters on a plane approximately 7 radiation 
lengths into the calorimeter at $z = \pm 246.0225~\cm$. This cut, 
corresponding to an acollinearity angle of about 10.4~mrad, effectively 
defines the acceptance for 
single-photon radiative events, thus reducing the sensitivity of the 
measurement to the detailed energy response of the calorimeter. The 
distribution of $\Delta R$ for the data taken in the year 2000 is shown in 
Fig.~\ref{fig:swlumi}(a).

Inner and outer radial acceptance cuts delimited a region between 38~mrad and 
52~mrad on one side of the calorimeter, while for the opposite calorimeter 
a wider zone between 34~mrad and 56~mrad was used. Two luminosity 
measurements were formed with the narrower acceptance on one side or the 
other side. The final measurement was the average of the two and has no first
order dependence on beam offsets or tilts.  The distributions of the radial 
coordinates of the clusters for the data taken in the year 2000 are shown in 
Fig.~\ref{fig:swlumi}(b,c).

\begin{figure}
\begin{center}
\epsfxsize=\textwidth
 %\epsfbox{pbimage.eps}
\epsfbox{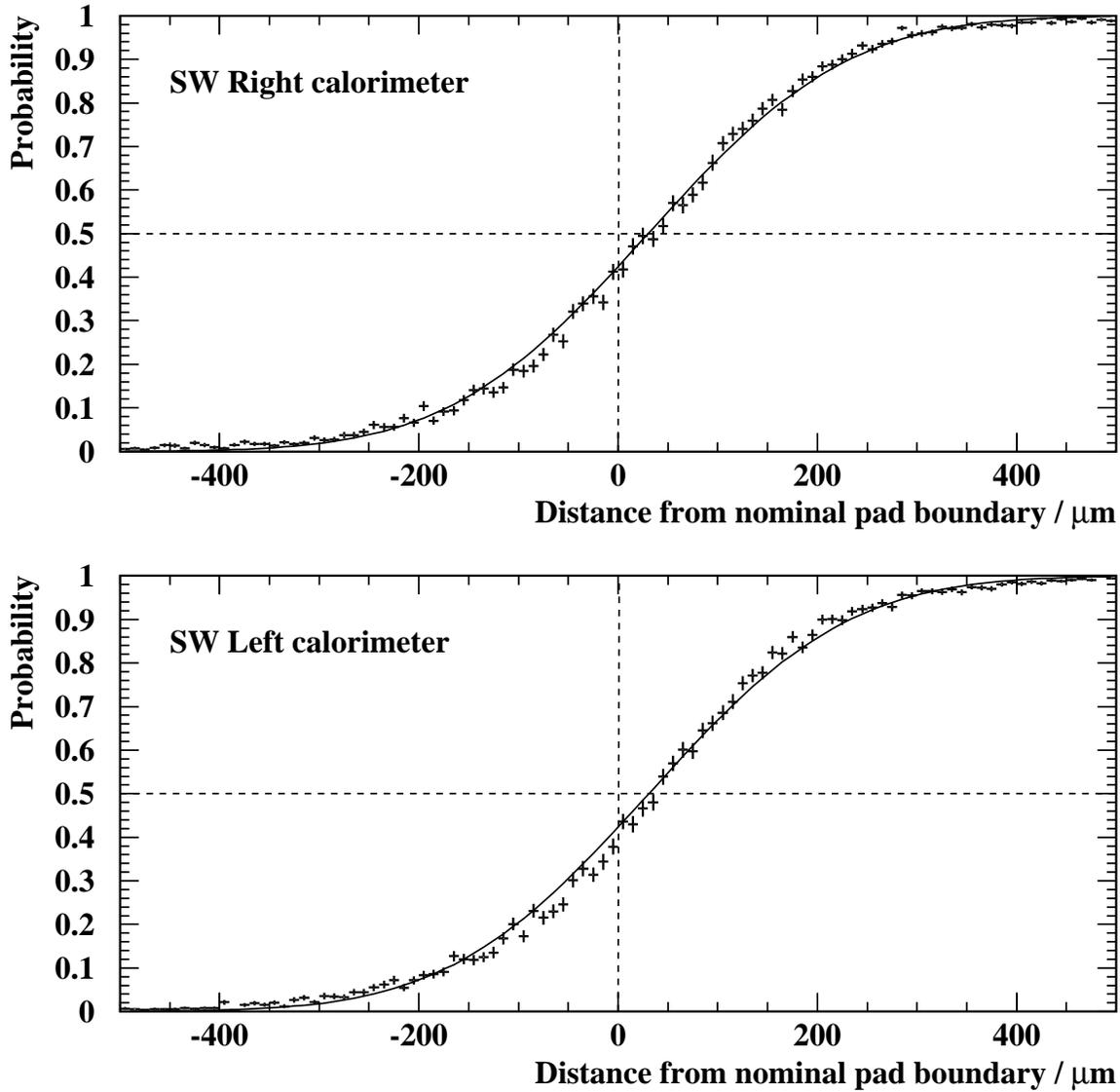}
\caption
{
  The pad boundary image at the inner acceptance cut ($R_{\rm in}$ = 9.45~cm) 
  used in the luminosity measurement. The nominal pad boundary is 
  conventionally set at zero. The points show the fraction of events with 
  pad maximum beyond the nominal cut as a function of distance from the pad 
  boundary for the layer located after 7 radiation lengths, for data taken
  in 2000. The solid curves show the fitted functions used to determine the 
  coordinate offsets. 
}
\label{fig:pbimage}
\end{center}
\end{figure}

The acceptance A of the luminosity measurement is affected by any change 
in the inner and outer edges of the acceptance as follows:

\begin{equation}
    \frac{\Delta A}{A} 
    \approx -
    \frac{\Delta R_{\mathrm{in}}}{21 ~\mu \mathrm{m}}\times 10^{-3}
\label{eq:rinner}
\end{equation}
    and
\begin{equation}
    \frac{\Delta A}{A} 
    \approx +
    \frac{\Delta R_{\mathrm{out}}}{51~\mu \mathrm{m}}\times 10^{-3},
\label{eq:router}
\end{equation}
where $R_{\rm in}$ and $R_{\rm out}$ denote the radial coordinates of the
inner and outer cuts. 
The coefficients in the expressions given above are determined by simple
analytic calculations, using the $1/\theta^3$ Bhabha spectrum, the nominal
half distance between the reference planes of the two calorimeters and the 
inner and outer acceptance radii (9.45~cm and 12.7~cm).
The residual bias on the inner and outer cut positions was estimated as at 
\LEPone, by a procedure called {\em anchoring}, which is fully explained in 
\cite{bib:OPAL-lumi}.
In this approach the fundamental tool is the radial position of the silicon 
pad with maximum signal in a given longitudinal layer. 
As the radial position of the incoming particles crosses a radial pad 
boundary in a single layer, the average pad-maximum moves rapidly from one 
pad to the next, giving an image of the pad boundary, as shown in 
Fig.~\ref{fig:pbimage}.
The coordinate offset at the inner cut position was determined to be 
about 30~$\mu$m in both the right and the left calorimeter; this offset
can be seen in Fig.~\ref{fig:pbimage}.
By applying Equation~(\ref{eq:rinner}) this is equivalent to an acceptance 
variation of +0.14~\%.
At \LEPone\ the absolute value of the 
coordinate offset at the inner cut was less than 10~$\mu$m.
The larger value in \LEPtwo\ data is attributed to the effects of about 
2~radiation lengths of preshowering material, consisting of cables and beam 
pipe support structures, in front of the central angular region of the
calorimeter including the position of the inner radial cut.
The estimated systematic correction was applied to the measurement, but the 
full size of the effect was conservatively kept as a systematic error.

The errors on the luminosity measurement at each energy are summarized in
Table~\ref{tab:lumi_errors}. The experimental systematic error is 
dominated by the uncertainty of the inner radial cut.
Among the other errors are trigger efficiency (0.06\%), energy response 
particularly in the low energy tail (0.03\%), beam parameters (0.02\%),
backgrounds (0.02\%) and Monte Carlo statistics (0.08\%), where all the 
numbers refer to year 2000 data but are fairly similar in the other data 
samples. The error on the theoretical prediction of the Bhabha cross-section 
of 0.12\% is taken from~\cite{bib:bhlumi_err}.

%The errors on the luminosity measurement at each energy are shown in
%Table~\ref{tab:lumi_errors}. The experimental systematic errors are 
%dominated by the uncertainty of the inner radial cut. In contrast to \LEPone, 
%this is located behind about 2~radiation lengths of preshowering material 
%consisting of cables and beam pipe support structures. This extra material 
%produces a bias of about 30~$\mu$m on the cut position, equivalent to an 
%acceptance variation of +0.14\%. This is corrected but the full size of the 
%effect is conservatively kept as a systematic error. Among the other errors 
%are trigger efficiency (0.06\%), energy response particularly in the low 
%energy tail (0.03~\%), beam parameters (0.02\%), backgrounds (0.02\%) 
%and Monte Carlo statistics (0.08\%), where all the numbers refer to year 
%2000 data but are fairly similar in the other data samples. The error on the
%theoretical prediction of the Bhabha cross-section of 0.12\% is taken
%from~\cite{bib:bhlumi_err}.

The errors on luminosity are included in the systematic errors on 
cross-section measurements presented in this paper. Correlations between 
cross-section measurements arising from common errors in the luminosity have 
been taken into account in the interpretation of the results.

%-----------------------------------------------------------------------
%\subsection{Hadronic Events}                            \label{sec:mh}
%-----------------------------------------------------------------------
%\input mh.tex
%-----------------------------------------------------------------------
\subsection{Hadronic Events}                              \label{sec:mh}
%-----------------------------------------------------------------------
\subsubsection{Event Selection}
\begin{figure}
\begin{center}
\epsfxsize=\textwidth
 %\epsfbox{l2mh189to207_pr.eps}
\epsfbox{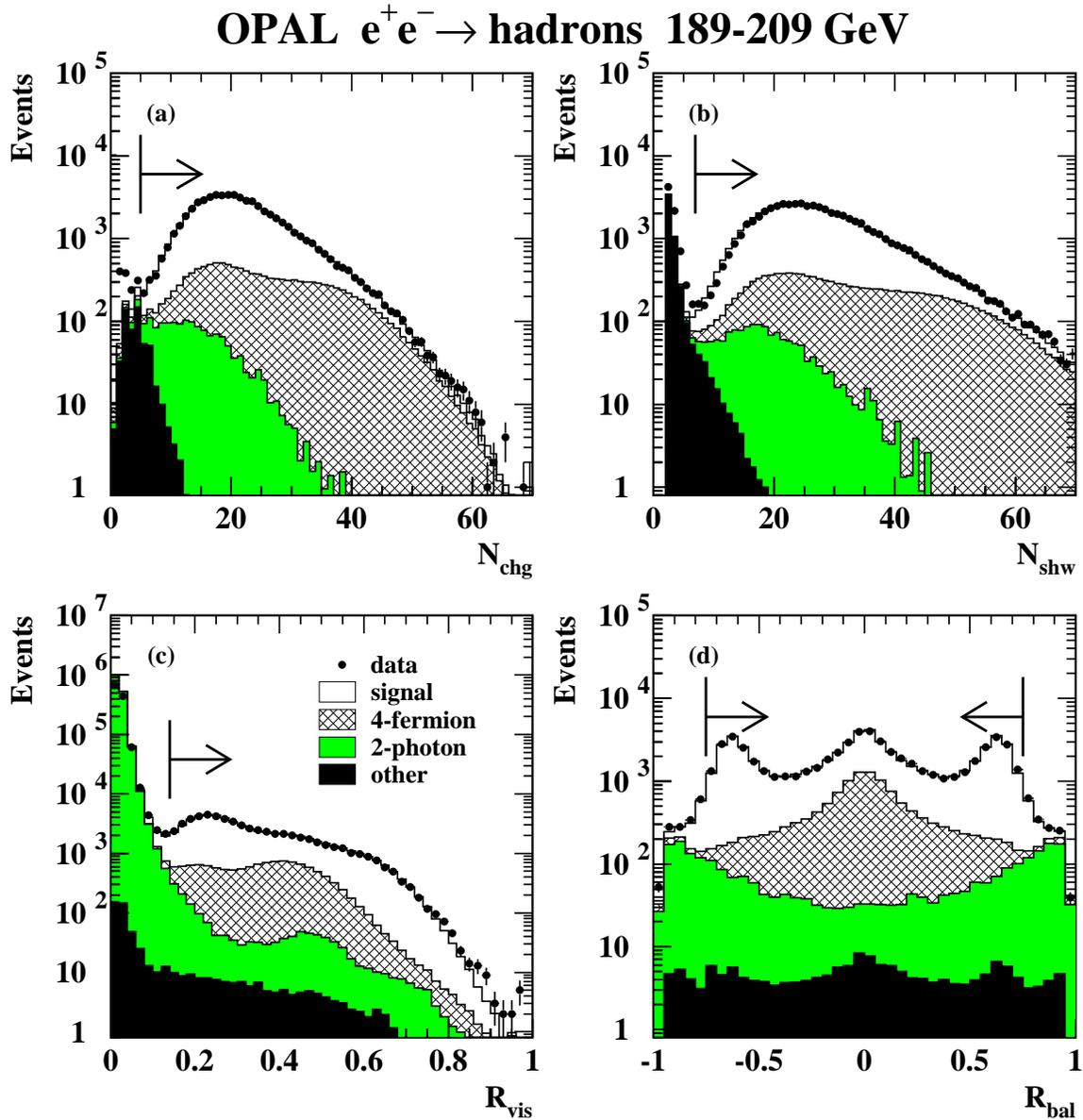}
\caption
{ Distributions of variables used in the selection of hadronic events: 
  (a) number of tracks, (b) number of electromagnetic calorimeter clusters, 
  (c) ratio of the visible energy to the centre-of-mass energy and 
  (d) energy balance along the beam direction.
  The points show the data for all centre-of mass energies combined and the 
  histograms the Monte Carlo predictions normalized to the integrated 
  luminosity of the data. In each case the distribution is shown after 
  the selection cuts associated with the other three variables have been
  applied. The positions of these cuts are indicated by the vertical bars,
  with the arrow pointing into the accepted region. The \WW\ rejection
  cuts have not been applied. Background labelled `other' is mainly 
  \tautau.
}
\label{fig:mh_dists}
\end{center}
\end{figure}

%A detailed description of the analysis of hadronic final states can be
%found in~\cite{bib:Chris}.
The selection of hadronic events is identical to previous
analyses~\cite{bib:OPAL-SM172,bib:OPAL-SM183,bib:OPAL-SM189}. For
both inclusive and non-radiative samples, the selection efficiency 
is typically $\sim$85\% and the purity is $\sim$92\%.
\begin{itemize}
 \item{}
  To reject leptonic final states, events were required to have high 
  multiplicity: at least 7~electromagnetic clusters and at least 5~tracks
  satisfying standard quality criteria~\cite{bib:OPAL-SM172}.
 \item{} Background from
  two-photon events was reduced by requiring a total energy deposited in 
  the electromagnetic calorimeter of at least 14\% of the centre-of-mass 
  energy: $ \Rvis \equiv {\Sigma \Eclus }/\sqrt{s} > 0.14, $
  where $\Eclus$ is the energy of each cluster.
 \item{} 
  Any remaining background from beam-gas and beam-wall interactions was 
  removed, and two-photon events further reduced,
  by requiring an energy balance along the beam direction which satisfied
  \mbox{$ \Rbal \equiv \mid \Sigma (\Eclus \cdot \cos \theta_{\rm clus}) \mid /
  \Sigma \Eclus < 0.75$}, where $\theta_{\rm clus}$ is the polar angle of 
  the cluster.
 \item
  At these centre-of-mass energies, the cross-section for production
  of \WW\ events is comparable to that for non-radiative $\qqbar$ events,
  and the above selection cuts have high efficiency for those
  \WW\ events with hadrons in the final state.
  Events selected as \WW\ candidates using the criteria described 
  in~\cite{bib:OPAL-WW189} (with reference histograms updated for the
  higher energy data) were therefore rejected. This cut also
  removes some of the (much smaller) contribution from $\PZ\PZ$ final states.
  As a cross-check, an analysis was also performed in which \WW\ candidates 
  were not rejected, but their expected contribution subtracted.
 \item
  The effective centre-of-mass energy, $\sqrt{s'}$, of the \epem\ collision,
  determined as described below, was required to satisfy $s'/s > 0.01$
  for the inclusive sample and $s'/s > 0.7225$ for the non-radiative 
  sample.
\end{itemize}
Distributions of selection variables, before applying the \WW\ rejection
and $s'$ cuts, are shown in Fig.~\ref{fig:mh_dists}. The Monte Carlo 
modelling of these variables is generally very good, except at very low 
multiplicities. 

\begin{figure}
  \epsfxsize=\textwidth
  %\epsfbox[0 0 567 567]{sprovers.eps}
\epsfbox{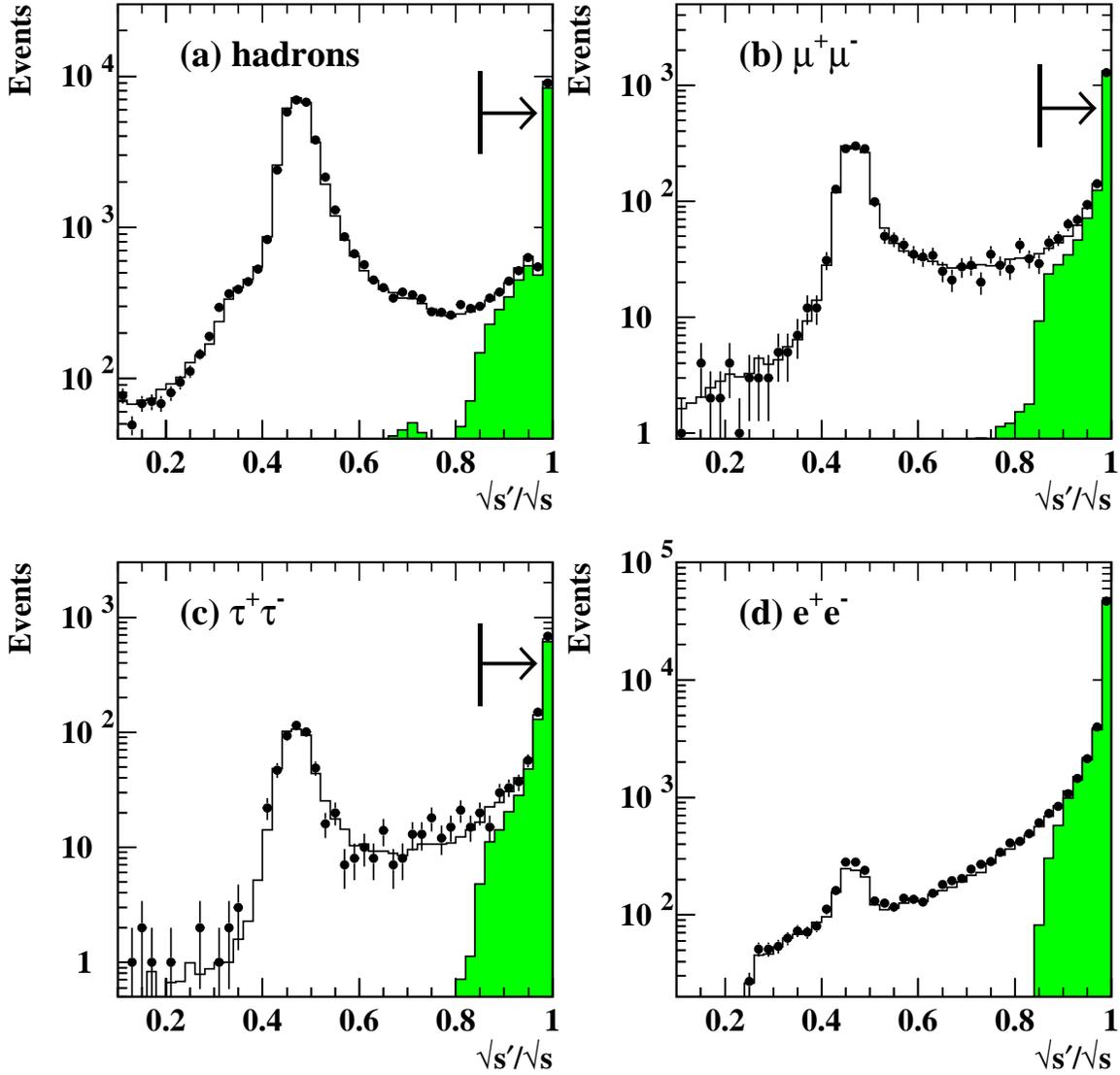}
  \caption
{
 The distributions of reconstructed $\protect\sqrt{s'/s}$ for 
 (a) hadronic events, (b) $\mumu$ events, (c) $\tautau$ events and
 (d) $\epem$ events with $\absctepem < 0.9$ and $\thacol < 170\degree$, 
 for all data combined. 
 In each case, the points show the data and the histogram the Monte Carlo 
 prediction, normalized to the integrated luminosity of the data, with the 
 contribution from events with true $s'/s > 0.7225$ shaded in (a), (b) and (c),
 and the contribution from events with $\thacol < 10\degree$ shaded in (d).
 The vertical bars in (a), (b) and (c) show the position of the cut used to
 select `non-radiative' events.
}
\label{fig:sp}
\end{figure}

The effective centre-of-mass energy $\sqrt{s'}$ of the \epem\ collision
was estimated as follows. The method is the same as that used in 
previous analyses~\cite{bib:OPAL-SM172,bib:OPAL-SM183,bib:OPAL-SM189}.
Isolated photons in the electromagnetic calorimeter, with a minimum
energy of 10~GeV, 
were identified, and the remaining tracks, electromagnetic and hadron
calorimeter clusters formed into jets using the Durham ($k_{T}$) 
scheme~\cite{bib:durham} with a jet resolution parameter $y_{\rm cut}=0.02$.
If more than four jets were found the number was forced to be four by
adjusting the jet resolution parameter. The jet energies and angles were 
corrected for double counting of energy using the 
algorithm described in~\cite{bib:MT}. The jets and observed photons 
were then subjected to a series of kinematic fits imposing the constraints of 
energy and momentum conservation, in which zero, one, or two additional 
photons emitted close to the beam direction were allowed. The fit with 
the lowest number of extra photons which gave an acceptable $\chi^2$ was 
chosen. The value of $\sqrt{s'}$ was then computed from the fitted 
four-momenta of the jets, i.e.\ excluding photons identified in the 
detector or those close to the beam direction resulting from the fit,
which were assumed to arise from initial-state radiation. If none of the 
kinematic fits gave an acceptable $\chi^{2}$, $\sqrt{s'}$ was estimated 
directly from the angles of the jets as in~\cite{bib:OPAL-SM130}. The 
distribution of $\sqrt{s'}$ for all energies combined is shown in 
Fig.~\ref{fig:sp}(a). 

The efficiency of the selection cuts was determined from Monte Carlo events
generated with the \KK\ program, without inclusion of interference between
initial- and final-state photon radiation, and corrected for the effect
of the four-fermion signal component as described in Section~\ref{sec:4f}.
The feedthrough of events with lower $s'$ into the non-radiative sample
and expected backgrounds were also determined from Monte Carlo. For both
efficiencies and backgrounds a linear fit to values at centre-of-mass energies 
between 189~GeV and 208~GeV was used to determine the value at the mean 
centre-of-mass energy of the data. The \KORALW\ Monte Carlo events used
to estimate the four-fermion background do not include complete
electroweak $\cal{O}(\alpha)$ corrections to the $\eetoWW$ process which
are now available in the program \KANDY~\cite{bib:kandy}. The backgrounds
calculated using \KORALW\ were corrected using samples of events generated
with \KANDY. These corrections are small for the standard analysis, where
\WW\ candidates are rejected, but significant for the cross-check analysis in
which \WW\ events are not rejected. In this case they increase the measured
inclusive and non-radiative cross-sections by 0.5\% and 1.2\% respectively.
Efficiencies and backgrounds are summarized in 
%Tables~\ref{tab:eff1} and~\ref{tab:eff2}.
Table~\ref{tab:eff1}.

To measure the angular distribution of the primary quark in the hadronic
events, we have used as an estimator the thrust axis for each event
determined from the observed tracks and clusters. The angular distribution
of the thrust axis was then corrected to the primary quark level using
bin-by-bin corrections determined from Monte Carlo events. For the bin
size chosen, the bin-to-bin migration of events is $\sim$10\% to either
side. No attempt was made to identify the charge in hadronic events, and thus 
we measured the folded angular distribution.

\subsubsection{Systematic Uncertainties}
The systematic errors on the hadronic cross-sections are summarized in
Table~\ref{tab:sys_all}, with a detailed breakdown at 200~GeV given in
Table~\ref{tab:mh_syserr}. Where no dependence on energy or year was expected
or seen, the values were determined by combining data at all energies. The 
resulting high statistics have resulted in a reduction of many contributions 
compared with previous analyses.

{\bf Initial-state radiation modelling.} Efficiencies were calculated using 
the \KK\ generator with ${\cal O}(\alpha^2)$ Coherent Exclusive Exponentiation
(CEEX) of radiation. To assess the effect of initial-state radiation on the 
selection efficiencies and $s'$ determination, the events were reweighted to 
${\cal O}(\alpha)$ CEEX. In accordance with the recommendations 
of~\cite{bib:KK2f}, half the difference between ${\cal O}(\alpha)$ and 
${\cal O}(\alpha^2)$ was assigned as the systematic error, reflecting the 
effects of missing higher order terms in the perturbative expansion.

{\bf Fragmentation modelling.} The effect of the hadronization model on
the selection efficiencies has been investigated by comparing the string model
implemented in \PYTHIA\ with the cluster model of \HERWIG~\cite{bib:herwig} and
the colour-dipole model of \ARIADNE~\cite{bib:ariadne}. To reduce the 
statistical errors on this comparison, the same primary quarks generated 
with \KK\ were fragmented according to each model in turn, and the selection 
efficiencies compared. The deviations of the two predictions from the \PYTHIA\ 
value were evaluated. Statistically significant differences were seen, and 
the larger of these was assigned as the systematic error. In addition, the 
effects on the efficiencies of changing the cuts on the number of tracks and 
clusters by one unit were also taken into account, to cover imperfections in 
the modelling of low multiplicity jets. 

{\bf Detector effects.} The selection of inclusive events is mainly based on 
the electromagnetic calorimeter, and is thus sensitive to the energy scale of
the calorimeter, and any angular dependence of the energy scale. For
non-radiative events, the selection is sensitive to jet and photon energies,
angles, and their errors, and jet masses, which are used as input to the 
kinematic fits used to determine $s'$. Studies of 
calibration data taken at the Z peak have been used to determine small 
year-dependent corrections to these parameters in the Monte Carlo simulations.
Variations of these corrections by their errors were used to assign 
corresponding systematic errors on the cross-sections. The uncertainty in the 
energy scale of the electromagnetic calorimeter leads to an error on
the inclusive cross-sections of about 0.2\%. The largest effect on the 
non-radiative cross-sections (0.12\%) arises from the jet energy scale.

{\bf \boldmath $s'$ determination.}
Possible systematic effects in the determination of $s'$ not already covered
by the studies of initial-state radiation modelling, fragmentation modelling
and detector effects were studied using two alternative methods of calculating
$s'$. Firstly the default algorithm was modified to allow only a single 
radiated photon, either in the electromagnetic calorimeter or along the beam 
axis. Alternatively the cuts defining photon candidates in the detector were 
varied. The differences, averaged over all centre-of-mass energies, were
not statistically significant, and the statistical precision of this test
was included as a systematic error associated with the $s'$ determination.

\begin{figure}
\begin{center}
\epsfxsize=\textwidth
 %\epsfbox{w420_189to207_nrplots_310703.eps}
\epsfbox{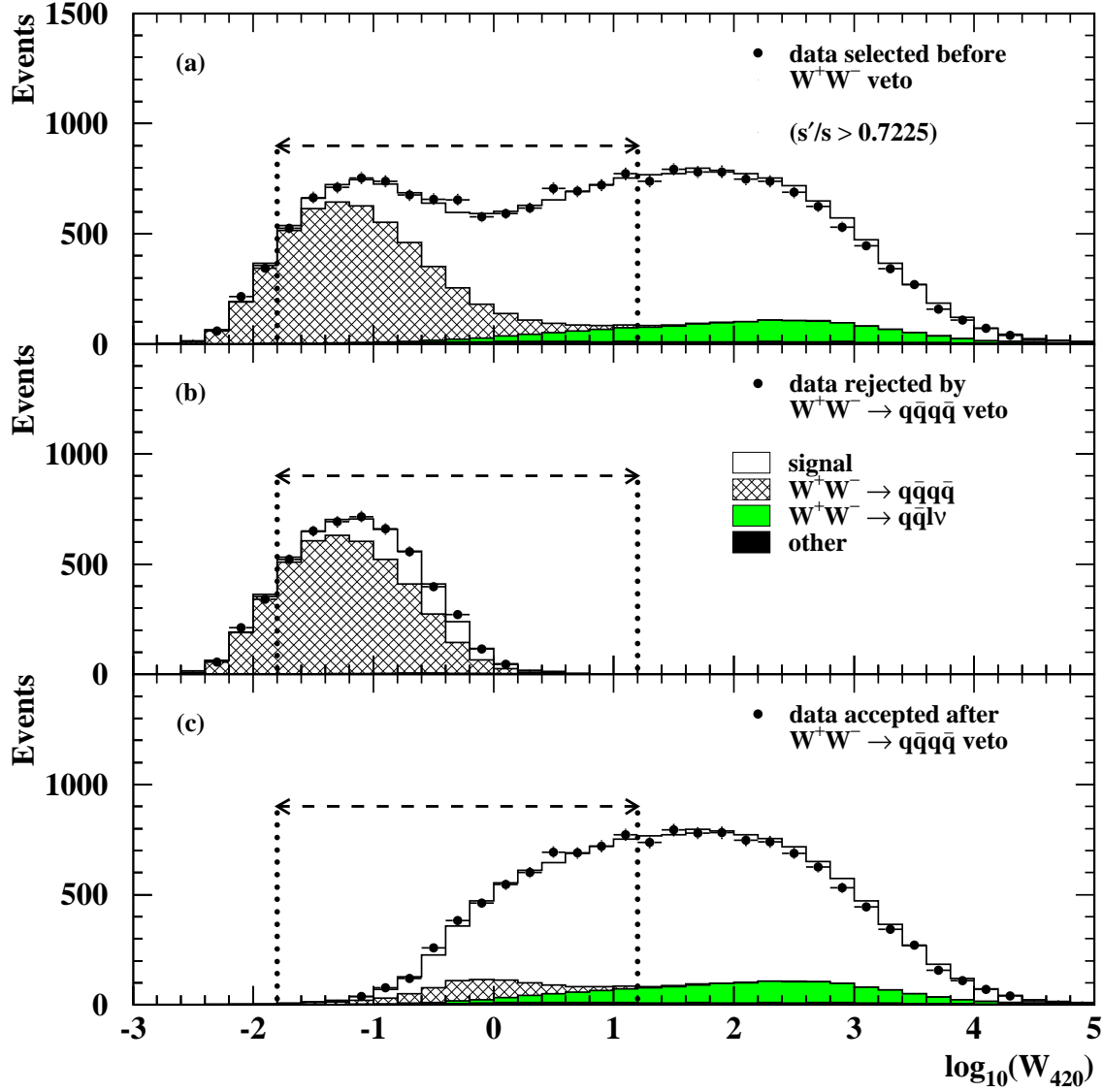}
\caption
{ Distributions of the QCD matrix element for four-jet production
  $W_{420}$ for events (a) passing the hadronic event selection 
  and the non-radiative $s'$ cut before applying the \WW\ veto,
  (b) additionally failing the $\WW\rightarrow\qqbar\qqbar$ veto or 
  (c) additionally passing the $\WW\rightarrow\qqbar\qqbar$ veto.
  The points show the data for all centre-of-mass energies combined
  and the histograms the Monte Carlo predictions normalized to the
  integrated luminosity of the data. The contributions from signal
  events and from the various sources of background are indicated,
  while the fit region (discussed in the text) is shown by the arrows.
}
\label{fig:wwveto}
\end{center}
\end{figure}

{\bf \boldmath \WW\ rejection cuts.} The effect of the \WW\ rejection cuts 
on the signal efficiency was studied using distributions of variables which 
distinguish \WW\ events from $\qqbar$ events. In the case of 
$\WW\rightarrow\qqbar\qqbar$ events the QCD matrix element for four-jet 
production, $W_{420}$~\cite{bib:w420} was used. This is an event weight
formed from the ${\cal O}(\alphas^2)$ matrix elements for the four-jet
production processes
$\epem\rightarrow\qqbar\rightarrow\qqbar\qqbar,\qqbar{\rm gg}$.
The distribution of $W_{420}$ after all event selection cuts except the 
\WW\ veto is shown in Fig.~\ref{fig:wwveto}(a) for non-radiative events 
from the combined data sample. A clear separation between signal events and 
background events is seen. Figures~\ref{fig:wwveto}(b) and (c) show the 
distributions for those
events rejected by and passing the \WW\ veto respectively. The distribution 
obtained from the data, before applying the \WW\ rejection, was fitted with
the sum of expected signal and background contributions, allowing the absolute 
normalization of both to vary. The fit region was chosen to include the 
majority of signal events rejected by the \WW\ veto and a majority of 
background events which pass the veto, as illustrated in Fig.~\ref{fig:wwveto}.
The resulting scale factor for the signal, or its statistical error,
was applied to the rejected signal cross-section to estimate the corresponding
uncertainty in efficiency. In the case of 
$\WW\rightarrow\qqbar\ell\nu$ events, a similar procedure was applied to
the distribution of the magnitude of the vector sum of transverse momenta 
for all visible particles. In all cases, the scale factors were found to
be consistent with unity.

{\bf Backgrounds.} The uncertainty in the \WW\ background was estimated from 
the fits to $W_{420}$ and transverse momentum distributions (for 
$\WW\rightarrow\qqbar\qqbar$ and $\WW\rightarrow\qqbar\ell\nu$ respectively) 
described above for the \WW\ rejection cuts. In addition, the
effects of initial-state radiation modelling and fragmentation modelling on
the four-fermion background were investigated by varying these models. In the
inclusive sample, the largest background uncertainty arises from tagged 
two-photon events. This was investigated by comparing the predictions of 
different Monte Carlo generators: either a combination of \HERWIG\ for 
single-tagged events plus \PHOJET\ for double-tagged events, or \TWOGEN\ for 
both. An average of the two prescriptions was found to give the best 
representation of the data at low $s'$, and was used in the cross-section 
determination, with half the difference between the two predictions taken as 
the systematic error. The (small) differences between the \PYTHIA\ and \PHOJET\
programs were used to assess the systematic uncertainty in the untagged 
two-photon background. Similarly, the small uncertainties in the $\tautau$ 
background were estimated by comparing the predictions of \KK\ and \KORALZ.   

{\bf Interference.} The error arising from the subtraction of interference
between initial- and final-state photon radiation was estimated from the
difference between the two methods of determination as described in
Section~\ref{sec:ifsr}.

%\subsubsection{Results}
%The numbers of selected events and measured cross-sections are shown in
%%Tables~\ref{tab:xsec1} and~\ref{tab:xsec2}. 
%Table~\ref{tab:xsec1}. 
%The measured differential cross-sections are shown in 
%Table~\ref{tab:qq_angdis}. 
The results of the cross-check analysis, in which events identified
as \WW\ were not removed, but the expected contribution subtracted,
are in excellent agreement with those of the primary analysis, with 
slightly larger total errors.

%-----------------------------------------------------------------------
%\subsection{Muon Pairs}                                \label{sec:mumu}
%-----------------------------------------------------------------------
%\input mumu.tex
%-----------------------------------------------------------------------
\subsection{Muon Pairs}                                \label{sec:mumu}
%-----------------------------------------------------------------------
\subsubsection{Event Selection}
\begin{figure}
\begin{center}
\epsfxsize=\textwidth
 %\epsfbox{mu_dists.eps}
\epsfbox{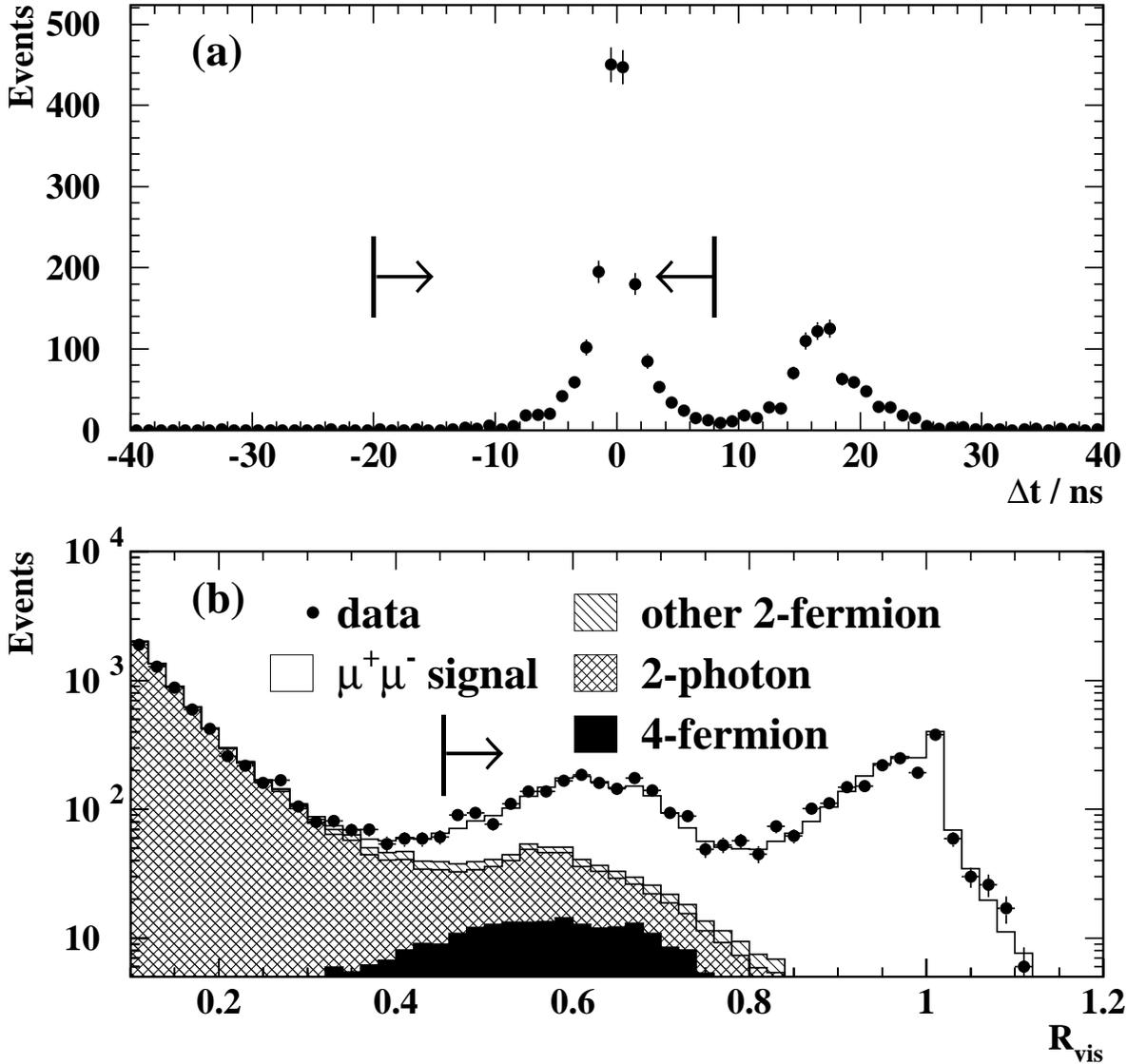}
\caption
{
  (a) Time difference between back-to-back hits in the time-of-flight
      counters in the barrel region. Events in the combined data
      sample which pass all $\mumu$ selection criteria except for
      the cosmic veto are included, if they have back-to-back TOF hits.
  (b) Ratio of the visible energy, defined as the sum of the muon
      momenta plus the energy of the highest energy electromagnetic
      calorimeter cluster, to the centre-of-mass energy, for $\mumu$
      candidates passing all cuts except those on the visible energy and the 
      mass of the muon pair. The points show the combined data
      and the histograms show the Monte Carlo expectation, normalized
      to the integrated luminosity of the data, with the background
      contributions as indicated. The vertical bars indicate the positions 
      of the cuts (for a centre-of-mass energy of 200~GeV in (b)), with the 
      arrow pointing into the accepted region in each case.
}
\label{fig:mu_dists}
\end{center}
\end{figure}

%A detailed description of the analysis of $\mumu$ events can be
%found in~\cite{bib:George}.
The selection of $\mumu$ events is essentially the same as in previous
analyses~\cite{bib:OPAL-SM189}, except that a small improvement has been made
in the rejection of cosmic ray events leading to a reduction in the uncertainty
associated with this background. The efficiency of the selection cuts is
typically $\sim$74\% for inclusive events and $\sim$88\% for non-radiative
events.  The corresponding purities of the selected samples are
$\sim$90\% and $\sim$97\% respectively. 
\begin{itemize}
 \item{}
 Muon pair events were required to have at least two tracks with momentum 
 greater than 6~GeV, 
 %transverse momentum greater than 1.65\% of the beam energy, 
 $\absct < 0.95$, separated in azimuthal angle by more than 320~mrad,
 and identified as muons. These tracks must have at least 20 hits
 in the central tracking chambers and the point of closest approach to the 
 nominal beam axis must lie less than 1~cm in the $r$--$\phi$ plane 
 and less than 50~cm along the beam axis from the nominal interaction
 point. To be identified as a muon, a track had to satisfy any of the 
 following conditions:
 \begin{itemize}
  \item
  At least 2 muon chamber hits associated with the track within an
  azimuthal angular range
  \hbox{$\Delta\phi=(100+100/p)$~mrad,} with the momentum~$p$ in~GeV;
  \item
  At least 4 hadron calorimeter strips associated with the track within
  an azimuthal angular range 
  \hbox{$\Delta\phi=(20+100/p)$~mrad,} with $p$ in~GeV. The average number of 
  strips per layer, taken over all layers with at least one hit, had to be 
  less than 2 to discriminate against hadrons. For \mbox{$|\cos\theta|<0.65$},
  where tracks traverse all 9 layers of strips in the barrel calorimeter, 
  a hit in one of the last 3~layers of strips was required;
  \item
  Momentum $p>15$~GeV and less than 3~GeV electromagnetic energy associated to 
  the track within a cone of half-angle 200~mrad.
 \end{itemize}
 If more than one pair of tracks satisfied the above conditions, the pair
 with the largest scalar sum of momenta was chosen. No requirement was
 made that the tracks have opposite charge.   
 \item{}
 Background from high multiplicity events was rejected by requiring that 
 there be no other track in the event with a transverse momentum (relative
 to the beam axis) greater than 1.65\% of the beam energy.
 \item{}
 Background from cosmic ray events was removed using the time-of-flight
 (TOF) counters and vertex cuts. Figure~\ref{fig:mu_dists}(a) shows the 
 distribution of the time difference, $\Delta t$, between pairs of 
 back-to-back TOF counters for $\mumu$ candidates, clearly showing one peak 
 at the origin from muon pairs and a second peak at about 15~ns from cosmic 
 rays. Events were accepted if they had $-20~{\rm ns} < \Delta t < 8~{\rm ns}$
 and at least one of the time measurements was within 10~ns of that expected 
 for a particle coming from the interaction point. If only one TOF hit was 
 recorded, it had to be within 10~ns of the expected time. In addition, 
 events were required to pass loose cuts on the matching of the central 
 detector tracks to the interaction vertex. Events without a good TOF hit
 were required to pass tight vertex criteria.
 \item{}
 Background from two-photon events was rejected by placing a cut on the
 total visible energy, $\Evis$, defined as the scalar sum of the momenta of the
 two muons plus the energy of the highest energy cluster in the 
 electromagnetic calorimeter:
 \[
    \Rvis \equiv \Evis / \sqrt{s} > 0.5 (\mPZ^{2} / s) + 0.35.
 \] 
 The value of this cut is 0.15 below the expected value of \Rvis\ for
 muon pairs in radiative return events where the photon escapes detection,
 visible as a secondary peak in Fig.~\ref{fig:mu_dists}(b). Furthermore,
 for inclusive events, if the ratio of the visible energy to the centre-of-mass
 energy was less than $0.5 (\mPZ^{2} / s) + 0.75$ the muon pair invariant mass 
 was required to be greater than 70~GeV. For all non-radiative events the 
 muon pair invariant mass was required to be greater than 
 $\sqrt{(\mPZ^{2} + 0.1 s)}$.
 \item{}
 The effective centre-of-mass energy $\sqrt{s'}$ of the \epem\ collision,
 determined as described below, was required to satisfy $s'/s > 0.01$
 for the inclusive sample and $s'/s > 0.7225$ for the non-radiative 
 sample.
\end{itemize}

Roughly 10\% of selected events have a photon detected in the electromagnetic
calorimeter with an energy above 30~GeV, separated from the nearest muon by
at least 20\degree. If such an event was planar, i.e.\ the sum of the
angles between the three particles (two muons plus photon) was greater than 
358\degree, the photon was assumed to be initial-state radiation and
$s'$ was calculated from the angles of the two muons and the photon using 
three-body kinematics. For all other events the value of 
$s'$ was estimated from the polar angles $\theta_{1}$ and $\theta_{2}$ of 
the two muons, assuming massless three-body kinematics to calculate the 
energy of a possible undetected initial-state photon along the beam direction 
as 
\begin{equation}
\label{equ:sp}
E_{\gamma} = \sqrt{s}\cdot |\sin(\theta_{1} + \theta_{2})| / 
         (|\sin(\theta_{1} + \theta_{2})| + \sin\theta_{1} +\sin\theta_{2}). 
\end{equation}
The observed distribution of $\sqrt{s'}$ for all data combined
is shown in Fig.~\ref{fig:sp}(b). 
 
The selection efficiencies and feedthrough of events from lower $s'$
into the non-radiative samples were determined from Monte Carlo events
generated with \KK\ without interference between initial- and final-state
radiation, corrected for the four-fermion contribution as discussed in
Section~\ref{sec:4f}. Backgrounds were also determined from Monte 
Carlo simulations. Efficiencies and backgrounds at each energy 
are summarized in 
%Tables~\ref{tab:eff1} and~\ref{tab:eff2}.
Table~\ref{tab:eff1}.

In approximately 2\% of $\mumu$ events the two muon tracks have the same 
charge; for the asymmetry and angular distribution measurements this charge
ambiguity was resolved using the acoplanarity of track segments reconstructed
in the muon chambers. Acoplanarity is defined as 
$|\phi_1 - \phi_2| - 180\degree$ where $\phi_1$ and $\phi_2$ are the
azimuthal angles of the muon segments. Bending of the charged particle
trajectories in the magnetic field results in positive or negative
acoplanarity depending on the charge of the particle with the lower
value of $\phi$.
In the measurement of both the angular distributions 
and asymmetries, the final values were obtained by averaging the distribution 
measured using the negative muon with that using the positive muon; although 
this averaging does not reduce the statistical errors on the measurements, it 
is expected to reduce most systematic effects. The forward-backward 
asymmetries at each energy were obtained by counting the numbers of events in 
the forward and backward hemispheres, after correcting for background and
efficiency. The asymmetries were corrected to the full angular range by 
applying a multiplicative correction obtained from \ZFITTER\ to the asymmetry 
measured within the acceptance of the selection cuts ($\absct < 0.95$).

\subsubsection{Systematic Uncertainties}
Systematic errors on the $\mumu$ cross-sections are summarized in 
Table~\ref{tab:sys_all}, with a detailed breakdown at 200~GeV given
in Table~\ref{tab:mu_syserr}. The main contributions are discussed below.

{\bf Efficiency.} The systematic uncertainty in the efficiency was evaluated
using high statistics \LEPone\ data and Monte Carlo samples. The \mumu\
cross-section at the Z peak is well known: it has been measured with a
systematic uncertainty of about 0.2\%~\cite{bib:PR328}.
The $\mumu$ selection cuts were applied to the \LEPone\ data and Monte
Carlo samples. A statistically significant difference between the number of 
data events selected and the number expected from Monte Carlo was 
observed, and this difference was used to estimate the systematic error 
associated with the efficiency. Most kinematic cuts
are a function of $\sqrt{s}$ and scale smoothly to the Z peak; for this
comparison it was necessary only to relax the cut on the visible 
energy so that the efficiency for events on the Z peak remained high.
\LEPtwo\ events have a different angular distribution from \LEPone\
events, and in particular radiative events are boosted towards the
endcap regions of the detector. The agreement between data and Monte Carlo 
was therefore checked as a function of \ct, and the results were reweighted 
to the angular distribution of the high energy data to obtain the systematic 
error on the total cross-section. 
To check for possible changes of the detector response with time,
this procedure was repeated with the calibration data taken at
the Z during 1998--2000; the observed difference between data and Monte Carlo
was consistent with that determined from the \LEPone\ study, but with poorer 
statistical precision.

{\bf Initial-state radiation modelling.} The systematic error on efficiency
derived from the \LEPone\ data does not include the effects of uncertainties
in the modelling of initial-state radiation. As for hadronic events, this
uncertainty was estimated by reweighting \KK\ events from 
${ \cal O}(\alpha^{2})$ to ${\cal O}(\alpha)$ CEEX and taking half the 
predicted change in efficiency.

{\bf Feedthrough.} The uncertainty in the feedthrough of events with lower
$s'$ into the $s'/s > 0.7225$ sample was estimated by comparing the
prediction of \KK\ with that of \KORALZ.

{\bf Cosmic background.} The uncertainty due to any remaining cosmic
background in the muon pairs was estimated from the vertex distribution
of events after relaxing some of the time-of-flight and vertex criteria.

{\bf Other backgrounds.} The main backgrounds in the muon pairs arise
from various leptonic four-fermion final states and from tau pairs.
The four-fermion backgrounds are principally channels including at
least two muons, and include a significant contribution from production 
of $\epem\mumu$ final states via two-photon processes. 
Backgrounds were studied by considering distributions of 
selection variables after loosening some of the selection cuts. The 
numbers of events in data and Monte Carlo were compared for a region enriched 
in a particular background, and the difference, or its statistical error, 
whichever was greater, used to estimate the systematic error from that 
background source. For example, the two-photon \epem\mumu\ background was 
studied using the distribution of visible energy after removing the cuts on 
visible energy and muon-pair mass, shown in Fig.~\ref{fig:mu_dists}(b);
the comparison was made in the visible energy range between 10\% and 40\%
of the centre-of-mass energy, which is completely dominated by this
background. 

{\bf Interference.} The uncertainty arising from the removal of the
contribution from interference between initial- and final-state
radiation was estimated as described in Section~\ref{sec:ifsr}.

{\bf Asymmetry.}
Systematic uncertainties in the asymmetry measurement were assessed by 
comparing results obtained using different combinations of tracking and 
muon chambers to measure the muon angles. The change in asymmetry when 
same-sign events were excluded from the sample was included as a systematic 
error. Other small contributions arise from the efficiency and background 
correction and subtraction of interference between initial- and final-state 
radiation.

%\subsubsection{Results}
%The numbers of selected events and resulting cross-sections are shown in
%%Tables~\ref{tab:xsec1} and~\ref{tab:xsec2}. 
%Table~\ref{tab:xsec1}. 
%The forward-backward asymmetries are shown in 
%%Tables~\ref{tab:afb1} and~\ref{tab:afb2}. 
%Table~\ref{tab:afb1}. 
%The differential cross-sections are given in Table~\ref{tab:mu_angdis}. 

%-----------------------------------------------------------------------
%\subsection{Tau Pairs}                                 \label{sec:tautau}
%-----------------------------------------------------------------------
%\input tautau.tex
%-----------------------------------------------------------------------
\subsection{Tau Pairs}                                 \label{sec:tautau}
%-----------------------------------------------------------------------
\subsubsection{Event Selection}
%A detailed description of the analysis of $\tautau$ events can be
%found in~\cite{bib:George}.
The selection of $\eetotautau$ events is based on that used in previous 
analyses~\cite{bib:OPAL-SM189}, using information from the central tracking 
detectors and electromagnetic calorimetry to identify events with two 
collimated, low multiplicity jets. However, the cuts have been tightened
to improve the background rejection at higher energies.
The efficiency of the selection cuts is typically $\sim$33\% for inclusive 
events and $\sim$48\% for non-radiative events. The corresponding purities 
of the selected samples are $\sim$88\% and $\sim$92\% respectively. 
 
Tracks and electromagnetic clusters, each treated as separate particles
with no attempt to correct for double-counting of energy,
were combined into jets in the following way.
First the highest energy particle in the event was selected and
a cone with a half angle of 35$\degree$ was defined around it.
The particle with the next highest energy inside the cone was
combined with the first.
The momenta of the combined particles were added and the direction
of the sum was used to define a new cone,
inside which the next highest
energy particle was again sought. This procedure was repeated
until no more particles were found inside the cone.
Similarly, starting with the highest energy particle among the
remainder, a new cone was initiated and treated in the same way.
This process continued until finally all
the particles in the event had been assigned to a cone.

The following cuts were applied to select $\tautau$ candidates.

\begin{figure}
\begin{center}
\epsfxsize=\textwidth
 %\epsfbox{tau_dists.eps}
\epsfbox{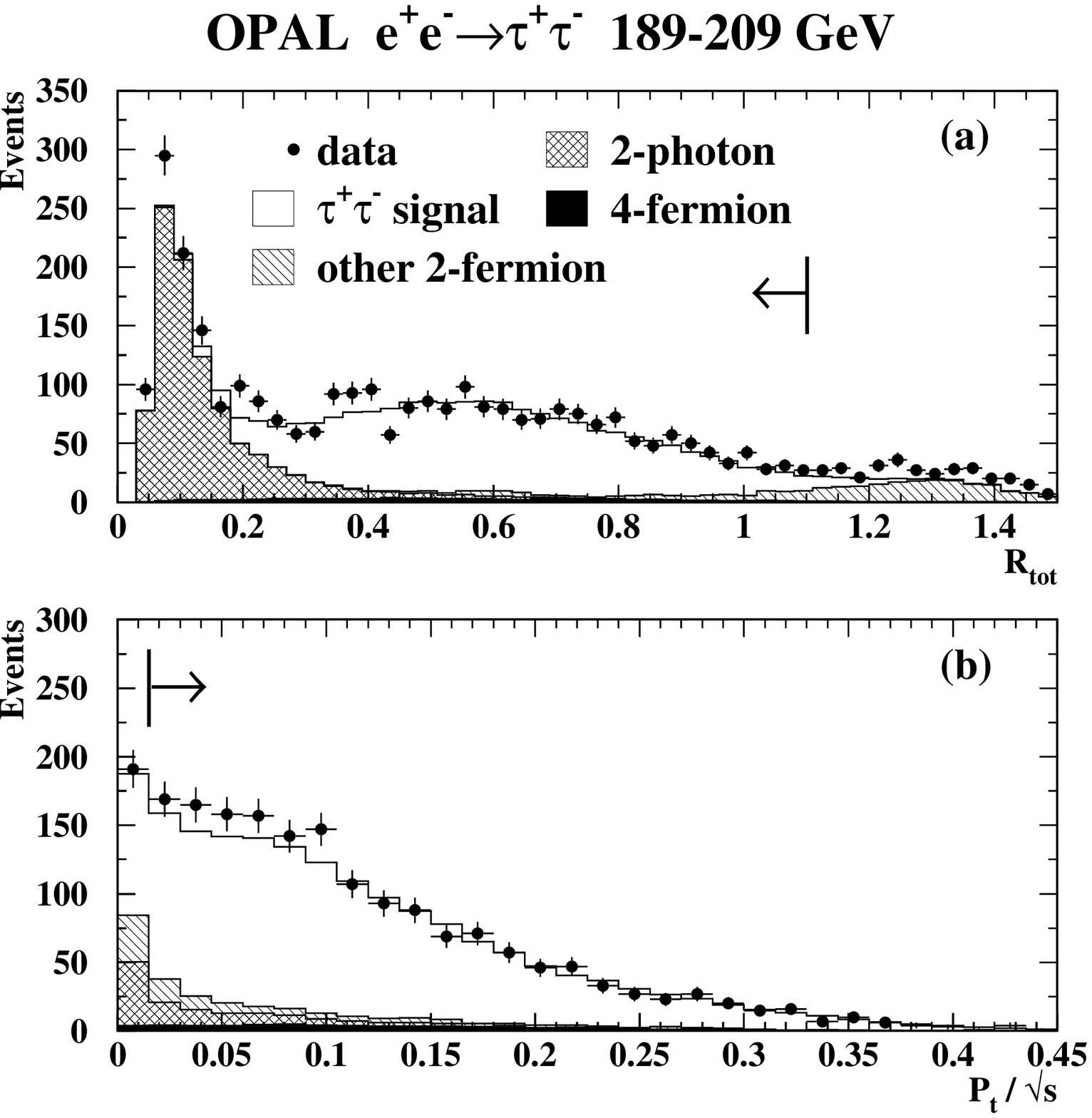}
\caption
{
  (a) Ratio of the total event energy, defined as the scalar sum of track
      momenta plus the energy of electromagnetic calorimeter clusters, 
      to the centre-of-mass energy, for $\tautau$ candidates passing all  
      cuts except those on the visible energy and its track and cluster
      components.
  (b) The missing momentum, calculated using electromagnetic calorimeter 
      clusters, divided by the centre-of-mass energy for $\tautau$
      events passing all cuts except those on the missing momentum and
      the cosine of the polar angle of its direction. 
      In each case, the points show the combined data and the histograms 
      show the Monte Carlo expectations, normalized to the integrated 
      luminosity of the data, with the background contributions as
      indicated. 
      The vertical bars indicate the positions of the cuts, with the arrow 
      pointing into the accepted region in each case. Note that, in the 
      case of the total event energy in (a), further cuts are placed on the 
      separate track and cluster components at both low and high values.
}
\label{fig:tau_dists}
\end{center}
\end{figure}

\begin{itemize}
\item{}
Hadronic events were rejected by demanding low multiplicity: the
number of tracks reconstructed in the central tracking detectors had
to be at least two and at most six, and the sum of the number of tracks 
plus the number of electromagnetic clusters not more than~15.
\item{}
The total energy of an event was restricted in order to reject events from 
$\eetoee(\gamma)$, $\mumu(\gamma)$ and two-photon processes. The total 
event energy, $E_{\rm tot}$, defined as the scalar sum of all track momenta 
plus all electromagnetic calorimeter energy, was required to be less than 
1.1$\roots$. The total electromagnetic calorimeter energy was required to be 
between 0.02$\roots$ and 0.7$\roots$ and the scalar sum of track momenta less 
than 0.8$\roots$.
Either the total electromagnetic calorimeter energy or the scalar sum of
track momenta was required to be greater than 0.2$\roots$. 
The distribution of $E_{\rm tot}/\sqrt{s}$, after all other cuts 
have been applied, is shown in Fig.~\ref{fig:tau_dists}(a) for all 
centre-of-mass energies combined. The agreement between data and simulation
is good in the region dominated by the $\tautau$ signal, but poor in
regions dominated by background; this discrepancy is used to estimate
the systematic uncertainty in the background.
\item{} Background from two-photon events was further reduced by cuts on the 
missing momentum and its direction. The missing momentum in the plane
transverse to the beam axis, calculated using the electromagnetic calorimeter, 
was required to exceed 0.015$\roots$, and the polar angle of the missing 
momentum was required to satisfy $\absct < 0.99$. Fig.~\ref{fig:tau_dists}(b)
shows the distribution of the missing momentum 
%calculated using electromagnetic clusters 
after all other cuts have been applied, for all centre-of-mass energies 
combined.
\item{}
Vertex and TOF cuts were imposed to remove cosmic ray events, as for
\mumu\ events. In addition, events identified as $\eetomumu$ using the 
criteria described in Section~\ref{sec:mumu} were removed.
\item{}
Cones formed from tracks and clusters as described above were classified as 
either charged or neutral. A charged cone was required to contain at least 
one charged particle with transverse momentum greater than 100~MeV and one 
electromagnetic cluster with energy greater than 100~MeV, and the sum of the 
energy in the electromagnetic calorimeter and the track momenta in the cone 
had to be more than 1\% of the beam energy. Neutral cones were required to 
contain no charged particle and an energy in the electromagnetic calorimeter 
of at least 1\% of the beam energy. Cones failing these criteria were 
discarded. Events which had exactly two charged cones  
were selected as $\eetotautau$ candidates. The direction of each $\tau$ was 
approximated by that of the total momentum vector of its cone of particles.
Events were accepted if both cones satisfied $\absct < 0.9$. To remove
events with poor momentum reconstruction, the event was rejected if the 
azimuthal angle of either cone, determined using tracks only, lay within
0.5\degree\ of an anode plane of the central tracking chamber. In addition 
to the two charged cones, an event may contain any number of neutral cones.
\item{}
To suppress electron- and muon-pair events further, we reject events
with cone energies or momenta compatible with these final states. 
Assuming that the final state consists only of two leptons plus a single 
unobserved photon along the beam direction, the values of the polar angles 
of the two $\tau$ cones were used to calculate the expected energy of each 
lepton $X_1,X_2$. It was required that
        \[ 0.02 < \sqrt{(E_1^2 + E_2^2)/(X_1^2 + X_2^2)} < 0.8,
        \]
        and
        \[ \sqrt{(P_1^2 + P_2^2)/(X_1^2 + X_2^2)} < 0.8,
        \]
where $E_1,E_2$ and $P_1,P_2$ are the total electromagnetic calorimeter 
energies and scalar sums of track momenta, respectively, in each $\tau$ cone.
%These cuts are designed to remove both electron and muon pairs.
\item{}
Remaining background from $\eetoee (\gamma)$ and $\epem\epem$ events was 
reduced by rejecting events if the ratio of the electromagnetic energy to 
the track momentum in each of the $\tau$ cones was consistent with that
expected for an electron.
\item{}
Most of the remaining background from two-photon processes was rejected
by a cut on the acollinearity and acoplanarity angles of the two $\tau$
cones: the acollinearity angle, in degrees, was required to satisfy
\[
  \thacol < (180\degree - 2\tan^{-1}(2\mPZ\sqrt{s} / (s - \mPZ^{2}))) + 
             10\degree
\]
and the acoplanarity angle was required to be less than 30$^{\circ}$.
The value of the cut on acollinearity was chosen such as to include the
peak from radiative return events at each energy; it is $92\degree$ at
200~GeV. The acoplanarity cut was not applied to events with a photon
observed in the detector with energy above 30 GeV, if the event was 
planar (i.e.\ the sum of the opening angles between the three particles
was greater than 358\degree).
\item{}
Events classified as \WW\ candidates according to the criteria 
in~\cite{bib:OPAL-WW189} were rejected.
\item{}
After the above cuts, the region of $\sqrt{s'}$ between the radiative return
and full energy peaks, $110~\GeV<\sqrt{s'}<0.85\sqrt{s}$, still contains
a significant fraction of background from two-photon events. To reduce this 
background, a likelihood for the process $\epem\rightarrow\tautau$ was 
formed from four variables: the missing momentum 
calculated using electromagnetic clusters, the scalar sum of the track
momenta, the invariant mass of the two $\tau$ cones and the difference
between the electromagnetic calorimeter energy in the two $\tau$ cones. 
The value of this likelihood was required to be greater than 0.5.  
\item{}
 The effective centre-of-mass energy $\sqrt{s'}$ was determined in
 an identical manner to the determination for muon pairs. The observed 
 distribution of $\sqrt{s'}$ for all energies combined is shown in 
 Fig.~\ref{fig:sp}(c). Inclusive events were required to satisfy 
 $s'/s > 0.01$ and non-radiative events were required to have
 $s'/s > 0.7225$. 
\end{itemize}

%The effective centre-of-mass energy $\sqrt{s'}$ was determined in a
%similar manner to the determination in muon pairs. For those events 
%with a photon detected in the electromagnetic calorimeter with an energy 
%above 30~GeV, $s'$ was calculated from the angles of the two $\tau$ cones 
%and the photon, if the event was planar (i.e.\ the sum of the opening angles 
%was greater than 358\degree). 
%For the remaining events the value of $s'$ was estimated from the polar 
%angles of the two taus assuming a single photon along the beam axis, as 
%described in Section~\ref{sec:mumu}. The observed distribution of 
%$\sqrt{s'}$ for all energies combined is shown in Fig.~\ref{fig:sp}(c). 

The selection efficiencies and feedthrough of events from lower $s'$
into the non-radiative samples were determined from Monte Carlo events
generated with \KK\ without interference between initial- and final-state
radiation, corrected for the four-fermion contribution as discussed in
Section~\ref{sec:4f}. Backgrounds were also determined from Monte 
Carlo simulations. Efficiencies and backgrounds at each energy 
are summarized in 
%Tables~\ref{tab:eff1} and~\ref{tab:eff2}.
Table~\ref{tab:eff1}.

For the measurement of the angular distributions and asymmetries, the
small ($\sim$2\%) fraction of events where the two $\tau$ cones have the same
charge (as determined from the sum of the charges of the tracks in the
cone) was not used. The final values were obtained by averaging the 
distribution measured using the negative $\tau$ with that using the positive 
$\tau$, as for the muon pairs. The forward-backward asymmetries at
each energy were obtained by counting the numbers of events in the 
forward and backward hemispheres, after correcting for background and
efficiency. The asymmetries were corrected to the full angular range by 
applying a multiplicative correction obtained from \ZFITTER\ to the asymmetry 
measured within the acceptance of the selection cuts ($\absct < 0.9$).

\subsubsection{Systematic Uncertainties}
Systematic errors on the $\tautau$ cross-sections are summarized in 
Table~\ref{tab:sys_all}, with a detailed breakdown at 200~GeV given in
Table~\ref{tab:tau_syserr}. The main contributions are discussed below.

{\bf Efficiency.} The systematic error on the efficiency was evaluated
using high statistics \LEPone\ data and Monte Carlo samples, as for the
muon pairs. The \tautau\ cross-section at the Z peak has been measured with 
a systematic uncertainty of about 0.5\%~\cite{bib:PR328}. As in the
case of muon pairs, a statistically significant difference between the
observed and expected numbers of events was seen, and the difference
was assigned as the systematic error associated with the selection cuts. 

{\bf Initial-state radiation modelling.} The systematic error on efficiency
derived from the \LEPone\ data does not include the effect of uncertainties
in the modelling of initial-state radiation. As for hadronic events and
muon pairs, this was estimated by reweighting \KK\ events from 
${\cal O}(\alpha^{2})$ to ${\cal O}(\alpha)$ CEEX and taking half the 
predicted change in efficiency.

{\bf Feedthrough.} The uncertainty in the feedthrough of events with lower
$s'$ into the $s'/s > 0.7225$ sample was estimated by comparing the
prediction of \KK\ with that of \KORALZ.

{\bf Backgrounds.} The largest background in the tau pairs arises from 
Bhabha events. Other important backgrounds arise from \epem\epem\ and 
\epem\tautau\ final states. As for the muon pairs, systematic errors on
each background channel were assessed by comparing data and Monte Carlo
distributions of selection variables, after loosening selection cuts, 
in a region enriched in the particular background under study. For example, 
the \epem\mumu\ background was studied using the distribution of 
total event energy. 
For small backgrounds which cannot be studied in this way,
we conservatively assume an error of 50\%.

{\bf Interference.} The uncertainty arising from the removal of the
contribution from interference between initial- and final-state
radiation was estimated as described in Section~\ref{sec:ifsr}.

{\bf Asymmetry.} Systematic errors on the asymmetry measurement were assessed 
by comparing different methods of determining the asymmetry: using tracks, 
electromagnetic clusters or both to determine the $\tau$ angles.

%\subsubsection{Results}
%The numbers of selected events and resulting cross-sections are shown in
%%Tables~\ref{tab:xsec1} and~\ref{tab:xsec2}. 
%Table~\ref{tab:xsec1}. 
%The forward-backward asymmetries are shown in 
%%Tables~\ref{tab:afb1} and~\ref{tab:afb2}. 
%Table~\ref{tab:afb1}. 
%The differential cross-sections are given in Table~\ref{tab:tau_angdis}. 

%-----------------------------------------------------------------------
%\subsection{Electron Pairs}                            \label{sec:ee}
%-----------------------------------------------------------------------
%\input ee.tex
%-----------------------------------------------------------------------
\subsection{Electron Pairs}                            \label{sec:ee}
%-----------------------------------------------------------------------
The production of electron pairs is dominated by $t$-channel photon
exchange, for which a definition of $s'$ as for the other channels is
less meaningful. In addition, the increased probability for final-state 
radiation relative to initial-state radiation renders the separation between 
initial- and final-state photons more difficult.
Events with little radiation were therefore selected by a cut 
on $\thacol$, the acollinearity angle between electron and positron. 
%A cut 
%of $\thacol < 10^{\circ}$ roughly corresponds to a cut on the effective 
%centre-of-mass energy of $s'/s > 0.8$, for the $s$-channel contribution.
We measure cross-sections for three different acceptance regions, defined
in terms of the angle of the electron, $\theta_{\mathrm{e^-}}$, or positron,
$\theta_{\mathrm{e^+}}$, with respect to the incoming electron direction,
and the acollinearity angle: 
\begin{itemize}
\item {\bf A:} $\absctem < 0.9$, $\absctep < 0.9$, $\thacol < 170^{\circ}$;
      this is a loose `inclusive' measurement;
\item {\bf B:} $\absctem < 0.7$, $\thacol < 10^{\circ}$; this acceptance
      region is enriched in the $s$-channel contribution, 
      and is used for asymmetry measurements;
\item {\bf C:} $\absctem < 0.96$, $\absctep < 0.96$, $\thacol < 10^{\circ}$; 
      this `large acceptance' region is enriched in the $t$-channel
      contribution and acts as a check on the luminosity measurements.
\end{itemize} 
In addition, we measure the electron angular distribution in the region:
\begin{itemize}
\item {\bf D:} $\absctem < 0.9$, $\thacol < 10^{\circ}$.
\end{itemize}
In all cases, measurements are corrected to correspond to electron and 
positron energies each greater than 0.2~GeV. 

\subsubsection{Event Selection}
\begin{figure}
\begin{center}
\epsfxsize=0.89\textwidth
 %\epsfbox{ee_dists.eps}
\epsfbox{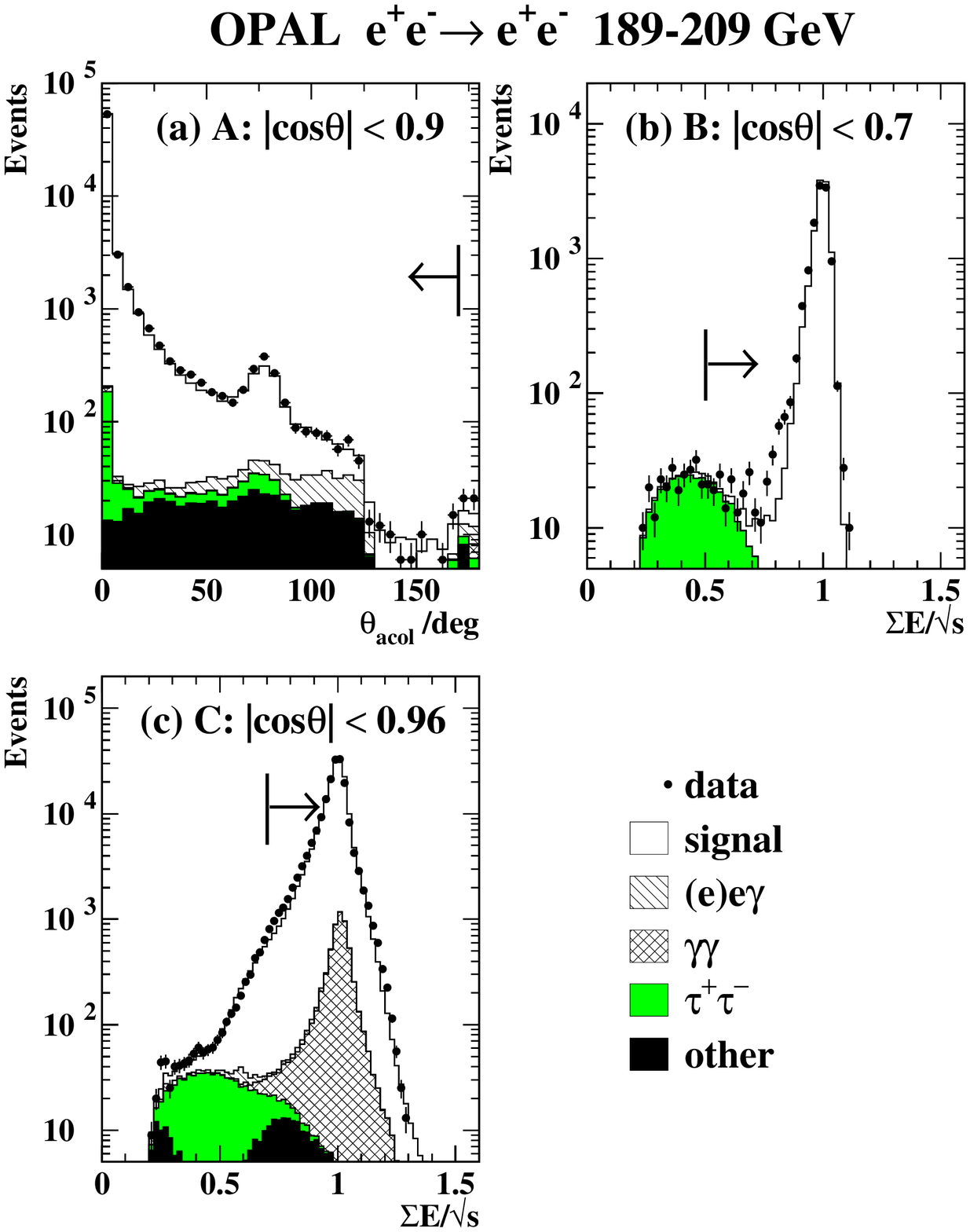}
\caption
{
  (a) The acollinearity angle distribution for events satisfying the
  inclusive $\eetoee$\ selection, in the acceptance region A, 
  $\absctepem < 0.9$.
  (b) The distribution of the ratio of total electromagnetic calorimeter
  energy to the centre-of-mass energy for $\eetoee$ events in 
  acceptance region B, $\absctem < 0.7$ and $\thacol < 10\degree$. 
  (c) The same distribution for the large acceptance region, C,
  $\absctepem < 0.96$ and $\thacol < 10\degree$. 
  Distributions are shown after all cuts except the one on the 
  variable plotted.
  %(d) Observed distribution of $\ct$ of the outgoing electron in \epem\
  %events with $\thacol < 10\degree$. 
  In each case, the points show the combined data and the 
  histograms the Monte Carlo expectations, normalized to the integrated 
  luminosity of the data, with the background contributions as
  indicated.
  The vertical bars indicate the positions of the cuts in the displayed
  variable, with the arrow pointing into the accepted region in each case.
}

\label{fig:ee_dists}
\end{center}
\end{figure}

The selection of $\epem$ events is identical to previous 
analyses~\cite{bib:OPAL-SM172,bib:OPAL-SM183,bib:OPAL-SM189}.
The selection efficiencies are typically $\sim$98\%, and the purities
of the selected samples $\sim$98\%.
\begin{itemize}
 \item{}
 Events were required to have at least two and not more than eight clusters 
 in the electromagnetic calorimeter, and not more than eight tracks in the 
 central tracking chambers. 
 \item{}
 At least two clusters were required to have an energy exceeding 20\% of the 
 beam energy, and the total energy deposited in the electromagnetic 
 calorimeter was required to be at least 50\% of the centre-of-mass energy. 
 For the large acceptance selection, C, which has no requirement on the 
 association of tracks to clusters, the total electromagnetic energy was 
 required to be at least 70\% of the centre-of-mass energy. Distributions of 
 total electromagnetic calorimeter energy, after all other cuts, are shown
 in Fig.~\ref{fig:ee_dists}(b) and (c) for acceptance regions B and C for the 
 data from all years combined. There is reasonable agreement between data 
 and Monte Carlo. The apparent slight excess of data over Monte Carlo at 
 about 80\% of the centre-of-mass energy in acceptance B results from poor 
 modelling of the energy resolution in the region $0.6 < \absct < 0.7$. 
 The degraded energy resolution in acceptance region C arises from the 
 increased amount of material in front of the electromagnetic calorimeter at 
 large $\absct$, where the events are concentrated. The detailed modelling
 of the electromagnetic calorimeter energy resolution has very little
 effect on the selection efficiency for \epem\ events. 
 \item{}
 For selections A, B and D, at least two of the three highest energy clusters 
 were required to have an associated central detector track. If a cluster had 
 more than one associated track, the one with the highest momentum was chosen.
 If all three clusters had an associated track, the two highest energy 
 clusters were chosen to be the electron and positron. For the large 
 acceptance selection, C, no requirement was placed on the association of 
 tracks to clusters.
 \item{}
 For the measurement of the forward-backward asymmetry and the angular 
 distribution, the two tracks were required to have opposite charge. This 
 extra requirement reduces the efficiency by about 3.5\% in the region 
 $\absct < 0.9$. In addition, due to the extreme charge asymmetry for
 electrons in the forward direction, the problem of charge misassignment
 becomes severe for backward events at small angles. 
 In the measurement of the angular distribution we therefore 
 demanded that events with $\ctem < -0.8$ satisfy two extra criteria: both 
 electron and positron tracks must have momentum greater than 25\% of the beam
 momentum, and there must be only one good track associated with each cluster.
 These criteria significantly reduce the problem of charge misassignment, 
 reducing the contamination from wrong-sign events in this region from around 
 35\% to about 15\%.
 %\footnote{For the same-sign events used in the 
 %cross-section measurement in acceptance region B, the acoplanarity
 %of the clusters is used to distinguish electron from positron, as 
 %in~\cite{bib:PR328}. Acoplanarity is defined
 %as $|\phi_1 - \phi_2| - 180\degree$ where $\phi_1$ and $\phi_2$ are the
 %azimuthal angles of the two clusters. Bending of the charged particle
 %trajectories in the magnetic field results in positive or negative
 %acoplanarity depending on the charge of the particle with the lower
 %value of $\phi$.}.
 \item{}
 Acceptance cuts on acollinearity and \ct\ were made using the calorimeter
 clusters, with angles corrected for the position of the primary vertex.
 The acollinearity angle distribution for the inclusive selection, A, is 
 shown in Fig.~\ref{fig:ee_dists}(a), and we see good agreement between data 
 and Monte Carlo expectation, including the peak corresponding to $s$-channel 
 radiative return to the \PZ.
\end{itemize}

These cuts have a very high efficiency for \epem\ events while providing
excellent rejection of backgrounds, which either have high multiplicity or
lower energy deposited in the electromagnetic calorimeter. The efficiency of 
the selection cuts, and small acceptance corrections, have been determined 
using Monte Carlo events generated with the \BHWIDE~\cite{bib:bhwide} program. 
These were found to be almost independent of energy over the range considered 
here. Small corrections have been applied to the efficiencies derived from 
Monte Carlo simulations to account for tracking losses near the central jet 
chamber anode planes ($\sim 0.8\%$), and, in the case of the angular
distribution, to account for a discrepancy between data and Monte Carlo
in the fraction of events where both tracks have the same charge 
($\sim 0.5\%$).
Remaining backgrounds arise from \tautau\ events and, in the case of the 
loose acollinearity cut, also from electron pairs in two-photon events and 
from radiative Bhabha scattering events in which one electron is outside the 
detector acceptance but the photon is within the acceptance.
In the case of the large acceptance selection, C, which does not require
tracks, the main background arises from \gamgam\ final states. The efficiencies
and backgrounds at each energy are summarized in 
%Tables~\ref{tab:eff1} and~\ref{tab:eff2}.
Table~\ref{tab:eff1}.

The forward-backward asymmetries for the $\thacol < 10\degree$ sample at each 
energy within the angular range $\absctem < 0.7$ were evaluated by counting  
the numbers of events in the forward and backward $\ctem$ hemispheres,
after correcting for background and efficiency. For both the asymmetry
and angular distribution measurements, the positive or negative track
was used on alternate events to reduce systematic effects. 

In Fig.~\ref{fig:sp}(d) we show the distribution of $\sqrt{s'}$ for
the inclusive $\epem$ events for all energies combined.
The value of $s'$ for each event was estimated from the polar angles
of the two electrons assuming massless three-body kinematics to calculate 
the energy of a possible undetected initial-state photon along the beam 
direction as shown in Equation~(\ref{equ:sp}). 
For $\epem$, $s'$ is not really well-defined, but this calculation
gives an estimate of $s'$ for that part of the cross-section proceeding
via the $s$-channel.
Due to the dominance of the $t$-channel contributions, for electrons,
in contrast to the other final states, the radiative return peak forms
only a very small contribution.

\subsubsection{Systematic Uncertainties}
The systematic errors associated with the $\epem$ measurements
have generally been estimated in a similar manner to the previous
analysis~\cite{bib:OPAL-SM189}. They are summarized in 
Table~\ref{tab:sys_all}, with a detailed
breakdown at 200~GeV given in Table~\ref{tab:ee_syserr}. The most
significant change is that the systematic errors on the differential
cross-section measurements (acceptance D) 
have been estimated separately for three \ct\ regions, namely $\ctem < -0.7$, 
$\absctem < 0.7$ and $\ctem > +0.7$, rather than considering the whole 
distribution together. The systematic errors are not expected to be strongly 
dependent on centre-of-mass energy, so in general they have not been estimated
separately for each energy point. The most important ones are discussed below.

{\bf Four-fermion contribution.} The full size of the change in efficiency
arising from including $s$-channel four-fermion events in the signal definition
was included as a systematic error. This affects the inclusive selection
(selection A) only, and is negligible for events with a tight acollinearity
cut.

{\bf Multiplicity cuts.} The uncertainties arising from the requirement of
low multiplicity have been estimated from the change in the number of
selected events in data when varying the multiplicity cuts used by 
$\pm1$ unit. 

{\bf Calorimeter energy scale and resolution.}  A detailed
comparison between data and Monte Carlo has been made of the energy
scale and resolution of the electromagnetic calorimeter, and the results
of this study used to assess possible effects on the selection efficiency.
Typically the energy scale was varied by 0.3\% and the resolution by
10\% of its value.

{\bf Track requirements.} Matching between tracks and clusters has
been studied using events passing all selection cuts, except that only
one of the three highest energy clusters has an associated track. These
events are expected to be mainly $\epem\gamma$ final states where one electron
and the photon lie within the acceptance and $\gamma\gamma$ final states
where one photon has converted in the detector, with small contributions
from other final states. An excess of such events was seen in data 
compared with Monte Carlo expectation. Part of this excess is concentrated
in regions of $\phi$ near the anode planes of the central jet chamber, and 
arises from track reconstruction problems in this region. The rest 
could arise from track reconstruction problems, or could arise from problems 
modelling $\epem\gamma$ or $\gamma\gamma$ events. 
For each acceptance region we take the excess seen around the jet chamber 
anode planes plus half the difference between data and Monte Carlo in the
remaining region of $\phi$ as a correction to the efficiency. This correction
is typically around 0.8\%. Half the difference between data and Monte
Carlo in the regions of $\phi$ away from the jet chamber anode planes is
taken as the systematic error associated with track reconstruction.
In the case of the angular distribution, the two tracks in an event are
required to have opposite charge. The fraction of same-sign events in data
is roughly 0.5\% greater than in Monte Carlo. This difference is applied
as a correction to the efficiency derived from Monte Carlo, and the value
of the correction is included as a systematic error. The higher probability
of same-sign events in the data is also used to calculate a correction 
of (1.6$\pm$0.5)\% to the angular distribution in the region $\ctem < -0.7$ 
arising from charge misassignment.

{\bf Acceptance correction.} Because of the steepness of the angular 
distribution, uncertainties in the determination of $\theta$ are an 
important systematic error. These have been assessed by comparing 
measurements of $\theta$ in the electromagnetic calorimeter with those in the 
central tracking chambers and the muon chambers, using $\epem$ or $\mumu$
events as appropriate. These studies indicate a possible bias
in the $\theta$ reconstruction of electromagnetic clusters of 
$\sim$1~mrad in the endcap region of the detector. The effect of
the observed biases on the acceptance was calculated using Monte Carlo
events, and assigned as a systematic error associated with the
acceptance correction.

{\bf Background.}
If a tight acollinearity cut is applied, the dominant background in the 
selections including tracks is from \tautau\ events. With a loose
acollinearity cut, $\epem\gamma$ and \epem\epem\ events are also significant.
The systematic error arising from uncertainty in the background has
been assessed by comparing the numbers of events in data and Monte
Carlo which pass all cuts except the cut on total calorimeter energy;
these events are predominantly background. In each acceptance region 
the larger of the difference between data and Monte Carlo or the statistical
precision of the test was taken as the associated systematic error.
For the selection which does not use tracks, acceptance C, the only important 
background is from $\gamma\gamma$ final states; here we used the statistical
precision of the OPAL $\epem\rightarrow\gamma\gamma$ cross-section 
measurement~\cite{bib:OPAL-gg} to estimate the uncertainty in this 
background.

{\bf Asymmetry}.
Systematic uncertainties in the asymmetry measurement arise from the effects
of $\theta$ mismeasurement, charge misassignment and background and
efficiency corrections, and amount to 0.004.

%\subsubsection{Results}
%\label{sec:ee_results}
%The numbers of selected events and resulting cross-sections are shown in
%%Tables~\ref{tab:xsec1} and~\ref{tab:xsec2}. 
%Table~\ref{tab:xsec1}. 

%-----------------------------------------------------------------------
\subsection{Results}                            
%-----------------------------------------------------------------------
The numbers of selected events and measured cross-sections for all
channels are summarized in Table~\ref{tab:xsec1}. 
%Tables~\ref{tab:xsec1} and~\ref{tab:xsec2}.
Asymmetries for the leptonic final states are summarized in 
Table~\ref{tab:afb1},
while the measured differential cross-sections are given in 
Tables~\ref{tab:qq_angdis},~\ref{tab:mu_angdis},~\ref{tab:tau_angdis}
and~\ref{tab:ee_angdis} for hadrons, $\mumu$, $\tautau$ and $\epem$
respectively. 

%%-----------------------------------------------------------------------
%\section{Comparison With Standard Model Predictions}   \label{sec:SM}
%%-----------------------------------------------------------------------
%\input sm.tex
%-----------------------------------------------------------------------
\section{Comparison with Standard Model Predictions}    \label{sec:SM}
\begin{figure}
  \epsfxsize=0.96\textwidth
  %\epsfbox[0 0 567 680]{xs_mh.eps}
  \epsfbox{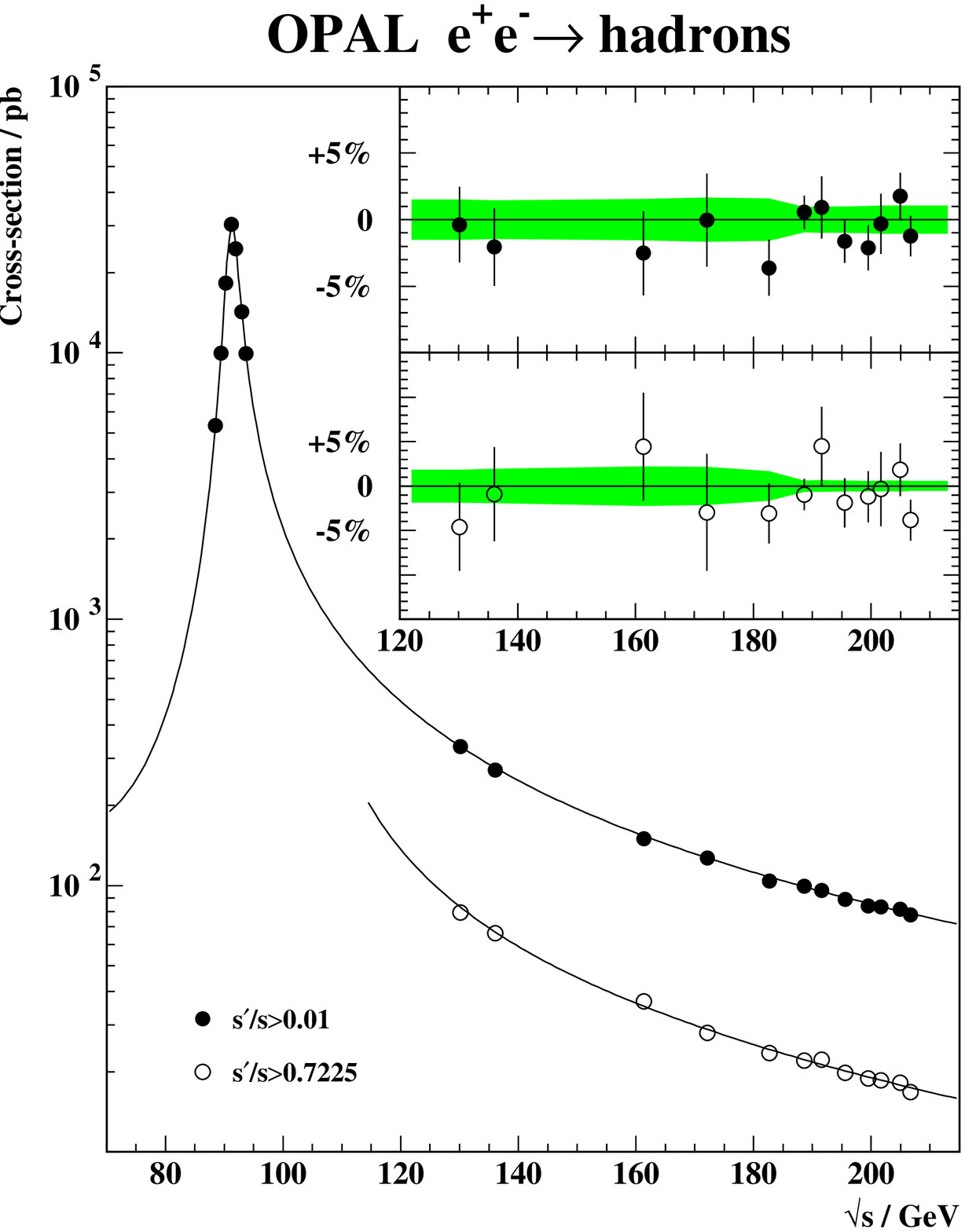}
  \caption
{
    Measured total cross-sections ($s'/s>0.01$) for hadronic events
    at lower energies~\cite{bib:PR328, bib:OPAL-SM172,bib:OPAL-SM183}
    and this analysis. Cross-section measurements for $s'/s>0.7225$ from 
    this analysis and from~\cite{bib:OPAL-SM172,bib:OPAL-SM183} are also 
    shown; the values at 161~GeV and 172~GeV have been corrected from 
    $s'/s > 0.8$ to $s'/s > 0.7225$ by adding the prediction of \ZFITTER\ for 
    this difference before plotting. The curves show the predictions of 
    \ZFITTER.
    The insets show the percentage differences between the measured
    values and the \ZFITTER\ predictions for the high energy points.
    %for (a) $s'/s>0.01$ and (b) $s'/s>0.7225$.
    The error bars on the differences represent statistical errors only; 
    the size of the experimental systematic error is indicated by the 
    shaded band.
}
\label{fig:mh_xsec}
\end{figure}

\begin{figure}
  \epsfxsize=0.96\textwidth
  %\epsfbox[0 0 567 680]{xs_mu.eps}
  \epsfbox{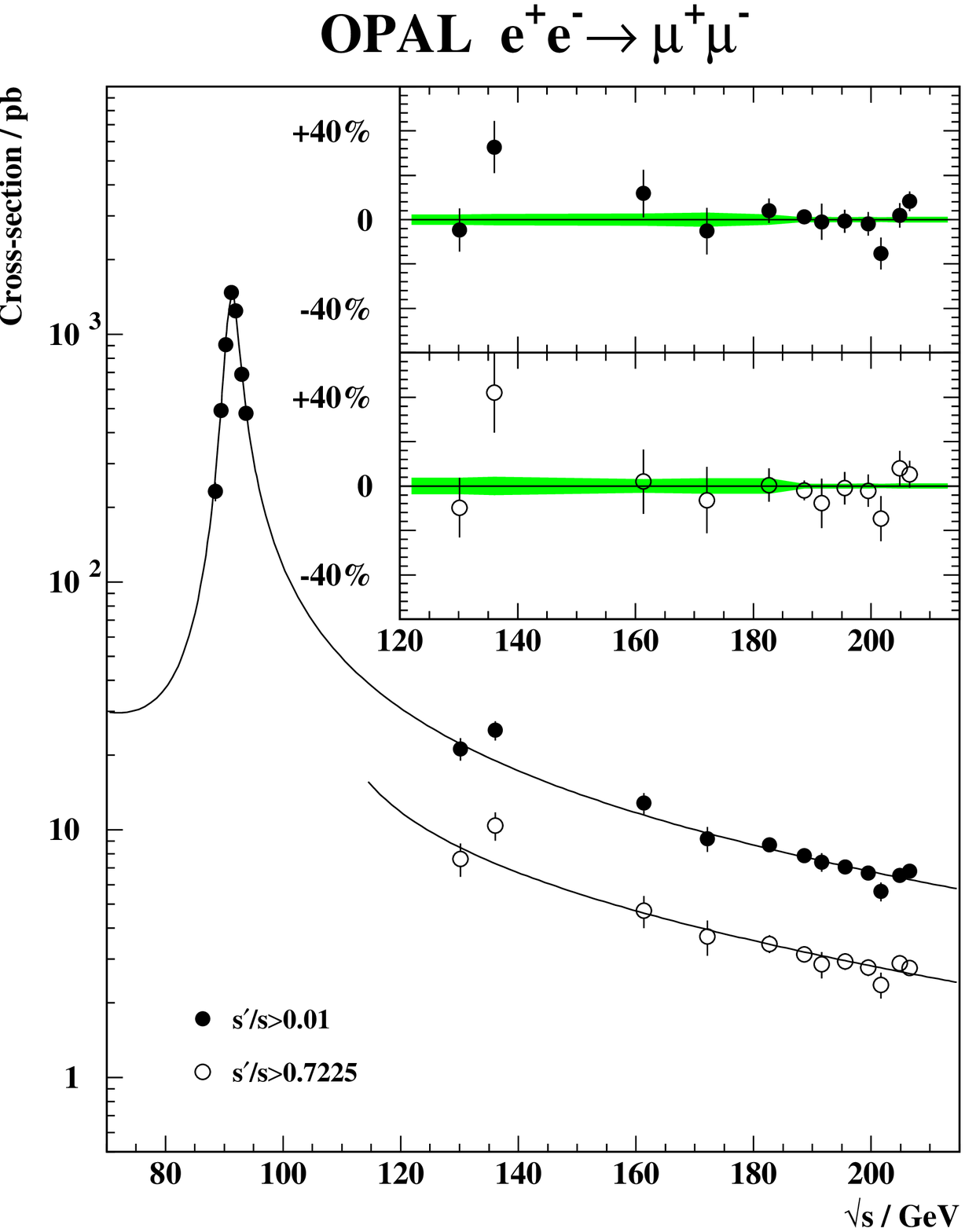}
  \caption
{
    Measured total cross-sections ($s'/s>0.01$) for $\mumu$ events
    at lower energies~\cite{bib:PR328,bib:OPAL-SM172,bib:OPAL-SM183}
    and this analysis. Cross-section measurements for $s'/s>0.7225$ from 
    this analysis and from~\cite{bib:OPAL-SM172,bib:OPAL-SM183} are also 
    shown; the values at 161~GeV and 172~GeV have been corrected from 
    $s'/s > 0.8$ to $s'/s > 0.7225$ by adding the prediction of \ZFITTER\ for 
    this difference before plotting. The curves show the predictions of 
    \ZFITTER.
    The insets show the percentage differences between the measured
    values and the \ZFITTER\ predictions for the high energy points.
    %for (a) $s'/s>0.01$ and (b) $s'/s>0.7225$.
    The error bars on the differences represent statistical errors only; 
    the size of the experimental systematic error is indicated by the 
    shaded band. 
}
\label{fig:mu_xsec}
\end{figure}

\begin{figure}
  \epsfxsize=0.96\textwidth
  %\epsfbox[0 0 567 680]{xs_tau.eps}
  \epsfbox{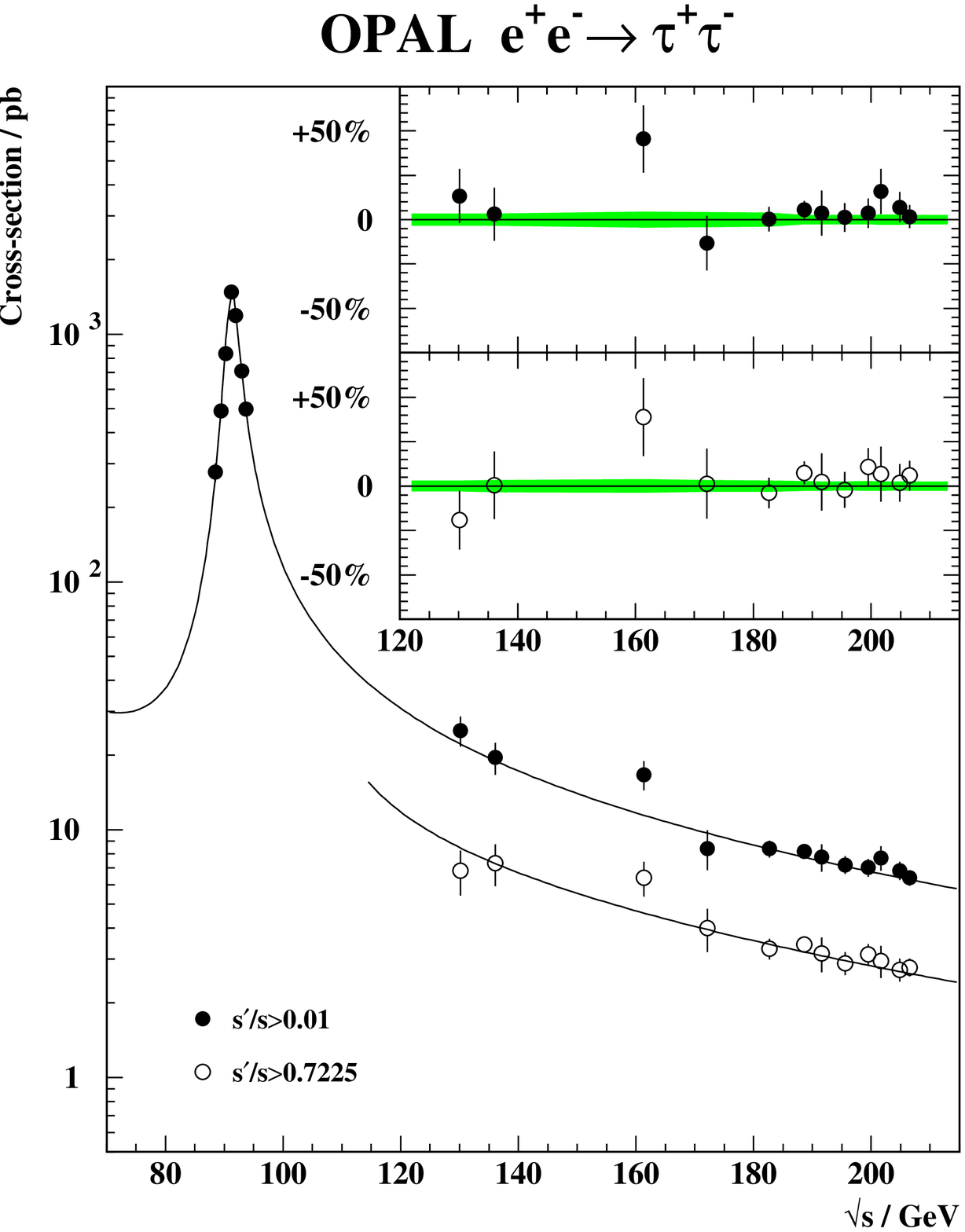}
  \caption
{
    Measured total cross-sections ($s'/s>0.01$) for $\tautau$ events
    at lower energies~\cite{bib:PR328,bib:OPAL-SM172,bib:OPAL-SM183}, 
    and this analysis. 
    Cross-section measurements for $s'/s>0.7225$ from this analysis and 
    from~\cite{bib:OPAL-SM172,bib:OPAL-SM183} are also shown; 
    the values at 161~GeV and 172~GeV have been corrected from $s'/s > 0.8$ to 
    $s'/s > 0.7225$ by adding the prediction of \ZFITTER\ for this difference 
    before plotting. The curves show the predictions of \ZFITTER.
    The insets show the percentage differences between the measured
    values and the \ZFITTER\ predictions for the high energy points.
    %for (a) $s'/s>0.01$ and (b) $s'/s>0.7225$.
    The error bars on the differences represent statistical errors only; 
    the size of the experimental systematic error is indicated by the 
    shaded band.
}
\label{fig:tau_xsec}
\end{figure}

\begin{figure}
  \epsfxsize=0.96\textwidth
  %\epsfbox[0 0 567 680]{xs_ee.eps}
  \epsfbox{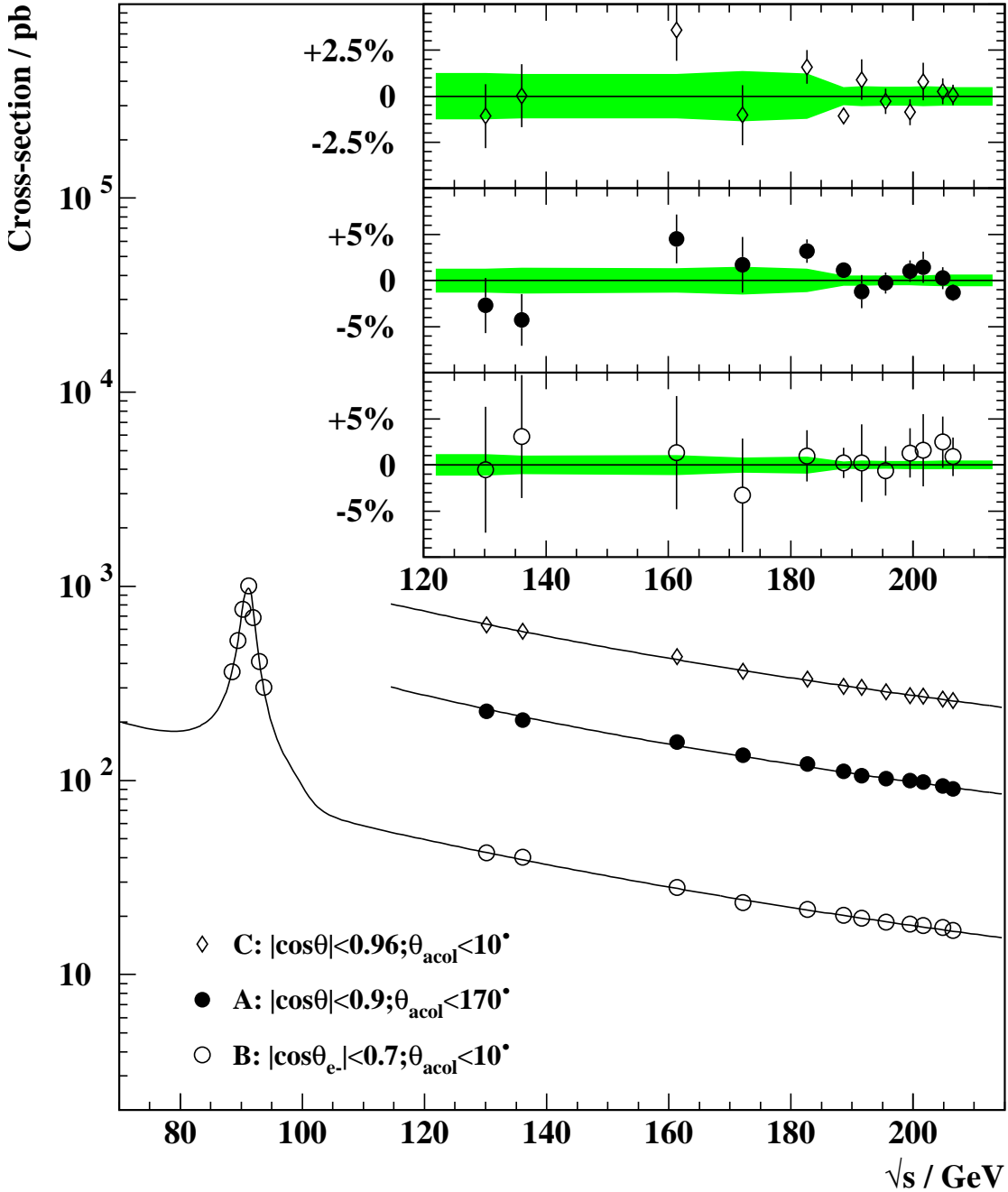}
  \caption
{
    Measured cross-sections for $\epem$ events at lower 
    energies~\cite{bib:PR328,bib:OPAL-SM172,bib:OPAL-SM183}, and this 
    analysis. The curves show the predictions of \BHWIDE.
    The insets show the percentage differences between the measured
    values and the \BHWIDE\ predictions for the high energy points.
    %for 
    %(a) $\absct < 0.96$, $\thacol < 10\degree$,
    %(b) $\absct < 0.9$, $\thacol < 170\degree$ and
    %(c) $\absctem < 0.7$, $\thacol < 10\degree$.
    The error bars on the differences represent statistical errors only; 
    the size of the experimental systematic error is indicated by the 
    shaded band.
}
\label{fig:ee_xsec}
\end{figure}

\begin{figure}
\epsfxsize=\textwidth
 %\epsfbox[0 0 567 680]{afb.eps}
\epsfbox{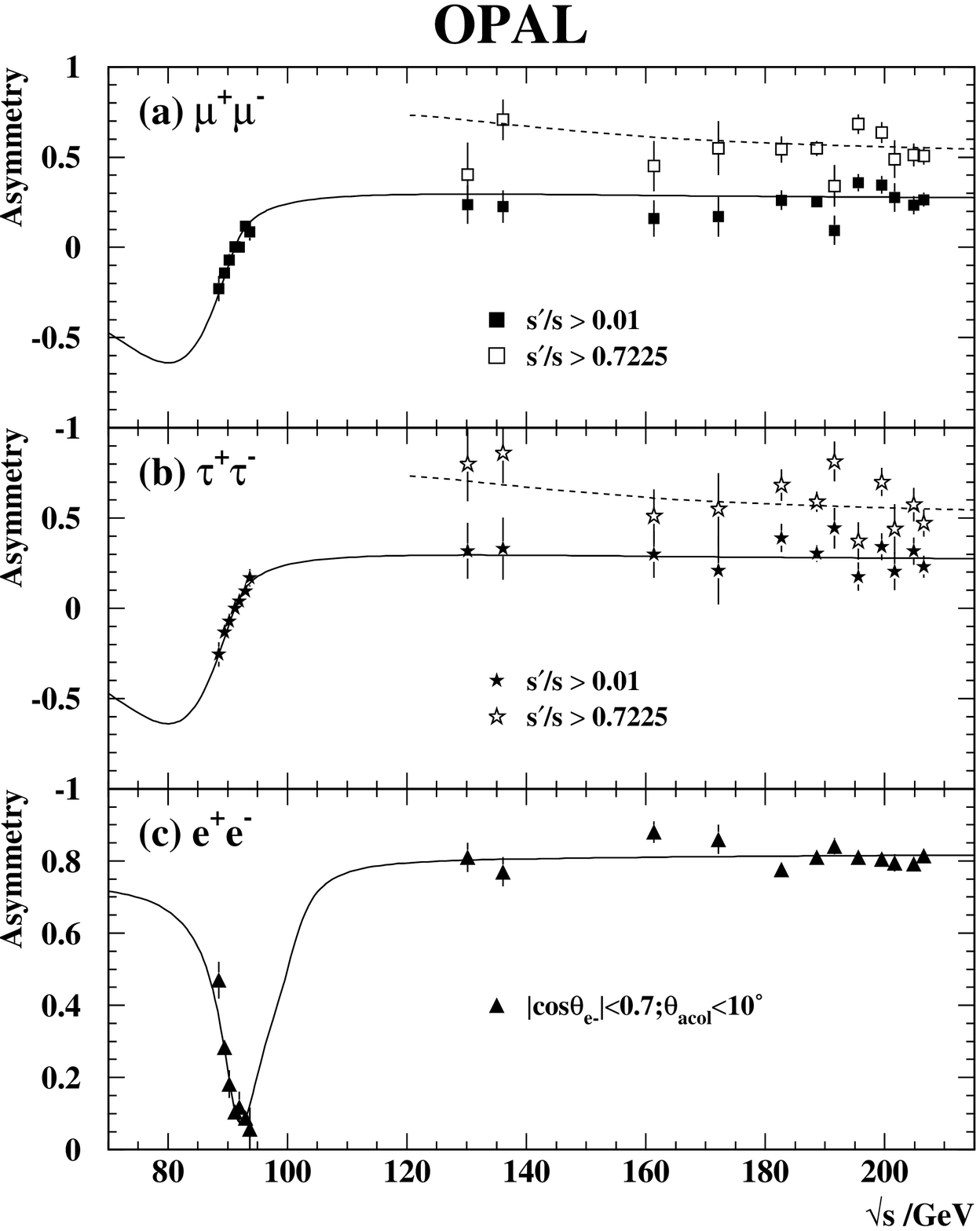}
\caption
{ (a) and (b) Measured asymmetries for inclusive ($s'/s>0.01$) and 
  non-radiative ($s'/s>0.7225$) samples as functions of $\protect\roots$
  for \Pgmp\Pgmm\ and \Pgtp\Pgtm\ events.  
  The curves show \ZFITTER\ predictions for $s'/s>0.01$ (solid) and
  $s'/s>0.7225$ (dashed).
  (c) Measured forward-backward asymmetry for \epem\ with 
  $|\cos\theta_{\Pem}|<0.7$ and $\thacol<10\degree$, as a function
  of $\protect\roots$. The curve shows the prediction of \BHWIDE.
  %Some points are plotted
  %at slightly displaced values of $\protect\roots$ for clarity.
  %as well as the Born-level expectation
  %without QED radiative effects (dashed). The expectation for $s'/s>0.7225$ 
  %lies very close to the Born curve, such that it appears 
  %indistinguishable on this plot. 
  Lower energy data values are taken 
  from~\cite{bib:PR328,bib:OPAL-SM172,bib:OPAL-SM183} for all channels.
}
\label{fig:afb}
\end{figure}

%-----------------------------------------------------------------------
The cross-section and asymmetry measurements at 189--207~GeV are compared
with the Standard Model predictions in Tables~\ref{tab:xsec1}--\ref{tab:afb2}. 
Figures~\ref{fig:mh_xsec}--\ref{fig:ee_xsec} show cross-sections, 
for both inclusive and non-radiative events, as a function of $\roots$, 
while Fig.~\ref{fig:afb} shows the measured asymmetry values.
The Standard Model predictions are calculated using \BHWIDE~\cite{bib:bhwide}
for the \epem\ final state and \ZFITTER~\cite{bib:zfitter} for all other 
final states; in this paper we use \ZFITTER\ version 6.30 with the following 
input parameters: $\mPZ$    = 91.1852~GeV~\cite{bib:PR328}, 
                  $\mtop$   = 174.3~GeV~\cite{bib:pdg2000},
                  $\mHiggs$ = 115~GeV,
                  $\Delta\alpha_{\rm had}^{(5)}$ = 0.02761~\cite{bib:dal5h} and
                  $\alphas(\mPZ^2)$  = 0.1185~\cite{bib:pdg2000}.
The theoretical uncertainties on the cross-section predictions are estimated 
to be 0.26\% for hadronic final states, 0.4\% for muon and tau final states, 
0.5\% for 
electron final states in the endcap region and 2.0\% for electron final 
states in the barrel region~\cite{bib:LEPMCWS}. In the fits described in
Section~\ref{sec:new_phys} we assign these values as the theoretical 
errors on the Standard Model cross-sections. For the non-radiative asymmetry
values we use a theoretical error of 0.004, derived from comparison of
the predictions of \ZFITTER\ and \KK. The agreement between the measured 
cross-sections and asymmetry values and the Standard Model predictions is 
generally good.

\begin{figure}
\begin{sideways}
\begin{minipage}[b]{\textheight}
\begin{center}
\begin{tabular}{cc}
\epsfxsize=0.46\textwidth\epsfbox[0 0 567 680]{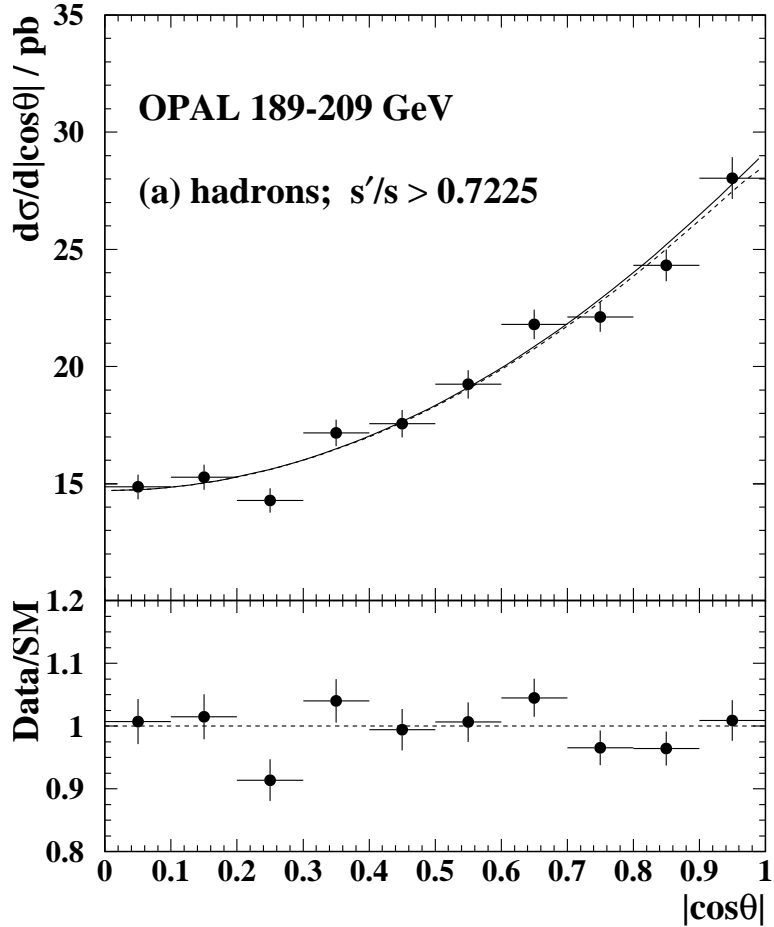} &
\epsfxsize=0.46\textwidth\epsfbox[0 0 567 680]{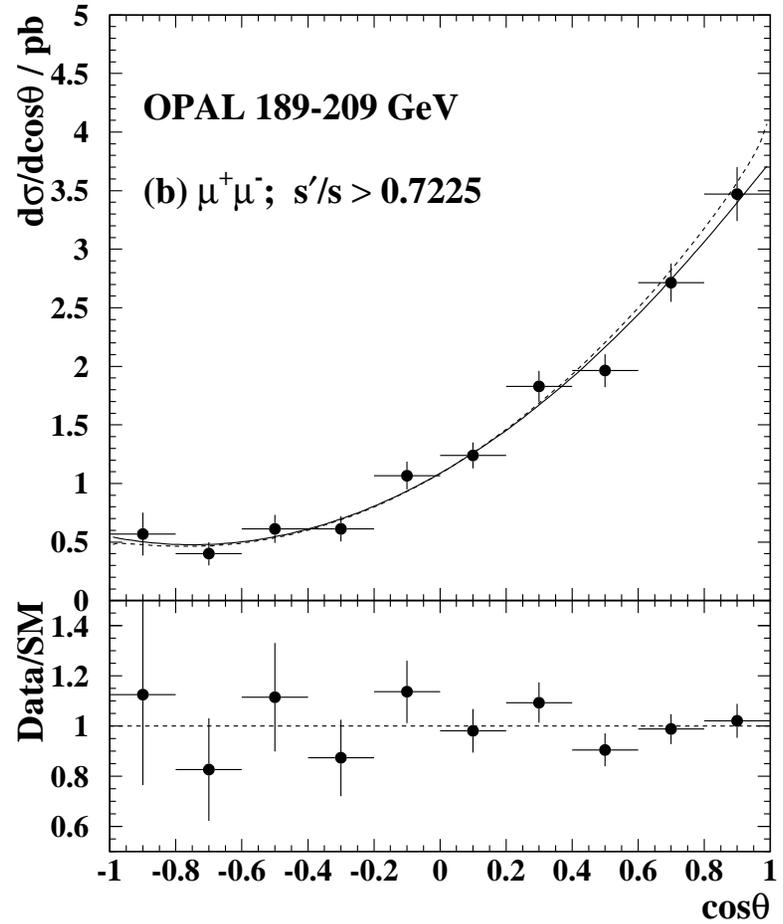} \\
\end{tabular}
\end{center}
\caption
{ Measured differential cross-sections for (a) hadronic events with 
  $s'/s > 0.7225$ and (b) \mumu\ events with $s'/s > 0.7225$. 
 The points show the luminosity-weighted average of all data from
 189~GeV to 207~GeV, corrected to no interference between initial- and 
 final-state radiation. The curves show the predictions of \ZFITTER\ without 
 interference between initial- and final-state radiation (solid) and with 
 interference (dashed). In each case the lower plot shows the ratio of
 the measurements to the Standard Model predictions (excluding interference).
}
\label{fig:angdis1}
\end{minipage}
\end{sideways}
\end{figure}

\begin{figure}
\begin{sideways}
\begin{minipage}[b]{\textheight}
\begin{center}
\begin{tabular}{cc}
\epsfxsize=0.46\textwidth\epsfbox[0 0 567 680]{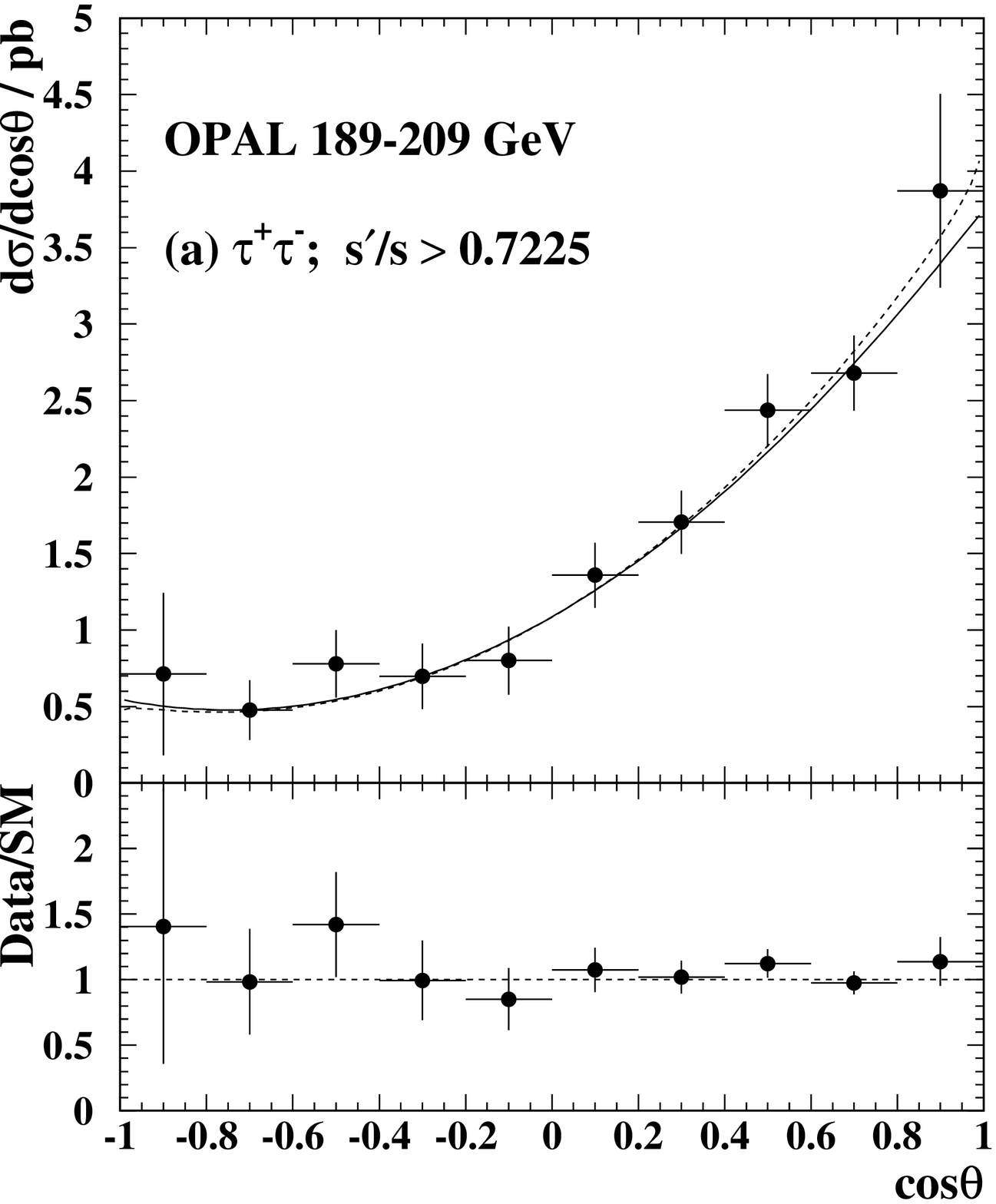} &
\epsfxsize=0.46\textwidth\epsfbox[0 0 567 680]{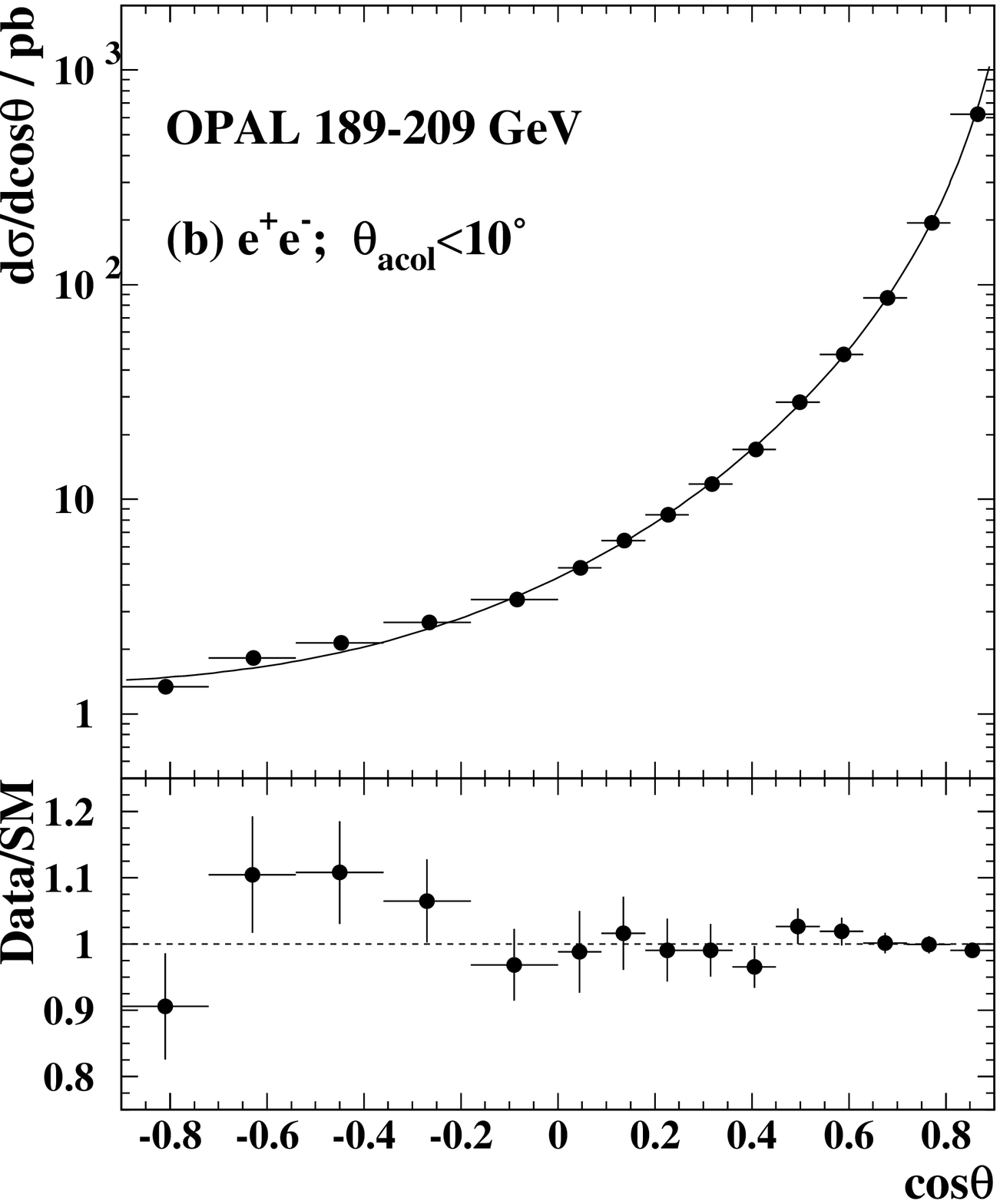} \\
\end{tabular}
\end{center}
\caption
{ Measured differential cross-sections for (a) \tautau\ events with 
  $s'/s > 0.7225$ and (b) \epem\ events with $\thacol < 10\degree$.
 The points show the luminosity-weighted average of all data from
 189~GeV to 207~GeV, corrected to no interference between initial- and 
 final-state radiation in (a). The curves in (a) show the predictions of 
 \ZFITTER\ without interference between initial- and final-state radiation 
 (solid) and with interference (dashed). The curve in (b) shows the
 prediction of \BHWIDE. In each case the lower plot shows the ratio of
 the measurements to the Standard Model predictions (excluding interference
 in (a)).
}
\label{fig:angdis2}
\end{minipage}
\end{sideways}
\end{figure}

The measured differential cross-sections at each energy are given in
Tables~\ref{tab:qq_angdis}--\ref{tab:ee_angdis}. The luminosity-weighted 
averages of data at all energies are compared with Standard Model predictions 
in Figs.~\ref{fig:angdis1} and~\ref{fig:angdis2}. The data are
well-described by the Standard Model curves.

In order to make a more quantitative test of the compatibility of our 
cross-sections and asymmetries with the Standard Model, we calculate a 
$\chi^2$ value between the measurements and the Standard Model predictions 
taking into account statistical and systematic errors and their correlations. 
Correlations between hadron and lepton cross-sections are very small, 
arising mainly
from the common luminosity measurements. Correlations between cross-sections
at different energies for the same channel arise from the systematic 
uncertainties in both efficiency and background, but amount at most to 
7.6\% for hadrons, 4.8\% for $\mumu$ and 11.4\% for $\tautau$. 
Correlations between cross-section and asymmetry measurements are generally 
negligible, amounting at most to about 1\% for $\tautau$, arising from
uncertainties in the (mainly Bhabha) background. 

The $\chi^2$ values for the hadronic cross-sections, and the $\mumu$ and 
$\tautau$ cross-sections and asymmetries, are shown in Table~\ref{tab:sm_fit}. 
Note that the `non-radiative' samples with $s'/s > 0.7225$ are a subset of 
the inclusive events with $s'/s > 0.01$, so the two $\chi^2$ values are not 
independent. All measurements are in agreement with the Standard Model 
expectations. 

The $\chi^2$ test would not necessarily reveal a discrepancy in the
overall scale of the cross-sections or asymmetries compared with the
Standard Model expectations. Therefore, as a further check, we have also 
calculated the average value of the ratio of the measurement to the 
Standard Model prediction, using a $\chi^2$ minimization technique including 
the experimental systematic errors. The results are shown in 
Table~\ref{tab:sm_fit}. All mean values are compatible with unity. 
The data are thus shown to be compatible with the Standard Model expectations 
to a precision of 1\% for hadrons and $\sim$3\% for leptons.
 
The cross-sections for $\epem$ events are dominated by the large \ct\
region. Rather than comparing the measured integrated cross-sections with
the Standard Model, we have calculated a $\chi^2$ for the differential
cross-sections, as presented in Table~\ref{tab:ee_angdis}. Correlations
between \ct\ bins and between energies are less than 10\%, except
for the region $\ct > 0.7$ where the systematic error is a significant
fraction of the total error; in this region the correlation between
measurements at different energies is 30\%--40\%. We find a $\chi^2$
%value of 49.7 for 63 degrees of freedom, showing excellent agreement.
value of 83.3 for 105 degrees of freedom, showing excellent agreement.

%%-----------------------------------------------------------------------
%\subsection{\boldmath Energy Dependence of \alphaem}      \label{sec:alphaem}
%%-----------------------------------------------------------------------
%\input alpha.tex
%%-----------------------------------------------------------------------
\subsection{\boldmath Energy Dependence of \alphaem}      \label{sec:alphaem}
%%-----------------------------------------------------------------------

\begin{figure}
\begin{sideways}
\begin{minipage}[b]{\textheight}
\begin{center}
\begin{tabular}{cc}
\epsfxsize=0.47\textwidth\epsfbox[0 0 567 567]{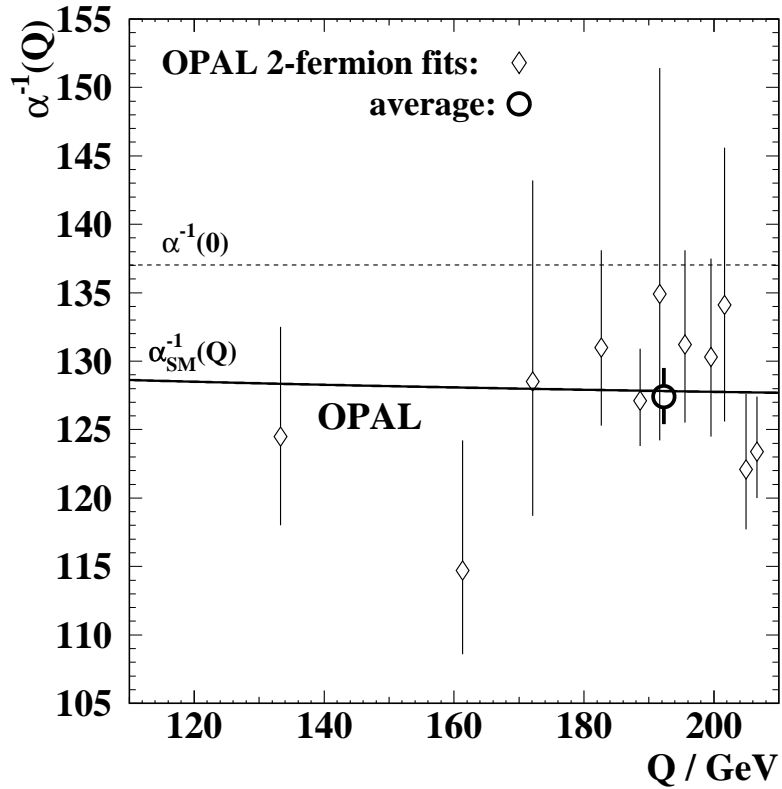} &
\epsfxsize=0.47\textwidth\epsfbox[0 0 567 567]{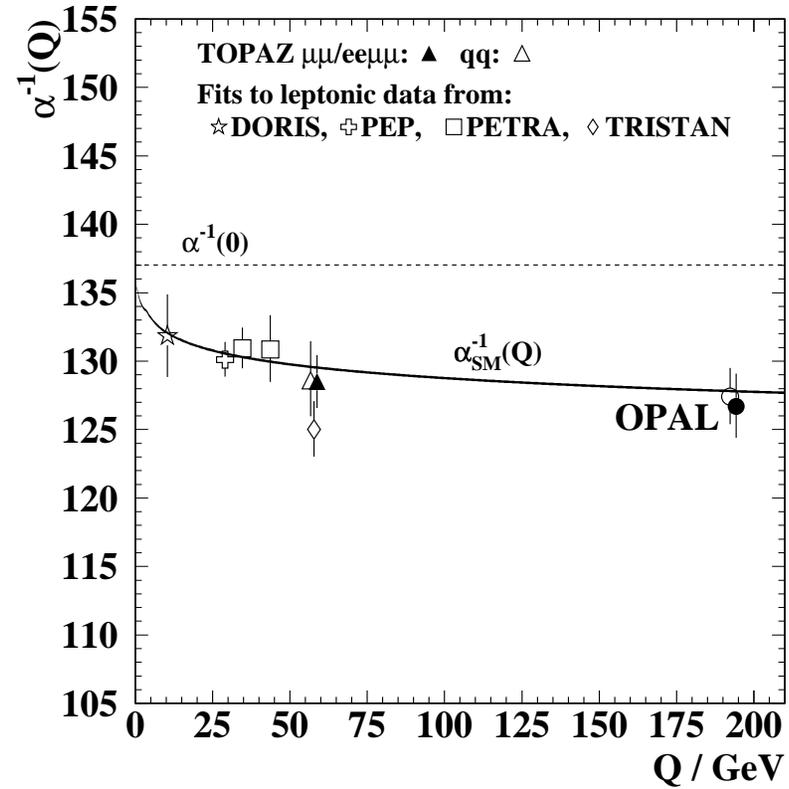} \\
\end{tabular}
\end{center}
\caption
{Fitted values of $1/\alphaem$ as a function of $Q$, which is
 $\protect \sqrt{s}$ for the OPAL fits. The left plot shows the
 results of fits to OPAL data at each centre-of-mass energy and
 of the combined fit in which \alphaem\ runs with a slope obtained from 
 fixing $1/\alphaem(0) = 137.036$. The right plot compares the results
 of the OPAL combined fits with values obtained by the TOPAZ 
 experiment~\cite{bib:alrun} and from fits to measurements of leptonic 
 cross-sections and asymmetries at the DORIS, PEP, PETRA and TRISTAN 
 \epem\ storage rings~\cite{bib:MK_alphaem}.
 Measurements shown by open symbols rely on assuming the Standard Model 
 running of \alphaem\ for $Q_{\mathrm{lumi}}$ below 4~GeV, whereas
 closed symbols indicate values derived from cross-section ratios which 
 do not depend on luminosity, as discussed in  Section~\ref{sec:alphaem}. 
 The solid line shows the Standard 
 Model expectation, with the thickness representing the uncertainty, 
 while the value of 1/$\alphaem(0)$ is shown by the dashed line. 
}
\label{fig:alphaem}
\end{minipage}
\end{sideways}
\end{figure}

In~\cite{bib:OPAL-SM172,bib:OPAL-SM183,bib:OPAL-SM189} we used non-radiative 
cross-section 
and asymmetry measurements to determine the electromagnetic coupling
constant \alphaem\ at \LEPtwo\ energies. We have repeated this fit
including the new measurements of hadronic, \mumu\ and \tautau\
cross-sections and the $\mumu$ and $\tautau$ asymmetry
values for $s'/s > 0.7225$ presented here. As before, we form
the $\chi^2$ between the measured values and the Standard Model 
predictions calculated as a function of $\alphaem(\sqrt{s})$ using
\ZFITTER, with all other \ZFITTER\ input parameters fixed~\cite{bib:zfitter}.
Correlations between measurements are fully taken into account. 
In Table~\ref{tab:alphaem1} we show the results of these fits. We 
perform fits to the data at each energy and also perform a fit to data at all
centre-of-mass energies in which $\alphaem$ runs with energy with a 
slope\footnote{If the Standard Model
running of $\alphaem$ is given by $\alpha_{\rm em}^{\rm SM} = \alphaem(0)/
(1-\Pi(Q))$, then we determine a constant $\kappa$ close to 1 such that
$\alpha_{\rm em}^{\rm fit} = \alphaem(0)/(1-\kappa\Pi(Q))$, i.e. the slope
${\rm d}\alphaem/{\rm d}\ln Q$ is multiplied by $\kappa$.}
obtained from fixing $1/\alphaem(0) = 137.036$.
As input to the combined fit we use the new measurements presented here 
together with the corresponding measurements at 130--183~GeV
from~\cite{bib:OPAL-SM172,bib:OPAL-SM183}.
For the combined fit the value of \alphaem\ is quoted at the 
centre-of-mass energy corresponding to the luminosity-weighted average of 
$1/s$. The errors on the fitted values of \alphaem\ arise from the errors on
the measurements; errors due to uncertainties in the \ZFITTER\ input
parameters are negligible.
The measured values of $\alphaem$ are shown in Fig.~\ref{fig:alphaem}.
They are consistent with the Standard Model expectations. 

The fits described above use measurements of cross-sections which 
depend on the measurement of luminosity. The luminosity measurement assumes 
the Standard Model running of $\alphaem$ from $Q^2 = 0$ to typically
$Q^2 = (4~\GeV)^2$, where\footnote{The hadronic vacuum 
polarization contribution used in the luminosity measurement is 
from~\cite{bib:bhlumi_dal5h}.} $1/\alphaem\simeq 134$. 
The fits therefore measure the running of $\alphaem$ only from 
$Q_{\rm lumi}\simeq 4~\GeV$ upwards. To become independent of the luminosity 
measurement, we have repeated the fits replacing the 
cross-sections for hadrons, muon and tau pairs with the ratios 
$\sigma(\mu\mu)/\sigma(\qqbar)$ and $\sigma(\tau\tau)/\sigma(\qqbar)$. 
This is possible since, above the Z peak, hadrons and leptons have very 
different sensitivity to \alphaem\ as discussed in~\cite{bib:OPAL-SM172}.
The results of these fits are also shown in Table~\ref{tab:alphaem1}.
The values of $1/\alphaem$ are close to those obtained from the cross-section 
fits but with somewhat larger errors. The value of $1/\alphaem$ obtained 
from the combined fit is
$1/\alphaem(193.2~\mathrm{GeV}) = 126.7^{+2.4}_{-2.3}$. 
This is about 4.3 standard deviations below the low energy limit of 
137.03599976(50)~\cite{bib:Mohr}, thus demonstrating the running of 
$\alphaem$ from $Q^2 = 0$ to \LEPtwo\ energies. 
This measurement of $\alphaem$ does not depend on calculations of low-mass 
hadronic loops and is nearly independent of the mass of the Higgs boson and 
$\alphas$; it can be scaled to the mass of the Z, giving 
$1/\alphaem(91.19~\mathrm{GeV}) = 127.9^{+2.1}_{-2.8}$, in good agreement
with the Standard Model prediction of 128.936$\pm$0.046~\cite{bib:dal5h}.

%%-----------------------------------------------------------------------
%\section{S-matrix Analysis}                            \label{sec:smat}
%%-----------------------------------------------------------------------
%\input smatrix.tex
%%-----------------------------------------------------------------------
\section{S-matrix Analysis}                            \label{sec:smat}
%%-----------------------------------------------------------------------
\subsection{Introduction}
Fermion-pair production cross-sections and asymmetries at \LEPone\ provide 
precise information about the Z resonance. The resonance can be described with
five parameters: the Z mass $\mPZ$, the Z width $\Gamma_{\rm Z}$, the
total hadronic cross-section at the peak, its ratio to the leptonic 
cross-section and the leptonic forward-backward asymmetry at the peak. 
Fitting the OPAL data for just these five parameters leads to a precise 
determination of the Z mass 
\mbox{$\mPZ =91.1852\pm 0.0030$ GeV}~\cite{bib:PR328}.
However, this fit assumes the contribution to the hadronic cross-section 
from \gZ\ interference behaves as predicted by the Standard Model. 
A more model-independent description of the Z lineshape is provided by
the S-matrix approach discussed in this section.
%A less biased description of the Z lineshape can be provided by a more model 
%independent parameterization.

The S-matrix formalism \cite{bib:smatrix} describes the process 
\mbox{$\epem\to \PZ / \gamma^\ast \to \ffbar$} assuming only 
the exchange of a combination of two neutral spin-1 bosons of which
one is massless. Contributions from boson exchange and interference are
explicitly allowed to vary independently.
The resulting parameterizations of the
fermion-pair cross-section $\sigma^0_{\rm tot}(s)$ and asymmetry
$A^0_{\rm fb}(s)$ in lowest order are: 

\begin{eqnarray}
 \sigma^0_{\rm tot}(s) = \frac{4}{3}\pi\alpha_{\rm em}^2\left[
    \frac{{g^{\rm tot}_{\rm f}}}{s}
 +  \frac{{j^{\rm tot}_{\rm f}}(s-\overline{m}^2_{\rm Z}) + 
          {r^{\rm tot}_{\rm f}}s}
         {(s-\overline{m}^2_{\rm Z})^2 + 
          \overline{m}^2_{\rm Z}\overline{\Gamma}^2_{\rm Z}}\right] 
\nonumber \\
  A^0_{\rm fb}(s) =  \frac{\pi\alpha_{\rm em}^2}{\sigma^0_{\rm tot}}\left[
    \frac{{g^{\rm fb}_{\rm f}}}{s} 
 +  \frac{{j^{\rm fb}_{\rm f}}(s-\overline{m}^2_{\rm Z}) + 
          {r^{\rm fb}_{\rm f}}s}
         {(s-\overline{m}^2_{\rm Z})^2 + 
          \overline{m}^2_{\rm Z}\overline{\Gamma}^2_{\rm Z}}\right]. 
\label{eq:smat}
\end{eqnarray}
%Here $\alphaem$ is the fine structure constant and $\sqrt{s}$ the 
%centre-of-mass energy. 
Besides the Z mass $\overline{m}_{\rm Z}$ and 
width $\overline{\Gamma}_{\rm Z}$ there are six
parameters per final state fermion $\rm f$, three for the cross-sections 
and three for the asymmetries. The photon exchange is described by 
$g^{\rm tot}_{\rm f}$ and $g^{\rm fb}_{\rm f} (\equiv 0)$ and
is assumed to be known. This leaves four parameters, namely 
$r^{\rm tot}_{\rm f}$, $r^{\rm fb}_{\rm f}$ describing the Z exchange and 
$j^{\rm tot}_{\rm f}$, $j^{\rm fb}_{\rm f}$ for the interference.
For the hadronic final state, the parameters are summed over all colours
and open flavours. Since the hadronic asymmetry is not measured,
$r^{\rm fb}_{\rm had}$ and $j^{\rm fb}_{\rm had}$ cannot be determined.
The lowest order expressions in Equation~(\ref{eq:smat}) serve to 
introduce the S-matrix parameters, but cannot be used directly to fit the
data without the inclusion of large QED radiative corrections.
To fit the data, expectations for cross-sections and asymmetries depending 
on S-matrix parameters including QED radiative corrections were calculated 
using the program \SMATASY~\cite{bib:smatasy} together with 
\ZFITTER~\cite{bib:zfitter}. These calculations also include very small
electroweak corrections to the photon couplings. 

In Equation~(\ref{eq:smat}) the $\PZ$ resonance is described with an 
$s$-independent width. Usually loop corrections to the $\PZ$ propagator are
absorbed in an $s$-dependent width via the transformation 
$\overline{\Gamma}_{\rm Z} \to  s \Gamma_{\rm Z} / {m}^2_{\rm Z}$. 
This results in a redefinition of the Z mass and width leading to
a numerical shift of: 
\begin{eqnarray}
\overline{m}_{\rm Z} & = & % m_{\rm Z} - \frac{\Gamma_{\rm Z}^2}{2 m_{\rm Z}}
m_{\rm Z} / \sqrt{1+ \Gamma_{\rm Z}^2 /  m_{\rm Z}^2}
\approx  m_{\rm Z} - \mbox{34 MeV} \; ,
\nonumber
\\
\overline{\Gamma}_{\rm Z} & = & % \Gamma_{\rm Z} - \frac{\Gamma_{\rm Z}^3}{2 m_{\rm Z}^2}
\Gamma_{\rm Z} / \sqrt{1+ \Gamma_{\rm Z}^2 /  m_{\rm Z}^2}
\approx  \Gamma_{\rm Z} - \mbox{1 MeV} \; .
\label{eq:mzcor}
\end{eqnarray}
Although the S-Matrix parameters are $\overline{m}_{\rm Z}$ and 
$\overline{\Gamma}_{\rm Z}$,
in this paper the above relations are used to
give numerical results for $m_{\rm Z}$ and $\Gamma_{\rm Z}$
to facilitate comparisons with other measurements.

A fit to \LEPone\ data alone leaving the parameters describing the 
interference free leads to a large uncertainty on the Z mass.
This is because most \LEPone\ data (about 88\%) are hadronic events taken
at three energy points. This results in effectively three very precise
measurements which dominate the determination of the $\PZ$ properties.
From these three measurements the four parameters $\mPZ$, $\GammaZ$,
$r^{\rm tot}_{\rm had}$ and $\jtoth$ cannot be determined simultaneously.  
The interference increases the cross-sections at energies above the peak 
and decreases them at lower energies. Therefore a change in this
contribution effectively shifts the position of the peak, which can be
interpreted as a change in the Z mass. This leads to a strong anti-correlation
of 96\% between the fit results of $\mPZ =91.1901\pm 0.0115$~GeV and 
$\jtoth=0.010\pm 0.650$ from \LEPone\ data alone~\cite{bib:PR328}. 

\LEPtwo\ data provide additional independent measurements to constrain the
contribution from interference. Using all OPAL measurements therefore leads 
to a determination of the Z mass which is less model-dependent than that
from the five parameter fit to \LEPone\ data, with an error that is only 
slightly larger. 
A similar analysis has been performed previously by other 
experiments~\cite{bib:L3-smat,bib:delphi-smat}.
Results presented here supersede the OPAL analysis at lower energies
\cite{bib:OPAL-SM172}.

\subsection{Fit Results}

To derive results for the S-matrix parameters a fit is performed comparing 
the predictions with OPAL measurements of fermion-pair production 
cross-sections and asymmetries at all LEP energies.
 
The \LEPone\ measurements are described in~\cite{bib:PR328}. The results for 
the hadronic, \epem, \mumu\ and \tautau\ cross-sections are given in 
Tables 8-12 of that paper. Leptonic asymmetries can be found in Tables 22-24.
The measurements used in this analysis are those already corrected
for the beam energy spread. 
In the analytical program used to calculate the S-matrix predictions 
$t$-channel
exchange is not implemented. To use Bhabha cross-sections and asymmetries
the $t$-channel contribution is corrected in the same 
way as described in~\cite{bib:PR328}. \ALIBABA~\cite{bib:alibaba} is used
to determine the Bhabha forward and backward cross-section for
the full $s+t$-channel and for the $s$-channel only. During the fit
the difference is added to the S-matrix predictions. 
The $t$-channel correction is parameterized as a function of
($\sqrt{s} - \mPZ$) and thus depends on the value of the
fitted Z mass. The systematic errors and their correlations are taken as 
described in~\cite{bib:PR328}. The treatment of the errors on the
 centre-of-mass energy and beam-energy spread is however simplified. 
For the fits described in~\cite{bib:PR328} the effects of 
these errors were determined iteratively during the fitting procedure. 
In this analysis the effect is calculated prior to
the fit according to the lineshape determined in~\cite{bib:PR328}. 
With this simplified error treatment the S-matrix result given 
in~\cite{bib:PR328} can be reproduced, with deviations of all fit parameters 
less than 3\% of their total error. 

The high energy measurements used are the non-radiative cross-sections
and asymmetries for hadrons, $\mumu$ and $\tautau$ presented here, together 
with the corresponding results from 
130--183~GeV~\cite{bib:OPAL-SM172,bib:OPAL-SM183}\footnote{At centre-of-mass
energies of 161~GeV and 172~GeV the cut is $s'/s > 0.8$.}. 
Bhabha cross-sections and asymmetries are not
included since even with a tight selection\footnote{Acollinearity 
$\theta_{\rm acol} < 10^\circ$ and angular range for the electron 
$|\cos{\theta_{\rm e^-}}| < 0.7$.} the observed total cross-section,
including $s+t$-channel, is about an order of magnitude larger than the 
$s$-channel contribution alone. Correlations between the \LEPtwo\
measurements are taken into account as discussed in Section~\ref{sec:SM}.
The systematic errors on the \LEPone\ and \LEPtwo\ data are basically 
uncorrelated.
The only common error is the systematic and theoretical error on the
luminosity determination. The resulting effect is found to be small and
has been neglected. 

Expectations for cross-sections and asymmetries depending on S-matrix 
parameters are calculated using the program \SMATASY~\cite{bib:smatasy} 
together with \ZFITTER~\cite{bib:zfitter}. 
A $\chi^2$ is calculated between the predictions and the measurements
of cross-sections and asymmetries. The results of fits to  
OPAL data taken at all LEP energies, with and without the assumption of 
lepton universality, are given in Table~\ref{tab:smat}.
The correlation matrices are given in Tables \ref{tab:smat_cor16} 
and~\ref{tab:smat_cor8}.
In principle the results for the S-matrix parameters do not depend on
the values of the Standard Model parameters. However, there is a small
effect due to the correction from the $s$-independent width parameters
$\overline{m}_{\rm Z}$, $\overline{\Gamma}_{\rm Z}$ to the usual parameters
$\mPZ$, $\GammaZ$ since the width $\GammaZ$ depends on the Standard Model
parameters. This effect is negligible compared with the overall errors
on the fitted parameters.
As a cross-check a fit is performed using only the \LEPtwo\ data with the
constraint of the S-matrix result given in~\cite{bib:PR328}. 
This leads to results very similar to the full fit, with differences of all 
fit parameters being less than 2\% of their error.

As can be seen in Table \ref{tab:smat}
the fitted Z mass gets smaller by 1.0 MeV when lepton universality is imposed.
This shift was already observed in a $C$-parameter fit described 
in Section 11.1 of~\cite{bib:PR328}, although it is larger in the S-matrix fit 
presented here. It is due to a subtle effect in the Bhabha 
$t$-channel correction, which gives an additional weak constraint on $\mPZ$
(see~\cite{bib:PR328} Appendix B).

\begin{figure}[p]
   \vspace*{-1cm}
   \begin{center} \mbox{
          \epsfxsize=0.98\textwidth
           %\epsffile{smat.eps}
           \epsffile{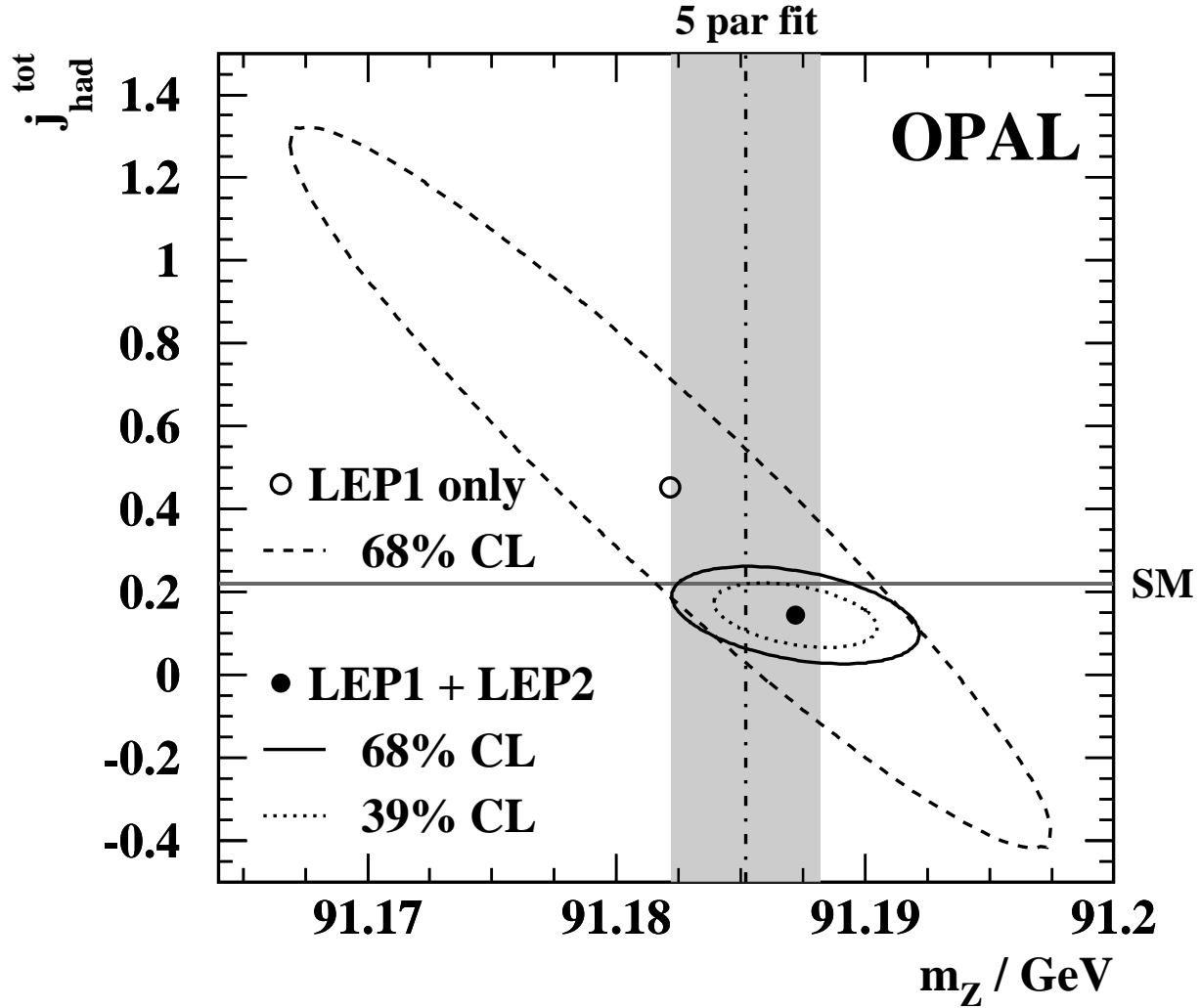}
           } \end{center}
           \vspace*{-1cm}
\caption[ ]{
 Confidence level contours in the $\mPZ$ - $\jtoth$ plane from the S-matrix 
 fits with lepton universality. The dashed curve shows the 68\% confidence
 level contour from the fit to \LEPone\ data alone, while the full and
 dotted curves show the 68\% and 39\% confidence level contours, respectively,
 from the fit to \LEPone\ and \LEPtwo\ data.
 The horizontal band indicates the Standard 
 Model value of $\jtoth$. The vertical band is the 1$\sigma$ error on the Z 
 mass from the five parameter fit~\cite{bib:PR328} which should be compared
 with the 39\% confidence level contour from the S-matrix fit.
}
\label{fig:smat}
\end{figure}

Figure~\ref{fig:smat} shows the correlation between $\mPZ$ and $\jtoth$ 
expressed as confidence level contours from fits with lepton universality 
imposed.
Using only \LEPone\ data the error on $\jtoth$ is 
large with a correlation coefficient of --0.96 to $\mPZ$.
Including the \LEPtwo\ data reduces the error on $\jtoth$ by a factor of 
five and 
the correlation is reduced to --0.39. As a result the error on $\mPZ$ is much 
improved from 11.5~MeV (\LEPone\ only) to 3.3~MeV. This is now
comparable to the error of 3.0~MeV obtained from the five parameter fit at 
\LEPone\ which assumes the \gZ\ interference according to the Standard Model.
Since the fit result for $\jtoth$ is close to the Standard Model expectation
the central values for $\mPZ$ are in good agreement.

%%-----------------------------------------------------------------------
\section{Limits on New Physics}    \label{sec:new_phys}
%%-----------------------------------------------------------------------
New physics could be revealed by deviations of the measured data from
Standard Model predictions. The generally good agreement seen between data 
and the Standard Model places severe constraints on the energy scale of such
new phenomena. The new data presented here have been combined with previous
measurements in order to provide updated limits on four-fermion contact 
interactions. In addition we present limits on the mass of a possible \zp\ 
boson. 

%\input ci.tex
%-----------------------------------------------------------------------
\subsection{Limits on Four-fermion Contact Interactions} 
\label{sec:ci}
%-----------------------------------------------------------------------

%A very general framework in which to search for and quantify the effect of 
%new physics is the four-fermion contact interaction. 
In the context of composite models of leptons and quarks, a four-fermion
contact interaction arises as a remnant of the binding force between the 
substructure of fermions.
Alternatively, a four-fermion contact interaction could be a good
description of deviations from the Standard Model due to the exchange
of a new very heavy boson of mass $m_{\rm X}$ if $m_{\rm X} \gg \sqrt{s}$.
More generally, the contact interaction is considered to be a convenient
parameterization to describe possible deviations from the Standard Model
which may be caused by some unknown new physics.

In this analysis we consider four-fermion contact interactions which
conserve flavour and helicity, and in which the 
$\rm SU(3) \times SU(2) \times U(1)$ gauge structure of the Standard
Model is valid. In this framework~\cite{bib:Eichten}
%In the framework of a contact interaction~\cite{bib:Eichten} 
the Standard Model Lagrangian for $\epem\to\ffbar$ is extended by a term 
describing a new effective interaction with an unknown coupling constant 
$g$ and energy scale $\Lambda$:
\begin{eqnarray}\label{eq-contact}
{\cal L}^{\mathrm{contact}} & = 
         & \frac{g^2}{(1 + \delta)\Lambda^2}
         \sum_{i,j=\mathrm{L,R}}\eta_{ij}[\bar{\Pe}_i\gamma^{\mu}{\Pe}_i]
                                       [\bar{\rm f}_j\gamma_{\mu}{\rm f}_j] ,
\end{eqnarray} 
where $\delta = 1$ for $\epem\to\epem$ and $\delta = 0$ otherwise.
Here $\mathrm{e_L} (f_\mathrm{L})$ and $\mathrm{e_R} (f_\mathrm{R})$ 
are chirality projections of electron (fermion) 
spinors, and $\eta_{ij}$ describes the chiral structure of 
the interaction.   
The parameters $\eta_{ij}$ are free in these models, but
typical values are between $-1$ and $+1$, depending on the type of theory 
assumed~\cite{bib:contacttable}. For example, a coupling of two 
right-handed currents is given by ($\eta_{\rm RR} = \pm 1, \eta_{\rm LL} =
\eta_{\rm LR} = \eta_{\rm RL} = 0$).
Here we consider the same set of 
models as in~\cite{bib:OPAL-SM172,bib:OPAL-SM183,bib:OPAL-SM189}.
The values of $\eta_{ij}$ which define these models are shown in
Table~\ref{tab:ccres}.

The inclusion of a contact interaction modifies both the total cross-section
and the angular distribution of fermion-pair production. In general, the
differential cross-section can be written in terms of a parameter 
$\varepsilon=(g^2/4\pi) / \Lambda^2$ as 
\begin{equation}\label{eq-sig_contact}
{{\rm d} \sigma \over {\rm d} \cos \theta } = 
   \sigma_{\rm SM}(s,t) + C_2^0(s,t) {\varepsilon} 
                        + C_4^0(s,t) {\varepsilon^2}\ .
\end{equation}
Here $t=-s(1-\cos\theta)/2$ and $\theta$ is the polar angle of the 
outgoing fermion with respect to the $\Pe^-$ beam direction. 
The $C_2^0$ term describes the interference between the Standard Model and 
the contact interaction, the $C_4^0$ term is the pure contact interaction 
contribution. The exact form of these terms depends on the type of fermion 
in the final state and the particular model chosen, and is given, 
for example, in~\cite{bib:OPAL-CI}\footnote{
Equation~(2) in~\cite{bib:OPAL-CI} has a typographical error:
the factor $4s$ on the left-hand side should be replaced by $2s$.}.
The interference term depends linearly on the $\eta_{ij}$
parameters, and thus can be positive or negative depending on their sign.
In fits to the data, the Standard Model cross-sections $
\sigma_{\rm SM}(s,t)$ were calculated
using \BHWIDE\ for the $\epem$ final state and \ZFITTER\ for all other
final states. Radiative corrections to the lowest order contact interaction
terms were taken into account as described in~\cite{bib:OPAL-SM172}.

We have fitted the measurements of the non-radiative cross-sections 
for $\eetoqq$, non-radiative cross-sections and asymmetries for $\eetomumu$ 
and $\eetotautau$ and the differential cross-sections for $\eetoee$ at 
189~GeV to 207~GeV presented here, together with the corresponding 
measurements at 130--183~GeV~\cite{bib:OPAL-SM172,bib:OPAL-SM183}. 
In all cases we use a $\chi^2$ fit, including the correlated systematic 
errors between the measurements and theoretical uncertainties in the 
Standard Model predictions as discussed in Section~\ref{sec:SM}. 
Fits are performed with the parameter $\varepsilon$
as the fitting parameter. The results for positive and negative interference
with the Standard Model (i.e.\ the sign of the $\eta_{ij}$ parameters) are
equivalent under the transformation $\varepsilon\leftrightarrow -\varepsilon$;
it is therefore sufficient to fit only for the case of positive interference
but to allow $\varepsilon$ to be both positive and negative.
Limits on the energy scale $\Lambda$ were extracted assuming $g^2/4 \pi = 1$. 
The 95\% confidence limits correspond to a change in $\chi^2$ of 3.84.

\begin{figure}
\begin{center}
\epsfxsize=0.86\textwidth
 %\epsfbox{ccfitbar.eps}
\epsfbox{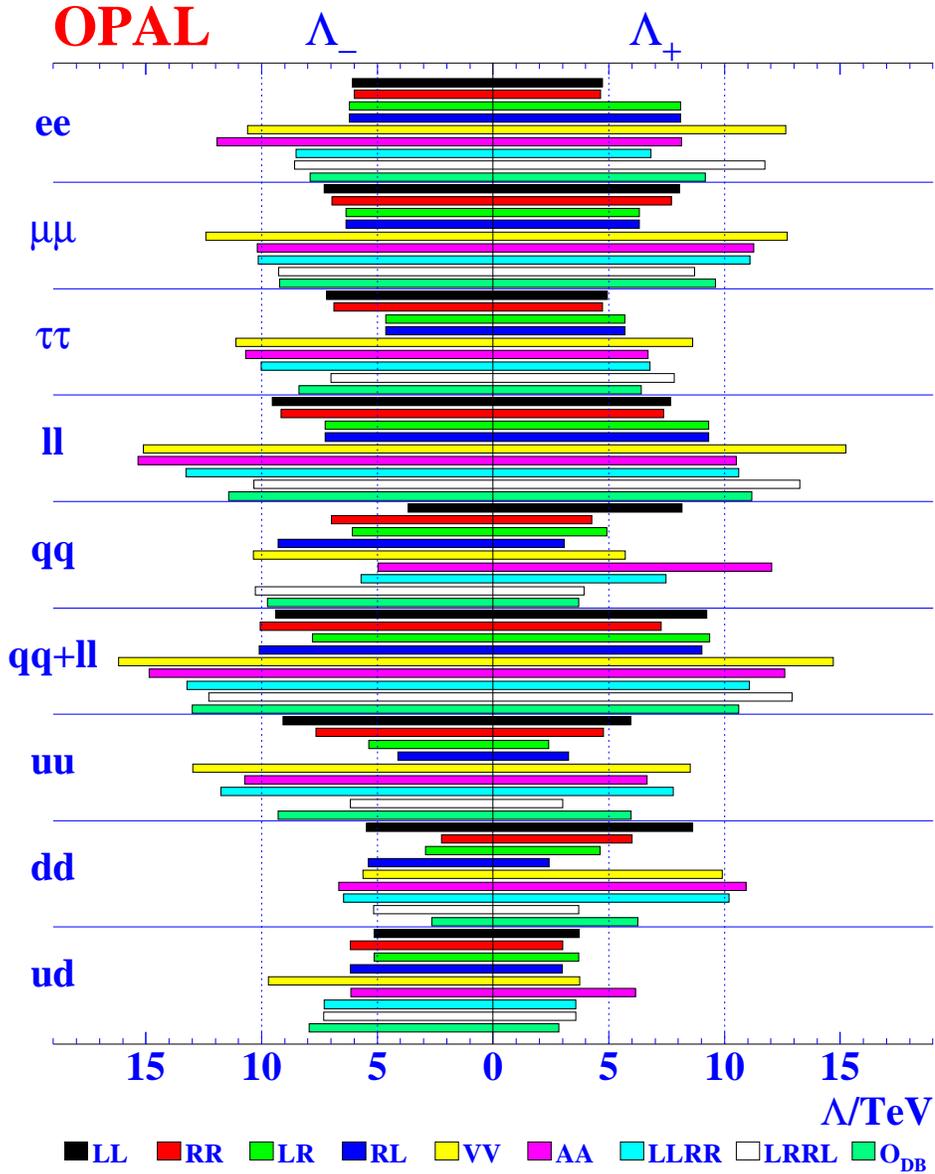}
\caption{95\% confidence level limits on the energy scale $\Lambda$
 resulting from the contact interaction fits to hadron and lepton-pair 
 data. For each channel, the bars from top to bottom indicate the results 
 for models LL to $\overline{\cal{O}}_{\mathrm{DB}}$ in the order given in 
 the key. The values for $\Lambda_+$ and $\Lambda_-$ correspond to the upper 
 and lower signs, respectively, of the $\eta_{ij}$ values which define 
 the models as given in Table~\ref{tab:ccres}.
}
\label{fig:ccres} 
\end{center}
\end{figure}

The results are shown in Table~\ref{tab:ccres} and illustrated
graphically in Fig.~\ref{fig:ccres}.  
The limits for \qqbar\ are derived from the hadronic cross-sections
assuming the new interaction couples to all flavours equally. 
Those for up-type quarks and down-type quarks are obtained by fitting 
the hadronic cross-sections assuming the new interaction couples only to one
flavour, whereas those for $\uubar + \ddbar$ assume a coupling to one 
generation only. The combined results include all leptonic channels and
the hadronic cross-sections.
The two sets of values $\Lambda_+$ and $\Lambda_-$ shown in 
Table~\ref{tab:ccres} correspond to positive and negative values of 
$\varepsilon$ respectively, reflecting the two possible signs 
of $\eta_{ij}$ in Equation~\eqref{eq-contact}.
The data are particularly sensitive to the 
VV and AA models; the combined data give lower limits on $\Lambda$ in the
range 13--16~TeV for these models.
For the other models the lower limits generally lie in the range 9--13~TeV.
The limits are typically 1~TeV higher than those for 130--189~GeV data 
alone~\cite{bib:OPAL-SM189}. 

Contact interactions involving quarks have also been studied in ep and
pp collisions, where limits comparable to our values are
found~\cite{bib:CI-ep,bib:CI-pp}. Atomic physics parity violation
experiments can place higher limits ($\simeq$ 15~TeV~\cite{bib:CI-atomic})
on models of eeuu and eedd contact interactions which violate parity.

%\input zp.tex
%-----------------------------------------------------------------------
\subsection{\boldmath Limits on a \zp\ Boson} 
\label{sec:zp}
%-----------------------------------------------------------------------
\subsubsection{\boldmath \zp\ Model Predictions}
Many theories predict a second heavy neutral vector boson $\rm Z'{^0}$ in 
addition to the Standard Model gauge boson $\rm Z^0$. 
The Z$^0$ has fermion couplings as predicted by the Standard Model,
whereas the axial and vector coupling constants of the $\rm Z'{^0}$ to
fermions are parameters of the particular model. In general the additional 
heavy boson $\rm Z'{^0}$ will mix with the Z$^0$ boson. The observed 
particles are the mass eigenstates Z and \zp~\cite{bib:altarelli}:
\begin{equation}
\rm
\left(\begin{array}{c}\rm  Z \\\rm Z' \end{array}\right) = 
\left(\begin{array}{rr} 
\cos{\tm} & \sin{\tm} \\ -\sin{\tm} & \cos{\tm}
\end{array}\right)
\left(\begin{array}{c}\rm  Z^0 \\\rm Z'{^0} \end{array}\right) \; .
\end{equation}
The mixing angle $\tm$ is a free parameter of the model.
Here and in the following $\rm Z^0$ and $\rm Z'{^0}$ denote gauge eigenstates 
whereas the mass eigenstates formed by mixing within the \zp\ model are 
denoted as Z and \zp .

In this paper we consider several \zp\ models.
In E6 GUT, the E(6) group may incorporate the Standard Model
groups of colour SU(3)$_{\rm C}$, weak isospin SU(2)$_{\rm L}$
and hypercharge U(1)$_{\rm Y}$ in the following way \cite{bib:zpGUT}:
\begin{eqnarray}\rm 
\rm  E(6)  &\to&\rm  SO(10) \times U(1)_\chi \nonumber \\
\rm SO(10) &\to&\rm  SU(5) \;\;\times U(1)_\psi \nonumber \\
\rm SU(5)  &\to&\rm  SU(3)_C \times SU(2)_L \times U(1)_Y
\; .
\end{eqnarray}
Thus the two additional gauge groups U(1)$_\chi$ and U(1)$_\psi$ are 
introduced, each related to a new gauge boson. In general, the $\rm Z'{^0}$ 
will be a mixed state of these two groups:
\begin{equation}
 \rm Z'{^0} = Z_\psi\; \sin{\te6} + Z_\chi\; \cos{\te6} \; .
\end{equation}
We derive limits on the \zp\ mass and mixing angle for all values of
$\te6$, with particular emphasis on three special cases: 
the two no-mixing models $\chi$ and $\psi$ 
($\te6 = 0$ and $\pi /2$) and 
the $\eta$ model with $\te6 = -\mbox{arctan}\sqrt{5/3} = -0.91$.
In this latter case the E6 group is broken by a non-Abelian discrete
symmetry to a rank-5 group which may occur in superstring theories
\cite{bib:zpGUT}. 

Another approach is the left-right symmetric model (LR)~\cite{bib:zpLR}.
In this scenario a symmetry group SU(2)$_{\rm R}$ is introduced whose three 
vector bosons couple to right-handed fermions. 
%The ratio between the 
%left- and right-handed coupling constants is given by
%\begin{equation}
%\alr = 
% \sqrt{\frac{\cos^2{\tw}}{\sin^2{\tw}}\frac{g^2_{\rm R}}{g^2_{\rm L}}-1}
% \; ,
%\end{equation}
% with $g_{\rm L} = g \cos{\tw}$.
The coupling constants of the $\rm Z'{^0}$ to fermions depend on one 
parameter $\alr$. This can take values $\sqrt{2/3} < \alr < 
\sqrt{(\cos^2{\tw} - \sin^2{\tw})/\sin^2{\tw}}$, where $\tw$ is the
weak mixing angle. For \mbox{$\alr =$ 1.53} (the upper limit of the allowed 
range) left and right-handed coupling constants are approximately the same. 
For $\alr = \sqrt{2/3}$ the LR model is equivalent to the E6 $\chi$ model.
We derive limits on the \zp\ mass and mixing angle as a function of $\alr$;
of particular interest is the symmetric case of equal left- and
right-handed couplings. 
The axial and vector coupling constants of the $ \rm Z'{^0}$ to 
fermions for the E6 and LR models can be found in~\cite{bib:zpGUT,bib:zefit}.

In addition to the models described above, we also present limits in the
case of a sequential Standard Model (SSM) \zp, which has the same couplings 
to fermions as the Standard Model $\PZ$. 

In a more model-independent approach, the \zp\ can be directly described in 
terms of its axial and vector couplings to fermions, $a_{f}'$ and 
$v_{f}'$~\cite{bib:zp_indep}. At energies far from the \zp\ resonance,
the data are sensitive to the normalized couplings
\[ 
a_{f}^{\rm N} \sim a_{f}'\sqrt{\frac{g^2}{4\pi}}\frac{\sqrt{s}}{\mzp},
\hspace{1cm}
v_{f}^{\rm N} \sim v_{f}'\sqrt{\frac{g^2}{4\pi}}\frac{\sqrt{s}}{\mzp}. 
\]
We present limits on the leptonic couplings, assuming lepton universality
and $g^2/4\pi = 1$.

In this analysis, cross-sections and asymmetries predicted by the \zp\
models are  obtained using \ZFITTER~\cite{bib:zfitter}
together with \ZEFIT~\cite{bib:zefit}.
Input parameters to the routine are the
model parameters: masses of Z and \zp , model angle $\te6$ or $\alr$ and
mixing angle $\tm$ between Z and \zp . In addition the usual \ZFITTER\
input parameters, as described in Section~\ref{sec:SM}, are used. 
For the \zp\ mass range relevant here (300 $\leq\mzp\leq$ 5000 GeV)
the difference between the fitted $\PZ$ mass within the Standard Model and 
the mass within the \zp\ model is less than 1 MeV, well within the 
experimental uncertainty of the OPAL measurement, 
$\mPZ$ = 91.1852 $\pm$ 0.0030 GeV~\cite{bib:PR328}. 
In the fits described below, $\mPZ$ is treated as a free parameter.
The parameters $\alphas$, $\mtop$ and $\mHiggs$ are fixed to the values
given in Section~\ref{sec:SM} unless otherwise stated.

\subsubsection{Analysis and Results}

Cross-sections and asymmetries measured at energies around the Z peak 
give a precise determination of the properties of the 
Z boson. If this particle is not the Standard Model Z$^0$ but a mixture with 
$\rm Z'{^0}$ the couplings to fermions will change. In particular,
the measured width of the Z is sensitive to the mixing angle $\tm$, and
this angle is therefore constrained by the \LEPone\ data. 
At energies above 130 GeV the interference between Z and \zp\ 
becomes increasingly important and the data are very sensitive to the mass 
of the \zp. Since changes to cross-sections and asymmetries arise from
interference terms, the precise form of these changes depends strongly
on the model.

To obtain limits on the \zp\ properties, cross-sections 
for the processes $\epem\to\mumu, \tautau, \qqbar$ and 
the forward-backward asymmetries for the leptonic processes
$\epem\to\mumu , \tautau$ at energies around the Z resonance
\cite{bib:PR328} and the non-radiative values at 
$\sqrt{s}$ = 130--183~GeV~\cite{bib:OPAL-SM172,bib:OPAL-SM183} and 
189--207~GeV presented here are compared 
to the predictions of the \zp\ models.
Cross-sections and asymmetries for b- and c-quark production are found to 
yield negligible additional sensitivity and are not used.
In calculating the $\chi^2$ between the predictions of the model and the
measurements the correlations of the experimental errors are taken into 
account as in the S-matrix fit.

A $\chi^2$ between model predictions and the measurements is calculated 
for different values of the \zp\ mass and the mixing angle $\tm$.
The difference between the minimum $\chi^2$ and the Standard Model $\chi^2$ 
is small for all models, being at most 0.8 for the LR model. 
%If a clear minimum inconsistent with 
%the Standard Model had been found, exclusion limits on the mixing angle 
%would have been better for large \zp\ masses where the overall $\chi^2$ is 
%larger, as is seen in other analyses \cite{bib:L3-zp,bib:delphi-zp}. This
%is not the case here.

%Results for the E(6) and LR models are presented in two different forms:
%\begin{itemize}
%\item Two-dimensional exclusion contours at 95\% confidence level in the
%      $\mzp$ -- $\tm$ plane for the E6 models $\chi , \psi , \eta$ 
%      and the LR model with $\alr =$ 1.53. These were 
%      obtained with $\chi^2 > \chi^2_{\min} + 5.99$. 
%\item One-dimensional limits at 95\% confidence level, obtained with 
%      $\chi^2 > \chi^2_{\min} + 3.84$, for both  $\mzp$ and $\tm$.
%      Results are presented for a scan over the E6 and LR model angles.
%      These are obtained with $\mHiggs$ 
%      fixed\footnote{A fit with a free Higgs mass is almost unconstrained, 
%      and therefore is numerically unstable.}.
%      Limits are presented for $\mPZ$, $\mtop$, $\alphas$ fixed or 
%      free but constrained by their experimental error.
%\end{itemize}
%In addition we present the lower mass limit in the SSM model.

\begin{figure}[htbp]
   \begin{center}
      \mbox{
          \epsfxsize=16.0cm
          %\epsffile{zp_contour.eps}
          \epsffile{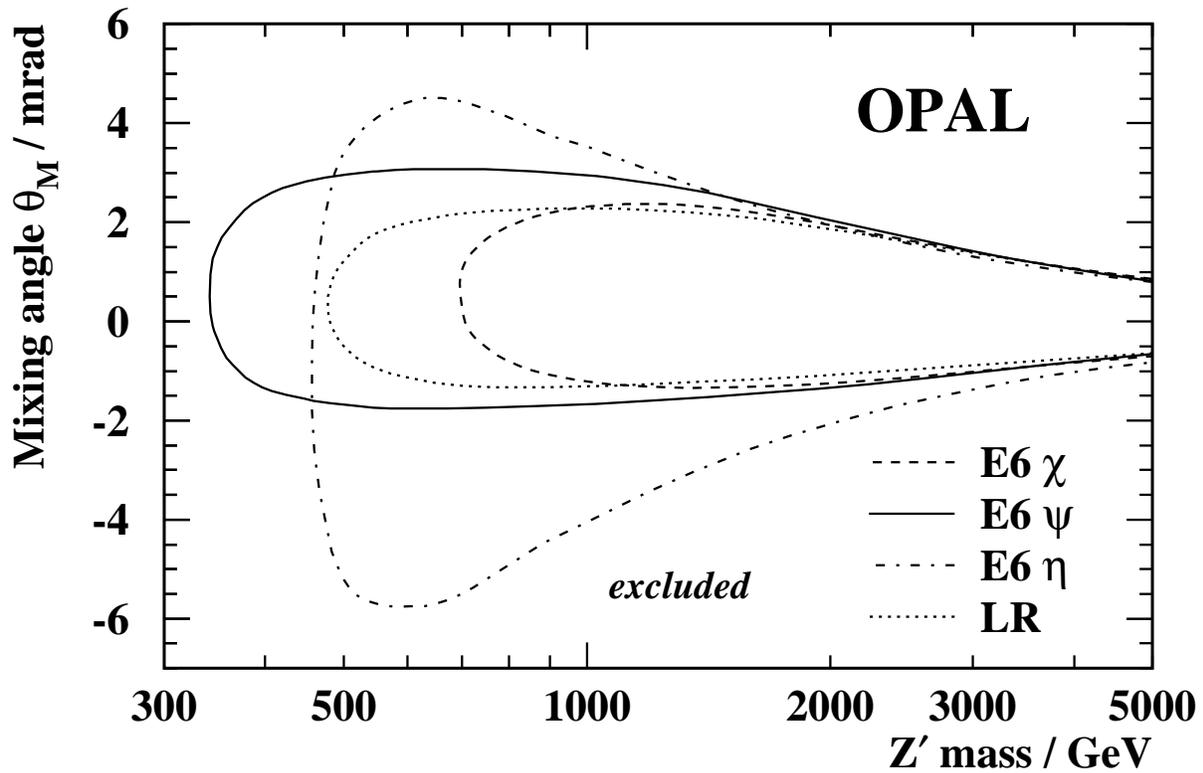}
           }
   \end{center}
\caption[ ]{
Exclusion contours in the \zp\ mass -- mixing angle plane at 95\% 
confidence level for four \zp\ models. The Z mass is free during the fit 
and the other three Standard Model parameters ($\alphas$, $\mtop$ and 
$\mHiggs$) are fixed at their default values.
Leaving $\mtop$ and $\alphas$ free in the fit would lead to an increase of 
the width in $\tm$ by less than 10~\%.
}
\label{fig:zpfig}
\end{figure}

In Fig.~\ref{fig:zpfig} we show the 95\% confidence level exclusion
contours in the $\mzp$ -- $\tm$ plane for the E6 models $\chi , \psi , \eta$ 
and the LR model with $\alr =$ 1.53. These 
%were obtained with Standard
%Model parameters fixed, and 
correspond to $\chi^2 > \chi^2_{\min} + 5.99$. 
The limit on the \zp\ mass depends strongly on the model. The allowed range 
for the mixing angle is approximately \mbox{--2 mrad $< \tm <$ 3 mrad}, and 
only for the E6 $\eta$ model is the contour much broader. For very large \zp\ 
masses the contours for the different models become similar. 

\begin{figure}[b]
   \begin{center}
      \mbox{
          \epsfxsize=13.5cm
          %\epsffile{zp_scan.eps}
          \epsffile{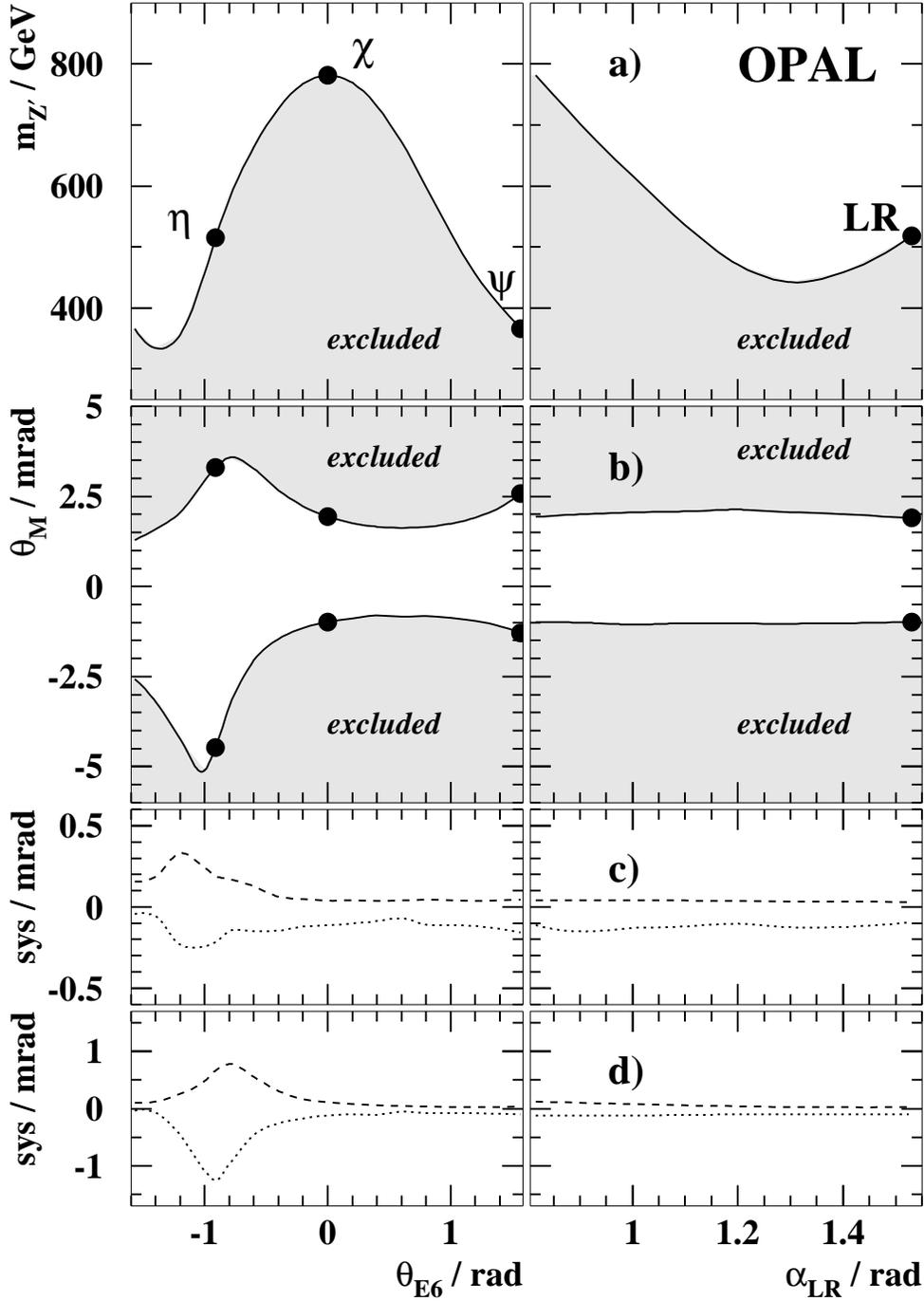}
           }
   \end{center}
\vspace{-1cm}
\caption[ ]{
 One dimensional limits at 95\% confidence level for E6 and LR 
 models as a function of the model angle. The particular cases of
 the $\eta$, $\chi$, $\psi$ and LR symmetric models are indicated
 by the dots.  (a) shows the limits 
 on the \zp\ mass, and (b) shows the upper and lower limits on 
 the mixing angle. They are obtained from a fit with $\alphas$, $\mtop$ 
 and $\mHiggs$ fixed but $\mPZ$ is free.
 (c) shows the absolute change in the limit on the mixing angle
 if $\alphas$ and $\mtop$ are free parameters but constrained by their
 experimental uncertainties.
 (d) shows the absolute change in the limit on the mixing angle if
 $\mHiggs$ = 250~GeV instead of the default value of 115~GeV. 
 In (c) and (d) the dashed curve denotes the change in the positive limit 
 and the dotted curve denotes the change in the negative limit.
}
\label{fig:zpfig2}
\end{figure}

In Fig.~\ref{fig:zpfig2}(a) and (b) we present one-dimensional limits on 
$\mzp$ and $\tm$ for a scan over the E6 and LR model angles. The 95\% 
confidence level limits correspond to $\chi^2 > \chi^2_{\min} + 3.84$. The 
Standard Model parameters were again fixed. Numerical values of the
limits on $\mzp$ and $\tm$ for parameters corresponding to the $\chi , \psi , 
\eta$ and symmetric LR models obtained from these one-dimensional fits are 
given in Table~\ref{tab:zptab}, where we also present corresponding limits 
on the sequential Standard Model \zp.
 
To assess the effect of fixing the Standard Model parameters, we have also 
performed fits in which the strong coupling constant and the top quark mass 
were treated as free parameters, but constrained by their experimental error.
The Higgs boson mass was fixed. The sensitivity of the \LEPone\ data to 
$\mHiggs$ arises mainly from the Z width, which changes in \zp\ models. 
The data cannot discriminate between \zp\ and Higgs effects, therefore a fit 
with a free Higgs boson mass would be numerically unstable. 
Figure~\ref{fig:zpfig2}(c)
shows the change in the limit on the mixing angle as a function of model
parameters. The limit on the mixing angle obtained with $\alphas$ and $\mtop$ 
free is typically 10\% less restrictive than that obtained when fixing these
parameters. Changes to the limits on the \zp\ mass are at most a few GeV.
Since the Higgs boson mass is unknown its influence on the limits has been 
studied by performing fits with a different fixed mass. 
Figure~\ref{fig:zpfig2}(d) 
shows the change in the limit on the mixing angle when the Higgs boson mass is
set to 250~GeV rather than its default value of 115~GeV. The change is
generally small, but amounts to almost 30\% in the region of the E6 $\eta$ 
model.

\begin{figure}
\begin{center}
\epsfxsize=\textwidth
 %\epsfbox{zp_coupl.eps}
\epsfbox{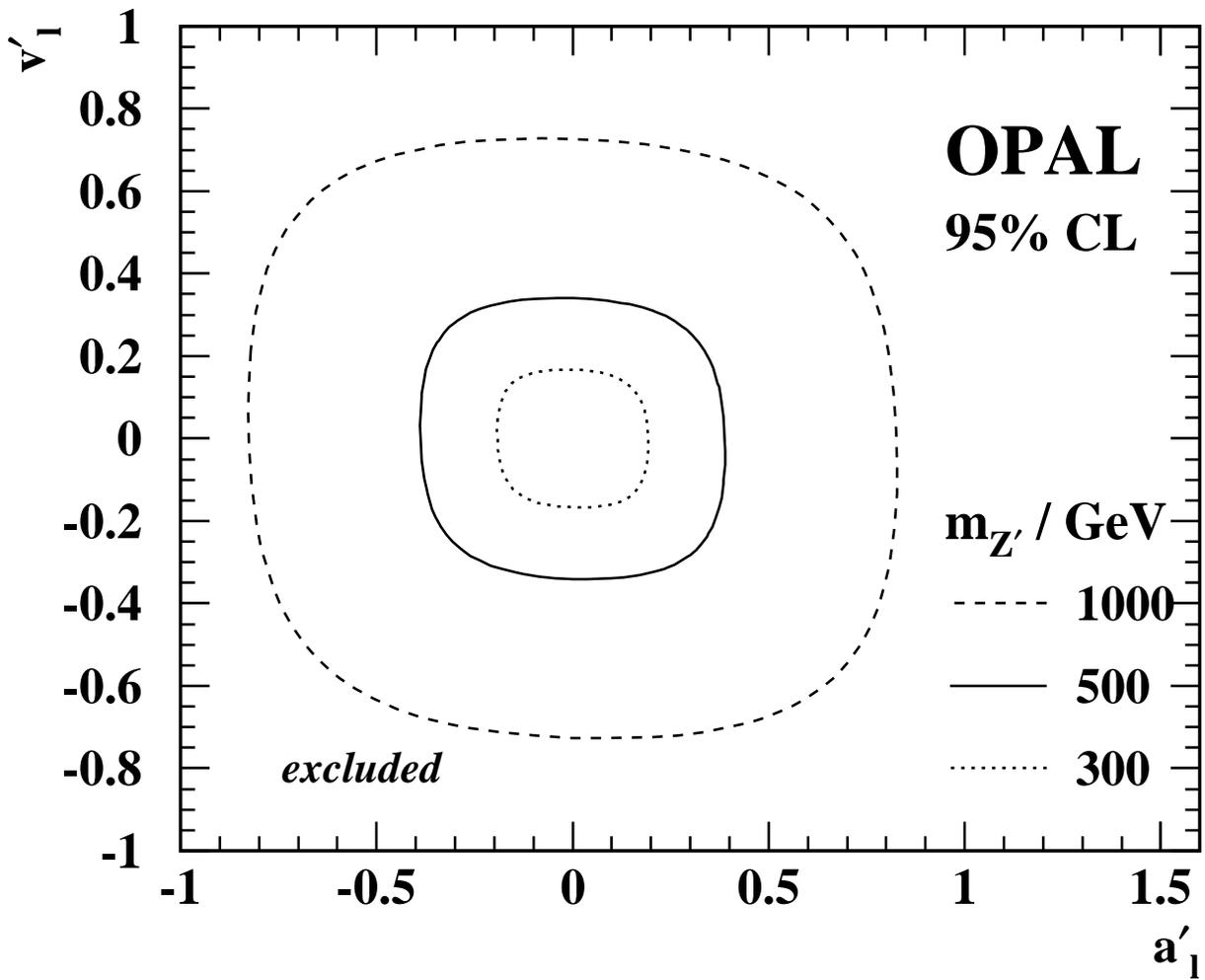}
\caption{95\% confidence level exclusion contours on the axial and vector 
couplings of a \zp\ to leptons, for three values of the \zp\ mass. 
}
\label{fig:zp_coupl} 
\end{center}
\end{figure}

Limits on the vector and axial-vector couplings of a \zp\ boson to leptons 
within the model-independent framework were derived with the mixing angle 
$\tm$ set to zero.
Therefore only the leptonic data at energies of 130~GeV and above were used 
in the fit, and the $\PZ$ mass was fixed. The couplings cannot be determined 
independently from the \zp\ mass, so we have determined limits for fixed 
masses of 300, 500 and 1000 GeV. The 95\% confidence level limits on the 
vector and axial-vector couplings, $v_{\ell}'$ and $a_{\ell}'$ are shown in 
Fig.~\ref{fig:zp_coupl}. The exclusion contours are roughly rectangular in 
shape. In terms of the normalized couplings, calculated at the 
luminosity-weighted mean centre-of-mass energy of 193.2~GeV, 
we find 95\% confidence level 
limits of $|a_{\ell}^{\rm N}| < 0.145$ and $|v_{\ell}^{\rm N}| < 0.127$.

%%-----------------------------------------------------------------------
\section{Conclusions}    \label{sec:sum}
%%-----------------------------------------------------------------------
We have presented new measurements of cross-sections and asymmetries for
hadronic and lepton-pair production in $\epem$ collisions at centre-of-mass
energies between 189~GeV and 209~GeV. At these energies, strong initial-state
radiation leads to excitation of the Z. We therefore distinguish two
kinematic regions depending on $s'$, the square of the centre-of-mass
energy of the \epem\ system after initial-state radiation: an `inclusive'
region with $s'/s > 0.01$ and a `non-radiative' region with $s'/s > 0.7225$.  
The results for both inclusive fermion-pair production and for non-radiative 
events are in good agreement with Standard Model expectations. 
From these and earlier measurements we derive a value for the 
electromagnetic coupling constant 
$1/\alphaem (193.2~\mathrm{GeV}) = 127.4^{+2.1}_{-2.0}$.
In addition, the results have been used together with OPAL measurements 
at 91--183~GeV within the S-matrix formalism assuming lepton universality
to determine the $\gamma$--\PZ\ interference term $\jtoth = 0.144\pm0.078$ 
and to make an almost model-independent measurement of the Z mass, 
$\mPZ = 91.1872\pm0.0033$~GeV. 

The measurements have also been used to place limits on new physics. In
the context of a four-fermion contact interaction we have improved the
limits on the energy scale $\Lambda$ from typically 3--13~TeV to 
5--16~TeV, assuming $g^2/4\pi = 1$. Lower limits on the mass of a
possible \zp\ boson in the range 334~GeV to 1018~GeV, depending on the
model, have been obtained.  

%\clearpage
%-----------------------------------------------------------------------
\section*{Acknowledgements}
%-----------------------------------------------------------------------
We particularly wish to thank the SL Division for the efficient operation
of the LEP accelerator at all energies
 and for their close cooperation with
our experimental group.  In addition to the support staff at our own
institutions we are pleased to acknowledge the  \\
Department of Energy, USA, \\
National Science Foundation, USA, \\
Particle Physics and Astronomy Research Council, UK, \\
Natural Sciences and Engineering Research Council, Canada, \\
Israel Science Foundation, administered by the Israel
Academy of Science and Humanities, \\
Benoziyo Center for High Energy Physics,\\
Japanese Ministry of Education, Culture, Sports, Science and
Technology (MEXT) and a grant under the MEXT International
Science Research Program,\\
Japanese Society for the Promotion of Science (JSPS),\\
German Israeli Bi-national Science Foundation (GIF), \\
Bundesministerium f\"ur Bildung und Forschung, Germany, \\
National Research Council of Canada, \\
Hungarian Foundation for Scientific Research, OTKA T-038240, 
and T-042864,\\
The NWO/NATO Fund for Scientific Research, the Netherlands.\\
\clearpage 
%---------------------------------------------------------------------
%       References
%---------------------------------------------------------------------

%-----------------------------------------------------------------------
%       Tables
%-----------------------------------------------------------------------
\clearpage
\begin{table}[htbp]%-------------------------------------------------------
\begin{sideways}
\begin{minipage}[b]{\textheight}
\begin{center}
\vspace*{2cm}
\begin{tabular}{|c|c|c|c|c|c|}
\hline%-----------------------------------------------------------------
\hline%-----------------------------------------------------------------
\multicolumn{6}{|c|}{\bf \boldmath Interference Corrections $s'/s>0.01$} \\
\hline%-----------------------------------------------------------------
  $\sqrt{s}$ / GeV
& $\Delta\sigma/\sigma_{\rm SM}$(\qqbar) (\%)  
& $\Delta\sigma/\sigma_{\rm SM}(\mu\mu)$ (\%)
& $\Delta\sigma/\sigma_{\rm SM}(\tau\tau)$ (\%)
& $\Delta\AFB(\mu\mu)$
& $\Delta\AFB(\tau\tau)$ \\ \hline
189 &+0.04$\pm$0.06 &--0.63$\pm$0.01 &--0.47$\pm$0.04 &--0.0070$\pm$0.0002
    &--0.0057$\pm$0.0001 \\
192 &+0.04$\pm$0.06 &--0.63$\pm$0.01 &--0.49$\pm$0.04 &--0.0070$\pm$0.0002
    &--0.0058$\pm$0.0002 \\
196 &+0.05$\pm$0.05 &--0.64$\pm$0.01 &--0.47$\pm$0.04 &--0.0071$\pm$0.0002
    &--0.0058$\pm$0.0002 \\
200 &+0.06$\pm$0.04 &--0.65$\pm$0.01 &--0.48$\pm$0.04 &--0.0072$\pm$0.0002
    &--0.0059$\pm$0.0002 \\
202 &+0.06$\pm$0.04 &--0.65$\pm$0.01 &--0.51$\pm$0.03 &--0.0072$\pm$0.0002
    &--0.0061$\pm$0.0001 \\
205 &+0.07$\pm$0.05 &--0.65$\pm$0.01 &--0.51$\pm$0.03 &--0.0073$\pm$0.0002
    &--0.0060$\pm$0.0002 \\
207 &+0.08$\pm$0.05 &--0.65$\pm$0.01 &--0.50$\pm$0.04 &--0.0073$\pm$0.0002
    &--0.0061$\pm$0.0001 \\
\hline%-----------------------------------------------------------------
\hline%-----------------------------------------------------------------
\multicolumn{6}{|c|}{\bf \boldmath Interference Corrections $s'/s>0.7225$} \\
\hline%-----------------------------------------------------------------
  $\sqrt{s}$ / GeV
& $\Delta\sigma/\sigma_{\rm SM}$(\qqbar) (\%)  
& $\Delta\sigma/\sigma_{\rm SM}(\mu\mu)$ (\%)
& $\Delta\sigma/\sigma_{\rm SM}(\tau\tau)$ (\%)
& $\Delta\AFB(\mu\mu)$
& $\Delta\AFB(\tau\tau)$ \\ \hline
189 &+0.14$\pm$0.24 &--1.46$\pm$0.12 &--0.95$\pm$0.02 &--0.0133$\pm$0.0015
    &--0.101$\pm$0.0009 \\
192 &+0.17$\pm$0.21 &--1.46$\pm$0.12 &--0.96$\pm$0.02 &--0.0135$\pm$0.0015
    &--0.102$\pm$0.0009 \\
196 &+0.22$\pm$0.18 &--1.46$\pm$0.12 &--0.93$\pm$0.02 &--0.0136$\pm$0.0015
    &--0.101$\pm$0.0008 \\
200 &+0.27$\pm$0.13 &--1.46$\pm$0.12 &--0.94$\pm$0.02 &--0.0137$\pm$0.0015
    &--0.103$\pm$0.0009 \\
202 &+0.30$\pm$0.11 &--1.46$\pm$0.12 &--0.96$\pm$0.02 &--0.0137$\pm$0.0015
    &--0.105$\pm$0.0009 \\
205 &+0.34$\pm$0.09 &--1.46$\pm$0.12 &--0.96$\pm$0.02 &--0.0139$\pm$0.0015
    &--0.105$\pm$0.0009 \\
207 &+0.37$\pm$0.10 &--1.46$\pm$0.11 &--0.95$\pm$0.02 &--0.0139$\pm$0.0015
    &--0.105$\pm$0.0009 \\
\hline%-----------------------------------------------------------------
\hline%-----------------------------------------------------------------
\end{tabular}
\end{center}
\caption[]{Corrections $\Delta\sigma$ and $\Delta\AFB$
 which have been applied to the measured cross-sections and 
 asymmetries in order to remove the contribution from interference
 between initial- and final-state radiation. Cross-section corrections are
 expressed as a percentage of the expected Standard Model cross-sections,
 calculated using \ZFITTER, while asymmetry corrections are given as 
 absolute numbers.
}
\label{tab:ifsr}
\end{minipage}
\end{sideways}
\end{table}%------------------------------------------------------------

\begin{table}[htbp]%-------------------------------------------------------
\centering
\begin{tabular}{|c|c|c|c|c|}
\hline%-----------------------------------------------------------------
\hline%-----------------------------------------------------------------
\multicolumn{3}{|c|}{$\sqrt{s}$ / GeV}  & & \\ \cline{1-3}
nominal &range &mean
&$\int{\cal L}\mathrm{d}t$ / pb$^{-1}$ 
&$\Delta{\cal L}/{\cal L}$ (\%) \\
\hline
189   &   &188.635$\pm$0.040 &185 &0.21 \\
\hline%-----------------------------------------------------------------
192   &   &191.590$\pm$0.042 &29  &0.32 \\
196   &   &195.526$\pm$0.042 &77  &0.26 \\
200   &   &199.522$\pm$0.042 &79  &0.27 \\
202   &   &201.636$\pm$0.042 &38  &0.30 \\
\hline%-----------------------------------------------------------------
205   &202.5 -- 205.5   &204.881$\pm$0.050  & 82    &0.26 \\
207   &205.5 -- 209.0   &206.561$\pm$0.050  &137    &0.24 \\
\hline%-----------------------------------------------------------------
\hline%-----------------------------------------------------------------
\end{tabular}
\caption[]{Centre-of-mass energy range, luminosity-weighted mean
           centre-of-mass energy~\cite{bib:ELEP}, approximate integrated 
           luminosity collected and total error on the luminosity measurement 
           at each nominal energy point. 
           %Data at 189~GeV were recorded
           %in 1998, data at 192~GeV to 202~GeV in 1999 (with a small amount 
           %at 200~GeV and 202~GeV added in 2000), those at higher energies 
           %in 2000. 
           The precise amount of data used in each analysis
           varies slightly from channel to channel.
}
\label{tab:lumi}
\end{table}%------------------------------------------------------------

\begin{table}[htbp]%-------------------------------------------------------
\centering
\begin{tabular}{|l|c|c|c|c|c|c|c|}
\hline%-----------------------------------------------------------------
\hline%-----------------------------------------------------------------
\multicolumn{1}{|c|}{$\sqrt{s}$ / GeV}
                        &189   & 192 & 196 & 200 & 202 & 205 & 207 \\ \hline
Experimental systematic &0.15  &0.20 &0.18 &0.19 &0.19 &0.18 &0.18 \\
Beam energy             &0.04  &0.04 &0.04 &0.04 &0.04 &0.05 &0.05 \\
Theory                  &0.12  &0.12 &0.12 &0.12 &0.12 &0.12 &0.12 \\ \hline
Data statistics         &0.09  &0.20 &0.13 &0.13 &0.19 &0.13 &0.10 \\ \hline
Total                   &0.21  &0.32 &0.26 &0.27 &0.30 &0.26 &0.24 \\
\hline%-----------------------------------------------------------------
\hline%-----------------------------------------------------------------
\end{tabular}
\caption[]{Errors on the luminosity measurement (in \%) at each nominal
           centre-of-mass energy.
}
\label{tab:lumi_errors}
\end{table}%------------------------------------------------------------

\begin{table}[htbp]%-------------------------------------------------------
\small
\centering
\begin{tabular}{|ll|c|c|c|}
\hline%-----------------------------------------------------------------
\hline%-----------------------------------------------------------------
\multicolumn{5}{|c|}{\bf \boldmath Efficiencies and backgrounds at 
                     $\sqrt{s}$ = 189~GeV}
\\
\hline%-----------------------------------------------------------------
Channel  & &Efficiency (\%)  &Background / pb &Feedthrough / pb \\ 
\hline%-----------------------------------------------------------------
\qqbar &$s'/s>0.01$ &86.6$\pm$0.3 &6.1$\pm$0.7 &-- \\
       &$s'/s>0.7225$ & 87.0$\pm$0.3 &1.55$\pm$0.06 &0.959$\pm$0.043 \\ 
\hline%-----------------------------------------------------------------
\Pgmp\Pgmm &$s'/s>0.01$   &74.7$\pm$0.8 &0.42$\pm$0.02 &-- \\
           &$s'/s>0.7225$ &87.9$\pm$0.9 &0.058$\pm$0.006 &0.044$\pm$0.005 \\
\hline%-----------------------------------------------------------------
\Pgtp\Pgtm &$s'/s>0.01$   &33.8$\pm$0.8 &0.32$\pm$0.03 &-- \\
           &$s'/s>0.7225$ &48.4$\pm$1.2 &0.123$\pm$0.014 &0.055$\pm$0.003 \\
\hline%-----------------------------------------------------------------
\Pep\Pem &A: $\absctepem<0.9$, $\thacol<170\degree$  &97.5$\pm$0.4 
                                               &1.8$\pm$0.3   &-- \\
         &B: $\absctem<0.7$, $\thacol<10\degree$ &99.0$\pm$0.3 
                                               &0.25$\pm$0.03 &-- \\
         &C: $\absctepem<0.96$, $\thacol<10\degree$  &98.5$\pm$0.4 
                                               &10.4$\pm$0.5  &-- \\
\hline%-----------------------------------------------------------------
\hline%-----------------------------------------------------------------
\multicolumn{5}{|c|}{\bf \boldmath Efficiencies and backgrounds at 
                     $\sqrt{s}$ = 192~GeV}
\\
\hline%-----------------------------------------------------------------
Channel  & &Efficiency (\%)  &Background / pb &Feedthrough / pb \\ 
\hline%-----------------------------------------------------------------
\qqbar &$s'/s>0.01$ &85.9$\pm$0.3 &6.2$\pm$0.7 &-- \\
       &$s'/s>0.7225$ & 86.7$\pm$0.3 &1.54$\pm$0.06 &0.972$\pm$0.044 \\ 
\hline%-----------------------------------------------------------------
\Pgmp\Pgmm &$s'/s>0.01$   &74.4$\pm$0.8 &0.43$\pm$0.02 &-- \\
           &$s'/s>0.7225$ &87.9$\pm$0.9 &0.062$\pm$0.005 &0.042$\pm$0.004 \\
\hline%-----------------------------------------------------------------
\Pgtp\Pgtm &$s'/s>0.01$   &33.6$\pm$0.8 &0.31$\pm$0.03 &-- \\
           &$s'/s>0.7225$ &48.4$\pm$1.2 &0.121$\pm$0.013 &0.052$\pm$0.003 \\
\hline%-----------------------------------------------------------------
\Pep\Pem &A: $\absctepem<0.9$, $\thacol<170\degree$  &97.5$\pm$0.4 
                                               &1.7$\pm$0.3   &-- \\
         &B: $\absctem<0.7$, $\thacol<10\degree$ &98.9$\pm$0.3 
                                               &0.24$\pm$0.02 &-- \\
         &C: $\absctepem<0.96$, $\thacol<10\degree$  &98.5$\pm$0.4 
                                               &10.0$\pm$0.5  &-- \\
\hline%-----------------------------------------------------------------
\hline%-----------------------------------------------------------------
\multicolumn{5}{|c|}{\bf \boldmath Efficiencies and backgrounds at 
                     $\sqrt{s}$ = 196~GeV}
\\
\hline%-----------------------------------------------------------------
Channel  & &Efficiency (\%)  &Background / pb &Feedthrough / pb \\ 
\hline%-----------------------------------------------------------------
\qqbar &$s'/s>0.01$ &85.1$\pm$0.3 &6.2$\pm$0.7 &-- \\
       &$s'/s>0.7225$ & 86.2$\pm$0.3 &1.52$\pm$0.06 &0.868$\pm$0.039 \\ 
\hline%-----------------------------------------------------------------
\Pgmp\Pgmm &$s'/s>0.01$   &74.0$\pm$0.8 &0.45$\pm$0.02 &-- \\
           &$s'/s>0.7225$ &87.9$\pm$0.9 &0.068$\pm$0.005 &0.040$\pm$0.004 \\
\hline%-----------------------------------------------------------------
\Pgtp\Pgtm &$s'/s>0.01$   &33.3$\pm$0.8 &0.31$\pm$0.03 &-- \\
           &$s'/s>0.7225$ &48.4$\pm$1.1 &0.118$\pm$0.012 &0.049$\pm$0.003 \\
\hline%-----------------------------------------------------------------
\Pep\Pem &A: $\absctepem<0.9$, $\thacol<170\degree$  &97.4$\pm$0.4 
                                               &1.7$\pm$0.2   &-- \\
         &B: $\absctem<0.7$, $\thacol<10\degree$ &98.9$\pm$0.3 
                                               &0.23$\pm$0.02 &-- \\
         &C: $\absctepem<0.96$, $\thacol<10\degree$  &98.5$\pm$0.4 
                                               & 9.7$\pm$0.5  &-- \\
\hline%-----------------------------------------------------------------
\hline%-----------------------------------------------------------------
\end{tabular}
\caption[]{
  Efficiency of selection cuts, accepted background, and feedthrough of events
  generated with lower $s'$ into the non-radiative samples, for each channel 
  at each energy. 
  The (very small) contribution of events with $s'/s<0.01$
  to the inclusive sample is included in the efficiency. 
  The errors include Monte Carlo statistics and systematic 
  effects. In the case of electron pairs, the efficiencies are
  effective values including the efficiency of selection cuts for events
  within the acceptance region and the effect of acceptance corrections.  
  An acceptance of $\absctepem<0.9$ (or 0.96) means that both electron and
  positron must satisfy this cut, whereas $\absctem<0.7$ means
  that only the electron need do so.
}
\label{tab:eff1}
\end{table}%------------------------------------------------------------

\begin{table}[htbp]%-------------------------------------------------------
\addtocounter{table}{-1}
\small
\centering
\begin{tabular}{|ll|c|c|c|}
\hline%-----------------------------------------------------------------
\hline%-----------------------------------------------------------------
\multicolumn{5}{|c|}{\bf \boldmath Efficiencies and backgrounds at 
                     $\sqrt{s}$ = 200~GeV}
\\
\hline%-----------------------------------------------------------------
Channel  & &Efficiency (\%)  &Background / pb &Feedthrough / pb \\ 
\hline%-----------------------------------------------------------------
\qqbar &$s'/s>0.01$ &84.2$\pm$0.4 &6.2$\pm$0.6 &-- \\
       &$s'/s>0.7225$ & 85.8$\pm$0.3 &1.50$\pm$0.06 &0.832$\pm$0.037 \\ 
\hline%-----------------------------------------------------------------
\Pgmp\Pgmm &$s'/s>0.01$   &73.7$\pm$0.8 &0.47$\pm$0.02 &-- \\
           &$s'/s>0.7225$ &87.9$\pm$0.9 &0.073$\pm$0.005 &0.038$\pm$0.004 \\
\hline%-----------------------------------------------------------------
\Pgtp\Pgtm &$s'/s>0.01$   &32.9$\pm$0.8 &0.30$\pm$0.02 &-- \\
           &$s'/s>0.7225$ &48.4$\pm$1.1 &0.115$\pm$0.011 &0.045$\pm$0.002 \\
\hline%-----------------------------------------------------------------
\Pep\Pem &A: $\absctepem<0.9$, $\thacol<170\degree$  &97.4$\pm$0.4 
                                               &1.6$\pm$0.3   &-- \\
         &B: $\absctem<0.7$, $\thacol<10\degree$ &98.9$\pm$0.3 
                                               &0.22$\pm$0.02 &-- \\
         &C: $\absctepem<0.96$, $\thacol<10\degree$  &98.4$\pm$0.4 
                                               & 9.3$\pm$0.5  &-- \\
\hline%-----------------------------------------------------------------
\hline%-----------------------------------------------------------------
\multicolumn{5}{|c|}{\bf \boldmath Efficiencies and backgrounds at 
                     $\sqrt{s}$ = 202~GeV}
\\
\hline%-----------------------------------------------------------------
Channel  & &Efficiency (\%)  &Background / pb &Feedthrough / pb \\ 
\hline%-----------------------------------------------------------------
\qqbar &$s'/s>0.01$ &83.8$\pm$0.4 &6.2$\pm$0.6 &-- \\
       &$s'/s>0.7225$ & 85.6$\pm$0.3 &1.49$\pm$0.06 &0.819$\pm$0.036 \\ 
\hline%-----------------------------------------------------------------
\Pgmp\Pgmm &$s'/s>0.01$   &73.5$\pm$0.8 &0.47$\pm$0.02 &-- \\
           &$s'/s>0.7225$ &87.9$\pm$0.9 &0.076$\pm$0.005 &0.037$\pm$0.004 \\
\hline%-----------------------------------------------------------------
\Pgtp\Pgtm &$s'/s>0.01$   &32.8$\pm$0.8 &0.29$\pm$0.02 &-- \\
           &$s'/s>0.7225$ &48.4$\pm$1.1 &0.114$\pm$0.011 &0.043$\pm$0.002 \\
\hline%-----------------------------------------------------------------
\Pep\Pem &A: $\absctepem<0.9$, $\thacol<170\degree$  &97.4$\pm$0.4 
                                               &1.6$\pm$0.3   &-- \\
         &B: $\absctem<0.7$, $\thacol<10\degree$ &98.9$\pm$0.3 
                                               &0.22$\pm$0.02 &-- \\
         &C: $\absctepem<0.96$, $\thacol<10\degree$  &98.4$\pm$0.4 
                                               & 9.1$\pm$0.5  &-- \\
\hline%-----------------------------------------------------------------
\hline%-----------------------------------------------------------------
\multicolumn{5}{|c|}{\bf \boldmath Efficiencies and backgrounds at 
                     $\sqrt{s}$ = 205~GeV}
\\
\hline%-----------------------------------------------------------------
Channel  & &Efficiency (\%)  &Background / pb &Feedthrough / pb \\ 
\hline%-----------------------------------------------------------------
\qqbar &$s'/s>0.01$ &83.1$\pm$0.4 &6.3$\pm$0.6 &-- \\
       &$s'/s>0.7225$ & 85.2$\pm$0.3 &1.48$\pm$0.05 &0.805$\pm$0.036 \\ 
\hline%-----------------------------------------------------------------
\Pgmp\Pgmm &$s'/s>0.01$   &73.2$\pm$0.8 &0.49$\pm$0.02 &-- \\
           &$s'/s>0.7225$ &87.9$\pm$0.9 &0.080$\pm$0.006 &0.035 $\pm$0.004  \\
\hline%-----------------------------------------------------------------
\Pgtp\Pgtm &$s'/s>0.01$   &32.5$\pm$0.8 &0.29$\pm$0.02 &-- \\
           &$s'/s>0.7225$ &48.4$\pm$1.1 &0.111$\pm$0.010 &0.040 $\pm$0.002  \\
\hline%-----------------------------------------------------------------
\Pep\Pem &A: $\absctepem<0.9$, $\thacol<170\degree$  &97.2$\pm$0.5 
                                               &1.5$\pm$0.3   &-- \\
         &B: $\absctem<0.7$, $\thacol<10\degree$ &98.8$\pm$0.4 
                                               &0.21$\pm$0.02 &-- \\
         &C: $\absctepem<0.96$, $\thacol<10\degree$  &98.4$\pm$0.4 
                                               & 8.8$\pm$0.4  &-- \\
\hline%-----------------------------------------------------------------
\hline%-----------------------------------------------------------------
\multicolumn{5}{|c|}{\bf \boldmath Efficiencies and backgrounds at 
                     $\sqrt{s}$ = 207~GeV}
\\
\hline%-----------------------------------------------------------------
Channel  & &Efficiency (\%)  &Background / pb &Feedthrough / pb \\ 
\hline%-----------------------------------------------------------------
\qqbar &$s'/s>0.01$ &82.7$\pm$0.4 &6.3$\pm$0.6 &-- \\
       &$s'/s>0.7225$ & 85.0$\pm$0.3 &1.47$\pm$0.05 &0.746$\pm$0.033 \\ 
\hline%-----------------------------------------------------------------
\Pgmp\Pgmm &$s'/s>0.01$   &73.0$\pm$0.8 &0.50$\pm$0.03 &-- \\
           &$s'/s>0.7225$ &87.9$\pm$0.9 &0.082$\pm$0.006 &0.034 $\pm$0.003  \\
\hline%-----------------------------------------------------------------
\Pgtp\Pgtm &$s'/s>0.01$   &32.4$\pm$0.8 &0.28$\pm$0.02 &-- \\
           &$s'/s>0.7225$ &48.4$\pm$1.1 &0.110$\pm$0.011 &0.039 $\pm$0.002  \\
\hline%-----------------------------------------------------------------
\Pep\Pem &A: $\absctepem<0.9$, $\thacol<170\degree$  &97.2$\pm$0.5 
                                               &1.5$\pm$0.3   &-- \\
         &B: $\absctem<0.7$, $\thacol<10\degree$ &98.8$\pm$0.4 
                                               &0.21$\pm$0.02 &-- \\
         &C: $\absctepem<0.96$, $\thacol<10\degree$  &98.4$\pm$0.4 
                                               & 8.7$\pm$0.4  &-- \\
\hline%-----------------------------------------------------------------
\hline%-----------------------------------------------------------------
\end{tabular}
\caption[]{Continued
%  Efficiency of selection cuts, background and feedthrough of events
%  with lower $s'$ into the non-radiative samples for each channel at 
%  200--207~GeV. 
%  The (very small) contribution of events with $s'/s<0.01$
%  to the inclusive sample is included in the efficiency. 
%  The errors include Monte Carlo statistics and systematic 
%  effects. In the case of electron pairs, the efficiencies are
%  effective values including the efficiency of selection cuts for events
%  within the acceptance region and the effect of acceptance corrections.  
%  An acceptance of $\absct<0.9$ (or 0.96) means that both electron and
%  positron must satisfy this cut, whereas $\absctem<0.7$ means
%  that only the electron need do so.
}
\label{tab:eff2}
\end{table}%------------------------------------------------------------

\begin{table}[htbp]%-------------------------------------------------------
\small
\centering
\begin{tabular}{|ll|c|r|l|c|}
\hline%-----------------------------------------------------------------
\hline%-----------------------------------------------------------------
\multicolumn{6}{|c|}{\bf \boldmath Cross-sections at $\sqrt{s}$ = 189~GeV} \\
\hline%-----------------------------------------------------------------
Channel   &       &$\int{\cal L}\mathrm{d}t$ / pb$^{-1}$ &  Events & 
    \multicolumn{1}{c|}{$\sigma$ / pb} &  $\sigma^{\mathrm{SM}}$ / pb \\
\hline%-----------------------------------------------------------------
\qqbar &$s'/s>0.01$   &185.9 &17146 & 99.5$\pm$0.8$\pm$0.9   & 98.9 \\
       &$s'/s>0.7225$ &      & 4019 & 22.0$\pm$0.4$\pm$0.1   & 22.2 \\
\hline%-----------------------------------------------------------------
\Pgmp\Pgmm &$s'/s>0.01$   &178.6 &1128 &7.85$\pm$0.25$\pm$0.09  & 7.75 \\
           &$s'/s>0.7225$ &      & 519 &3.14$\pm$0.15$\pm$0.03  & 3.21 \\
\hline%-----------------------------------------------------------------
\Pgtp\Pgtm &$s'/s>0.01$   &179.0 & 554 &8.17$\pm$0.39$\pm$0.21  & 7.74 \\
           &$s'/s>0.7225$ &      & 333 &3.45$\pm$0.21$\pm$0.09  & 3.21 \\
\hline%-----------------------------------------------------------------
\Pep\Pem &A: $\absctepem<0.9$, $\thacol<170\degree$  &185.9 &20538 &
                                              111.5$\pm$0.8$\pm$0.6 & 110.3 \\
         &B: $\absctem<0.7$, $\thacol<10\degree$ &      & 3758 &
                                               20.2$\pm$0.3$\pm$0.1 & 20.1 \\
         &C: $\absctepem<0.96$, $\thacol<10\degree$  &      &57669  &
                                              304.4$\pm$1.3$\pm$1.5 & 307.7 \\
\hline%-----------------------------------------------------------------
\hline%-----------------------------------------------------------------
\multicolumn{6}{|c|}{\bf \boldmath Cross-sections at $\sqrt{s}$ = 192~GeV} \\
\hline%-----------------------------------------------------------------
Channel   &       &$\int{\cal L}\mathrm{d}t$ / pb$^{-1}$ &  Events & 
    \multicolumn{1}{c|}{$\sigma$ / pb} &  $\sigma^{\mathrm{SM}}$ / pb \\
\hline%-----------------------------------------------------------------
\qqbar &$s'/s>0.01$   &29.6  &2617  & 95.9$\pm$2.0$\pm$0.9   & 95.0 \\
       &$s'/s>0.7225$ &      & 643  & 22.2$\pm$0.9$\pm$0.1   & 21.3 \\
\hline%-----------------------------------------------------------------
\Pgmp\Pgmm &$s'/s>0.01$   &29.0  &173  &7.40$\pm$0.61$\pm$0.09  & 7.47 \\
           &$s'/s>0.7225$ &      & 77  &2.86$\pm$0.34$\pm$0.03  & 3.10 \\
\hline%-----------------------------------------------------------------
\Pgtp\Pgtm &$s'/s>0.01$   &29.1  & 85  &7.74$\pm$0.95$\pm$0.20  & 7.47 \\
           &$s'/s>0.7225$ &      & 50  &3.17$\pm$0.50$\pm$0.08  & 3.10 \\
\hline%-----------------------------------------------------------------
\Pep\Pem &A: $\absctepem<0.9$, $\thacol<170\degree$  &29.5  &3084  &
                                              105.6$\pm$1.9$\pm$0.6 & 106.9 \\
         &B: $\absctem<0.7$, $\thacol<10\degree$ &      & 577  &
                                               19.5$\pm$0.8$\pm$0.1 & 19.5 \\
         &C: $\absctepem<0.96$, $\thacol<10\degree$  &      &9034   &
                                              301.0$\pm$3.3$\pm$1.6 & 298.3 \\
\hline%-----------------------------------------------------------------
\hline%-----------------------------------------------------------------
\multicolumn{6}{|c|}{\bf \boldmath Cross-sections at $\sqrt{s}$ = 196~GeV} \\
\hline%-----------------------------------------------------------------
Channel   &       &$\int{\cal L}\mathrm{d}t$ / pb$^{-1}$ &  Events & 
    \multicolumn{1}{c|}{$\sigma$ / pb} &  $\sigma^{\mathrm{SM}}$ / pb \\
\hline%-----------------------------------------------------------------
\qqbar &$s'/s>0.01$   &77.8  &6351  & 88.8$\pm$1.2$\pm$0.9   & 90.2 \\
       &$s'/s>0.7225$ &      &1509  & 19.8$\pm$0.6$\pm$0.1   & 20.2 \\
\hline%-----------------------------------------------------------------
\Pgmp\Pgmm &$s'/s>0.01$   &76.0  &435  &7.08$\pm$0.37$\pm$0.08  & 7.13 \\
           &$s'/s>0.7225$ &      &207  &2.93$\pm$0.22$\pm$0.03  & 2.96 \\
\hline%-----------------------------------------------------------------
\Pgtp\Pgtm &$s'/s>0.01$   &75.9  &206  &7.21$\pm$0.57$\pm$0.19  & 7.12 \\
           &$s'/s>0.7225$ &      &120  &2.89$\pm$0.30$\pm$0.07  & 2.96 \\
\hline%-----------------------------------------------------------------
\Pep\Pem &A: $\absctepem<0.9$, $\thacol<170\degree$  &77.7  &7879  &
                                              102.3$\pm$1.2$\pm$0.5 & 102.6 \\
         &B: $\absctem<0.7$, $\thacol<10\degree$ &      &1448  &
                                               18.6$\pm$0.5$\pm$0.1 & 18.7 \\
         &C: $\absctepem<0.96$, $\thacol<10\degree$  &      &22618  &
                                              285.7$\pm$2.0$\pm$1.5 & 286.5 \\
\hline%-----------------------------------------------------------------
\hline%-----------------------------------------------------------------
\end{tabular}
\caption[]{
  Measured cross-sections, integrated luminosity used in the analysis
  and numbers of selected events at each energy.
  For the cross-sections, the first error shown is statistical, the second
  systematic.  As in~\cite{bib:OPAL-SM172}, the cross-sections
  for hadrons, \Pgmp\Pgmm\ and \Pgtp\Pgtm\ 
  are defined to cover phase-space up to the limit imposed by the $s'/s$ cut, 
  with $\sqrt{s'}$ defined as the invariant mass
  of the outgoing two-fermion system {\em before} final-state
  radiation. The contribution of interference between initial-
  and final-state radiation has been removed.
  The last column shows the Standard Model cross-section predictions from
  \ZFITTER~\cite{bib:zfitter} (hadrons, \Pgmp\Pgmm, \Pgtp\Pgtm)
  and  \BHWIDE~\cite{bib:bhwide} (\Pep\Pem).
}
\label{tab:xsec1}
\end{table}%------------------------------------------------------------

\begin{table}[htbp]%-------------------------------------------------------
\addtocounter{table}{-1}
\small
\centering
\begin{tabular}{|ll|c|r|l|c|}
\hline%-----------------------------------------------------------------
\hline%-----------------------------------------------------------------
\multicolumn{6}{|c|}{\bf \boldmath Cross-sections at $\sqrt{s}$ = 200~GeV} \\
\hline%-----------------------------------------------------------------
Channel   &       &$\int{\cal L}\mathrm{d}t$ / pb$^{-1}$ &  Events & 
    \multicolumn{1}{c|}{$\sigma$ / pb} &  $\sigma^{\mathrm{SM}}$ / pb \\
\hline%-----------------------------------------------------------------
\qqbar &$s'/s>0.01$   &79.4  &6100  & 83.9$\pm$1.2$\pm$0.9   & 85.7 \\
       &$s'/s>0.7225$ &      &1468  & 18.9$\pm$0.5$\pm$0.1   & 19.1 \\
\hline%-----------------------------------------------------------------
\Pgmp\Pgmm &$s'/s>0.01$   &78.2  &423  &6.67$\pm$0.36$\pm$0.08  & 6.80 \\
           &$s'/s>0.7225$ &      &202  &2.77$\pm$0.21$\pm$0.03  & 2.83 \\
\hline%-----------------------------------------------------------------
\Pgtp\Pgtm &$s'/s>0.01$   &78.0  &205  &7.04$\pm$0.56$\pm$0.18  & 6.80 \\
           &$s'/s>0.7225$ &      &132  &3.14$\pm$0.30$\pm$0.08  & 2.83 \\
\hline%-----------------------------------------------------------------
\Pep\Pem &A: $\absctepem<0.9$, $\thacol<170\degree$  &79.4  &7819  &
                                               99.5$\pm$1.1$\pm$0.5 &  98.5 \\
         &B: $\absctem<0.7$, $\thacol<10\degree$ &      &1444  &
                                               18.2$\pm$0.5$\pm$0.1 & 17.9 \\
         &C: $\absctepem<0.96$, $\thacol<10\degree$  &      &22046  &
                                              272.8$\pm$1.9$\pm$1.4 & 275.2 \\
\hline%-----------------------------------------------------------------
\hline%-----------------------------------------------------------------
\multicolumn{6}{|c|}{\bf \boldmath Cross-sections at $\sqrt{s}$ = 202~GeV} \\
\hline%-----------------------------------------------------------------
Channel   &       &$\int{\cal L}\mathrm{d}t$ / pb$^{-1}$ &  Events & 
    \multicolumn{1}{c|}{$\sigma$ / pb} &  $\sigma^{\mathrm{SM}}$ / pb \\
\hline%-----------------------------------------------------------------
\qqbar &$s'/s>0.01$   &38.2  &2898  & 83.2$\pm$1.7$\pm$0.9   & 83.5 \\
       &$s'/s>0.7225$ &      & 692  & 18.5$\pm$0.8$\pm$0.1   & 18.6 \\
\hline%-----------------------------------------------------------------
\Pgmp\Pgmm &$s'/s>0.01$   &36.8  &171  &5.63$\pm$0.48$\pm$0.07  & 6.64 \\
           &$s'/s>0.7225$ &      & 82  &2.36$\pm$0.28$\pm$0.03  & 2.77 \\
\hline%-----------------------------------------------------------------
\Pgtp\Pgtm &$s'/s>0.01$   &36.9  &104  &7.69$\pm$0.84$\pm$0.20  & 6.63 \\
           &$s'/s>0.7225$ &      & 59  &2.95$\pm$0.43$\pm$0.07  & 2.77 \\
\hline%-----------------------------------------------------------------
\Pep\Pem &A: $\absctepem<0.9$, $\thacol<170\degree$  &38.2  &3697  &
                                               97.8$\pm$1.6$\pm$0.5 &  96.4 \\
         &B: $\absctem<0.7$, $\thacol<10\degree$ &      & 682  &
                                               17.8$\pm$0.7$\pm$0.1 & 17.6 \\
         &C: $\absctepem<0.96$, $\thacol<10\degree$  &      &10551  &
                                              271.7$\pm$2.7$\pm$1.4 & 269.5 \\
\hline%-----------------------------------------------------------------
\hline%-----------------------------------------------------------------
\multicolumn{6}{|c|}{\bf \boldmath Cross-sections at $\sqrt{s}$ = 205~GeV} \\
\hline%-----------------------------------------------------------------
Channel   &       &$\int{\cal L}\mathrm{d}t$ / pb$^{-1}$ &  Events & 
    \multicolumn{1}{c|}{$\sigma$ / pb} &  $\sigma^{\mathrm{SM}}$ / pb \\
\hline%-----------------------------------------------------------------
\qqbar &$s'/s>0.01$   &82.3  &6094  & 81.7$\pm$1.1$\pm$0.9   & 80.2 \\
       &$s'/s>0.7225$ &      &1458  & 18.2$\pm$0.5$\pm$0.1   & 17.8 \\
\hline%-----------------------------------------------------------------
\Pgmp\Pgmm &$s'/s>0.01$   &78.9  &418  &6.53$\pm$0.35$\pm$0.08  & 6.41 \\
           &$s'/s>0.7225$ &      &212  &2.88$\pm$0.21$\pm$0.03  & 2.67 \\
\hline%-----------------------------------------------------------------
\Pgtp\Pgtm &$s'/s>0.01$   &78.9  &199  &6.84$\pm$0.55$\pm$0.18  & 6.40 \\
           &$s'/s>0.7225$ &      &117  &2.72$\pm$0.28$\pm$0.07  & 2.67 \\
\hline%-----------------------------------------------------------------
\Pep\Pem &A: $\absctepem<0.9$, $\thacol<170\degree$  &82.3  &7613  &
                                               93.6$\pm$1.1$\pm$0.6 &  93.4 \\
         &B: $\absctem<0.7$, $\thacol<10\degree$ &      &1433  &
                                               17.4$\pm$0.5$\pm$0.1 & 17.0 \\
         &C: $\absctepem<0.96$, $\thacol<10\degree$  &      &21916  &
                                              261.8$\pm$1.8$\pm$1.3 & 261.1 \\
\hline%-----------------------------------------------------------------
\hline%-----------------------------------------------------------------
\multicolumn{6}{|c|}{\bf \boldmath Cross-sections at $\sqrt{s}$ = 207~GeV} \\
\hline%-----------------------------------------------------------------
Channel   &       &$\int{\cal L}\mathrm{d}t$ / pb$^{-1}$ &  Events & 
    \multicolumn{1}{c|}{$\sigma$ / pb} &  $\sigma^{\mathrm{SM}}$ / pb \\
\hline%-----------------------------------------------------------------
\qqbar &$s'/s>0.01$   &137.4 &9686  & 77.7$\pm$0.9$\pm$0.8   & 78.6 \\
       &$s'/s>0.7225$ &      &2260  & 16.8$\pm$0.4$\pm$0.1   & 17.5 \\
\hline%-----------------------------------------------------------------
\Pgmp\Pgmm &$s'/s>0.01$   &134.4 &739  &6.81$\pm$0.28$\pm$0.08  & 6.29 \\
           &$s'/s>0.7225$ &      &347  &2.77$\pm$0.16$\pm$0.03  & 2.63 \\
\hline%-----------------------------------------------------------------
\Pgtp\Pgtm &$s'/s>0.01$   &134.5 &318  &6.39$\pm$0.41$\pm$0.17  & 6.28 \\
           &$s'/s>0.7225$ &      &203  &2.78$\pm$0.22$\pm$0.07  & 2.63 \\
\hline%-----------------------------------------------------------------
\Pep\Pem &A: $\absctepem<0.9$, $\thacol<170\degree$  &137.7 &12335 &
                                               90.6$\pm$0.8$\pm$0.6 & 91.8 \\
         &B: $\absctem<0.7$, $\thacol<10\degree$ &      &2320  &
                                               16.9$\pm$0.4$\pm$0.1 & 16.7 \\
         &C: $\absctepem<0.96$, $\thacol<10\degree$  &      &35997  &
                                              257.1$\pm$1.4$\pm$1.3 & 256.9 \\
\hline%-----------------------------------------------------------------
\hline%-----------------------------------------------------------------
\end{tabular}
\caption[]{Continued
%  Integrated luminosity used in the analysis, numbers of selected events 
%  and measured cross-sections at 202--207~GeV.
%  For the cross-sections, the first error shown is statistical, the second
%  systematic.  As in~\cite{bib:OPAL-SM172}, the cross-sections
%  for hadrons, \Pgmp\Pgmm\ and \Pgtp\Pgtm\ 
%  are defined to cover phase-space up to the limit imposed by the $s'/s$ cut, 
%  with $\sqrt{s'}$ defined as the invariant mass
%  of the outgoing two-fermion system {\em before} final-state
%  radiation. The contribution of interference between initial-
%  and final-state radiation has been removed.
%  The last column shows the Standard Model cross-section predictions from
%  \ZFITTER~\cite{bib:zfitter} (hadrons, \Pgmp\Pgmm, \Pgtp\Pgtm)
%  and  \BHWIDE~\cite{bib:bhwide} (\Pep\Pem).
}
\label{tab:xsec2}
\end{table}%------------------------------------------------------------

\begin{table}[htbp]%-------------------------------------------------------
\centering
\begin{tabular}{|ll|r|r|c|c|}
\hline%-----------------------------------------------------------------
\hline%-----------------------------------------------------------------
\multicolumn{6}{|c|}{\bf \boldmath Asymmetries at $\sqrt{s}$ = 189~GeV} \\
\hline%-----------------------------------------------------------------
&&$N_{\mathrm{f}}$ &$N_{\mathrm{b}}$ &$\AFB$ &$\AFBSM$ \\ 
\hline%-----------------------------------------------------------------
\mumu    &$s'/s > 0.01$    &733    &395   &0.252$\pm$0.030$\pm$0.006 &0.281 \\
         &$s'/s > 0.7225$  &399.5  &119.5 &0.548$\pm$0.039$\pm$0.005 &0.569 \\
\hline%-----------------------------------------------------------------
\tautau  &$s'/s > 0.01$    &373    &170   &0.304$\pm$0.046$\pm$0.008 &0.281 \\
         &$s'/s > 0.7225$  &253.5  & 73.5 &0.591$\pm$0.054$\pm$0.012 &0.569 \\
\hline%-----------------------------------------------------------------
\epem    &$\absctem < 0.7$ &3332   &354   &0.811$\pm$0.010$\pm$0.004 &0.814 \\
         &and $\thacol < 10^\circ$ & & & &                \\ 
\hline%-----------------------------------------------------------------
\hline%-----------------------------------------------------------------
\multicolumn{6}{|c|}{\bf \boldmath Asymmetries at $\sqrt{s}$ = 192~GeV} \\
\hline%-----------------------------------------------------------------
&&$N_{\mathrm{f}}$ &$N_{\mathrm{b}}$ &$\AFB$ &$\AFBSM$ \\ 
\hline%-----------------------------------------------------------------
\mumu    &$s'/s > 0.01$    &100    &73    &0.095$\pm$0.080$\pm$0.006 &0.280 \\
         &$s'/s > 0.7225$  &52     &25    &0.341$\pm$0.115$\pm$0.005 &0.566 \\
\hline%-----------------------------------------------------------------
\tautau  &$s'/s > 0.01$    &62     &21    &0.444$\pm$0.111$\pm$0.008 &0.280 \\
         &$s'/s > 0.7225$  &42.5   & 6.5  &0.813$\pm$0.109$\pm$0.013 &0.565 \\
\hline%-----------------------------------------------------------------
\epem    &$\absctem < 0.7$ &518    &46    &0.841$\pm$0.023$\pm$0.004 &0.814 \\
         &and $\thacol < 10^\circ$ & & & &                \\ 
\hline%-----------------------------------------------------------------
\hline%-----------------------------------------------------------------
\multicolumn{6}{|c|}{\bf \boldmath Asymmetries at $\sqrt{s}$ = 196~GeV} \\
\hline%-----------------------------------------------------------------
&&$N_{\mathrm{f}}$ &$N_{\mathrm{b}}$ &$\AFB$ &$\AFBSM$ \\ 
\hline%-----------------------------------------------------------------
\mumu    &$s'/s > 0.01$    &305.5  &129.5 &0.358$\pm$0.048$\pm$0.005 &0.279 \\
         &$s'/s > 0.7225$  &172.5  &34.5  &0.683$\pm$0.055$\pm$0.005 &0.562 \\
\hline%-----------------------------------------------------------------
\tautau  &$s'/s > 0.01$    &125.5  &72.5  &0.175$\pm$0.077$\pm$0.008 &0.279 \\
         &$s'/s > 0.7225$  & 78    &36    &0.373$\pm$0.103$\pm$0.013 &0.561 \\
\hline%-----------------------------------------------------------------
\epem    &$\absctem < 0.7$ &1276   &137   &0.810$\pm$0.016$\pm$0.004 &0.815 \\
         &and $\thacol < 10^\circ$ & & & &                \\ 
\hline%-----------------------------------------------------------------
\hline%-----------------------------------------------------------------
\end{tabular}
\caption[]{Measured asymmetry values and numbers of forward ($N_{\mathrm{f}}$) 
  and backward ($N_{\mathrm{b}}$) events at each energy.
  The measured asymmetry values include corrections for background and 
  efficiency, and in the case of $\mumu$ and $\tautau$ are corrected to 
  the full 
  solid angle with interference between initial- and final-state radiation
  subtracted. The first error shown is statistical, the second systematic.
  The final column shows the Standard Model predictions of \BHWIDE\ for
  \epem\ and \ZFITTER\ for the other final states.
}
\label{tab:afb1}
\end{table}%------------------------------------------------------------

\begin{table}[htbp]%-------------------------------------------------------
\addtocounter{table}{-1}
\centering
\begin{tabular}{|ll|r|r|c|c|}
\hline%-----------------------------------------------------------------
\hline%-----------------------------------------------------------------
\multicolumn{6}{|c|}{\bf \boldmath Asymmetries at $\sqrt{s}$ = 200~GeV} \\
\hline%-----------------------------------------------------------------
&&$N_{\mathrm{f}}$ &$N_{\mathrm{b}}$ &$\AFB$ &$\AFBSM$ \\ 
\hline%-----------------------------------------------------------------
\mumu    &$s'/s > 0.01$    &294.5  &128.5 &0.346$\pm$0.049$\pm$0.005 &0.278 \\
         &$s'/s > 0.7225$  &164    &38    &0.637$\pm$0.059$\pm$0.005 &0.558 \\
\hline%-----------------------------------------------------------------
\tautau  &$s'/s > 0.01$    &142    &59    &0.341$\pm$0.074$\pm$0.005 &0.279 \\
         &$s'/s > 0.7225$  &107    &23    &0.700$\pm$0.077$\pm$0.009 &0.558 \\
\hline%-----------------------------------------------------------------
\epem    &$\absctem < 0.7$ &1290   &142   &0.805$\pm$0.016$\pm$0.004 &0.815 \\
         &and $\thacol < 10^\circ$ & & & &                \\ 
\hline%-----------------------------------------------------------------
\hline%-----------------------------------------------------------------
\multicolumn{6}{|c|}{\bf \boldmath Asymmetries at $\sqrt{s}$ = 202~GeV} \\
\hline%-----------------------------------------------------------------
&&$N_{\mathrm{f}}$ &$N_{\mathrm{b}}$ &$\AFB$ &$\AFBSM$ \\ 
\hline%-----------------------------------------------------------------
\mumu    &$s'/s > 0.01$    &114    &57    &0.277$\pm$0.080$\pm$0.006 &0.278 \\
         &$s'/s > 0.7225$  & 61    &21    &0.489$\pm$0.104$\pm$0.005 &0.556 \\
\hline%-----------------------------------------------------------------
\tautau  &$s'/s > 0.01$    & 67    &36    &0.205$\pm$0.105$\pm$0.006 &0.278 \\
         &$s'/s > 0.7225$  & 42.5  &16.5  &0.440$\pm$0.138$\pm$0.011 &0.556 \\
\hline%-----------------------------------------------------------------
\epem    &$\absctem < 0.7$ & 595   & 69   &0.795$\pm$0.024$\pm$0.004 &0.815 \\
         &and $\thacol < 10^\circ$ & & & &                \\ 
\hline%-----------------------------------------------------------------
\hline%-----------------------------------------------------------------
\multicolumn{6}{|c|}{\bf \boldmath Asymmetries at $\sqrt{s}$ = 205~GeV} \\
\hline%-----------------------------------------------------------------
&&$N_{\mathrm{f}}$ &$N_{\mathrm{b}}$ &$\AFB$ &$\AFBSM$ \\ 
\hline%-----------------------------------------------------------------
\mumu    &$s'/s > 0.01$    &270    &148   &0.234$\pm$0.051$\pm$0.006 &0.277 \\
         &$s'/s > 0.7225$  &160    & 52   &0.512$\pm$0.063$\pm$0.005 &0.553 \\
\hline%-----------------------------------------------------------------
\tautau  &$s'/s > 0.01$    &133    &58    &0.317$\pm$0.076$\pm$0.006 &0.277 \\
         &$s'/s > 0.7225$  & 87    &26    &0.575$\pm$0.092$\pm$0.011 &0.553 \\
\hline%-----------------------------------------------------------------
\epem    &$\absctem < 0.7$ &1248   &147   &0.792$\pm$0.016$\pm$0.004 &0.816 \\
         &and $\thacol < 10^\circ$ & & & &                \\ 
\hline%-----------------------------------------------------------------
\hline%-----------------------------------------------------------------
\multicolumn{6}{|c|}{\bf \boldmath Asymmetries at $\sqrt{s}$ = 207~GeV} \\
\hline%-----------------------------------------------------------------
&&$N_{\mathrm{f}}$ &$N_{\mathrm{b}}$ &$\AFB$ &$\AFBSM$ \\ 
\hline%-----------------------------------------------------------------
\mumu    &$s'/s > 0.01$    &488    &251   &0.264$\pm$0.038$\pm$0.006 &0.277 \\
         &$s'/s > 0.7225$  &261    & 86   &0.508$\pm$0.050$\pm$0.005 &0.552 \\
\hline%-----------------------------------------------------------------
\tautau  &$s'/s > 0.01$    &205    &105   &0.230$\pm$0.061$\pm$0.007 &0.277 \\
         &$s'/s > 0.7225$  &142    &53    &0.472$\pm$0.075$\pm$0.011 &0.551 \\
\hline%-----------------------------------------------------------------
\epem    &$\absctem < 0.7$ &2062   &216   &0.814$\pm$0.012$\pm$0.004 &0.816 \\
         &and $\thacol < 10^\circ$ & & & &                \\ 
\hline%-----------------------------------------------------------------
\hline%-----------------------------------------------------------------
\end{tabular}
\caption[]{Continued
%  The numbers of forward ($N_{\mathrm{f}}$) and backward
%  ($N_{\mathrm{b}}$) events and measured asymmetry values at each
%  202--207~GeV. The
%  measured asymmetry values include corrections for background and efficiency,
%  and in the case of muons and taus are corrected to the full solid angle.
%  The errors shown are the combined statistical and systematic errors.
%  The asymmetries for
%  \Pgmp\Pgmm, \Pgtp\Pgtm\ and for the combined \mumu\ and \tautau\ are
%  shown after the correction for interference between
%  initial- and final-state radiation.
%  The final column shows the Standard Model predictions of \BHWIDE\ for
%  \epem\ and \ZFITTER\ for the other final states.
}
\label{tab:afb2}
\end{table}%------------------------------------------------------------

\begin{table}[htbp]%-------------------------------------------------------
\begin{sideways}
\begin{minipage}[b]{\textheight}
\begin{center}
\vspace*{3cm}
\begin{tabular}{|c|r@{$\pm$}l|r@{$\pm$}l|r@{$\pm$}l|r@{$\pm$}l|r@{$\pm$}l|
                   r@{$\pm$}l|r@{$\pm$}l|}
\hline%-----------------------------------------------------------------
\hline%-----------------------------------------------------------------
\multicolumn{15}{|c|}{\bf \boldmath $\qqbar$} \\
\hline
$\absct$  &\multicolumn{14}{c|}{$\dsdabscc$ / pb} \\
\hline
&\multicolumn{2}{c|}{189~GeV}  &\multicolumn{2}{c|}{192~GeV} 
&\multicolumn{2}{c|}{196~GeV}  &\multicolumn{2}{c|}{200~GeV} 
&\multicolumn{2}{c|}{202~GeV}  &\multicolumn{2}{c|}{205~GeV} 
&\multicolumn{2}{c|}{207~GeV}  \\
\hline%-----------------------------------------------------------------
$[\;\;\;0.0,\;\;\;0.1]$  &17.4&1.0 &20.8&2.8  &16.4&1.5  
                         &11.8&1.3  
                         &13.6&2.0  &12.3&1.3  &13.0&1.0  \\
$[\;\;\;0.1,\;\;\;0.2]$  &17.9&1.0 &14.6&2.4  &14.4&1.5   
                         &15.8&1.5  
                         &16.1&2.2  &12.8&1.4  &13.4&1.1  \\
$[\;\;\;0.2,\;\;\;0.3]$  &17.8&1.0 &18.1&2.6  &14.4&1.5  
                         &11.9&1.3  
                         &15.4&2.2  &12.6&1.4  &10.7&1.0  \\
$[\;\;\;0.3,\;\;\;0.4]$  &18.3&1.1 &17.6&2.6  &16.1&1.6  
                         &16.2&1.5    
                         &20.5&2.5  &15.8&1.5  &16.6&1.2  \\
$[\;\;\;0.4,\;\;\;0.5]$  &18.6&1.1 &23.6&3.0  &19.2&1.7  
                         &17.4&1.6  
                         &16.3&2.2  &16.5&1.5  &15.0&1.1  \\
$[\;\;\;0.5,\;\;\;0.6]$  &21.3&1.1 &23.2&3.0  &20.5&1.7  
                         &18.0&1.6  
                         &14.4&2.1  &19.1&1.6  &17.1&1.2 \\
$[\;\;\;0.6,\;\;\;0.7]$  &24.8&1.2 &23.1&3.0  &23.4&1.8  
                         &20.8&1.7  
                         &18.6&2.4  &20.4&1.7  &18.9&1.3  \\
$[\;\;\;0.7,\;\;\;0.8]$  &25.8&1.2 &23.8&3.0  &21.3&1.7  
                         &21.4&1.7  
                         &21.1&2.5  &22.4&1.7  &17.6&1.2  \\
$[\;\;\;0.8,\;\;\;0.9]$  &26.6&1.3 &29.1&3.3  &23.0&1.8  
                         &27.6&2.0  
                         &23.4&2.6  &21.7&1.7  &20.9&1.3  \\
$[\;\;\;0.9,\;\;\;1.0]$  &31.0&1.7 &26.2&4.0  &28.9&2.6  
                         &27.7&2.5  
                         &25.3&3.4  &28.4&2.5  &24.7&1.8  \\
\hline%-----------------------------------------------------------------
\hline%-----------------------------------------------------------------
\end{tabular}     
\caption{
Differential cross-sections for $\qqbar$ production.
The values are for $s'/s > 0.7225$ and are corrected to no interference 
between initial- and final-state radiation.
Errors are statistical only; systematic errors are given in 
Table~\ref{tab:sys_all}.
}
\label{tab:qq_angdis}
\end{center}
\end{minipage}
\end{sideways}  
\end{table}%------------------------------------------------------------

\begin{table}[p]%-------------------------------------------------------
\begin{sideways}
\begin{minipage}[b]{\textheight}
\begin{center}
\vspace*{3cm}
\begin{tabular} {|c|r@{$\pm$}l|r@{$\pm$}l|r@{$\pm$}l|r@{$\pm$}l|r@{$\pm$}l|
                    r@{$\pm$}l|r@{$\pm$}l|}
\hline%-----------------------------------------------------------------
\hline%-----------------------------------------------------------------
\multicolumn{15}{|c|}{\bf \boldmath $\mumu$} \\
\hline
$\ct$  &\multicolumn{14}{c|}{$\dsdcc$ / pb} \\
\hline%-----------------------------------------------------------------
&\multicolumn{2}{c|}{189~GeV}  &\multicolumn{2}{c|}{192~GeV} 
&\multicolumn{2}{c|}{196~GeV}  &\multicolumn{2}{c|}{200~GeV}   
&\multicolumn{2}{c|}{202~GeV}  &\multicolumn{2}{c|}{205~GeV} 
&\multicolumn{2}{c|}{207~GeV}  \\
\hline%-----------------------------------------------------------------
$[-1.0,-0.8]$ &0.67&$^{0.32}_{0.25}$  &1.3&$^{1.3}_{0.8}$  
              &0.19&$^{0.39}_{0.19}$  &0.31&$^{0.45}_{0.28}$  
              &1.0&$^{1.0}_{0.7}$     &0.76&$^{0.58}_{0.42}$
              &0.44&$^{0.34}_{0.25}$  \\
$[-0.8,-0.6]$ &0.47&$^{0.18}_{0.14}$  &0.7&$^{0.7}_{0.4}$  
              &0.28&$^{0.26}_{0.16}$  &0.17&$^{0.23}_{0.13}$  
              &0.1&$^{0.4}_{0.1}$     &0.43&$^{0.30}_{0.20}$
              &0.49&$^{0.22}_{0.17}$  \\
$[-0.6,-0.4]$ &0.54&$^{0.20}_{0.16}$  &0.4&$^{0.6}_{0.3}$  
              &0.79&$^{0.38}_{0.29}$  &0.56&$^{0.34}_{0.24}$  
              &0.3&$^{0.5}_{0.2}$     &0.72&$^{0.35}_{0.26}$
              &0.72&$^{0.26}_{0.20}$  \\
$[-0.4,-0.2]$ &0.62&0.13              &0.5&$^{0.6}_{0.3}$  
              &0.54&$^{0.32}_{0.23}$  &0.60&$^{0.33}_{0.24}$  
              &0.5&$^{0.5}_{0.3}$     &0.60&$^{0.33}_{0.24}$
              &0.70&$^{0.25}_{0.20}$  \\
$[-0.2,\;\;\;0.0]$  &1.27&0.19             &1.9&$^{0.8}_{0.6}$  
                    &0.54&$^{0.30}_{0.21}$ &0.9&0.3   
                    &1.2&$^{0.6}_{0.5}$    &1.0&$^{0.4}_{0.3}$
                    &1.01&0.20  \\
$[\;\;\;0.0,\;\;\;0.2]$ &1.35&0.20          &0.8&$^{0.6}_{0.4}$ 
                        &1.7&0.3            &1.3&0.3                 
                        &0.9&$^{0.6}_{0.4}$ &1.5&0.3
                        &0.86&0.19     \\
$[\;\;\;0.2,\;\;\;0.4]$ &2.03&0.25          &2.0&$^{0.9}_{0.7}$ 
                        &1.6&0.3            &1.9&0.4                  
                        &0.8&$^{0.6}_{0.4}$ &2.2&0.4
                        &1.7&0.3                \\
$[\;\;\;0.4,\;\;\;0.6]$ &2.15&0.26          &1.9&$^{0.9}_{0.7}$ 
                        &2.2&0.4            &1.6&0.3                  
                        &1.4&$^{0.7}_{0.5}$ &1.8&0.4
                        &2.0&0.3               \\
$[\;\;\;0.6,\;\;\;0.8]$ &2.85&0.30          &2.7&$^{1.0}_{0.8}$ 
                        &3.0&0.5            &2.8&0.5                   
                        &2.8&$^{0.7}_{0.7}$ &2.2&0.4
                        &2.6&0.3                 \\
$[\;\;\;0.8,\;\;\;1.0]$ &3.77&0.42          &1.9&$^{1.3}_{0.9}$ 
                        &3.9&0.6            &3.8&0.6                    
                        &3.0&$^{1.2}_{1.0}$ &3.2&0.6
                        &3.3&0.5                \\
\hline%-----------------------------------------------------------------
\hline%-----------------------------------------------------------------
\end{tabular}     
\caption{
Differential cross-sections for \mumu\ production. 
The values are for $s'/s > 0.7225$ and are corrected to no interference 
between initial- and final-state radiation. Errors are statistical only; 
systematic errors are given in Table~\ref{tab:sys_all}.
}
\label{tab:mu_angdis}
\end{center}  
\end{minipage}
\end{sideways}  
\end{table}%------------------------------------------------------------

\begin{table}[p]%-------------------------------------------------------
\begin{sideways}
\begin{minipage}[b]{\textheight}
\begin{center}
\vspace*{3cm}
\begin{tabular} {|c|r@{$\pm$}l|r@{$\pm$}l|r@{$\pm$}l|r@{$\pm$}l|r@{$\pm$}l|
                    r@{$\pm$}l|r@{$\pm$}l|}
\hline%-----------------------------------------------------------------
\hline%-----------------------------------------------------------------
\multicolumn{15}{|c|}{\bf \boldmath $\tautau$} \\
\hline
$\ct$  &\multicolumn{14}{c|}{$\dsdcc$ / pb} \\
\hline%-----------------------------------------------------------------
&\multicolumn{2}{c|}{189~GeV}  &\multicolumn{2}{c|}{192~GeV} 
&\multicolumn{2}{c|}{189~GeV}  &\multicolumn{2}{c|}{200~GeV}  
&\multicolumn{2}{c|}{202~GeV}  &\multicolumn{2}{c|}{205~GeV} 
&\multicolumn{2}{c|}{207~GeV}  \\
\hline%-----------------------------------------------------------------
$[-1.0,-0.8]$ &1.1&$^{1.0}_{0.8}$ &--0.1&$^{1.6}_{0.0}$ &1.3&$^{1.9}_{1.2}$
              &0.4&$^{1.4}_{0.7}$  
              &2.5&$^{4.0}_{2.6}$ &--0.1&$^{0.6}_{0.0}$
              &0.2&$^{0.9}_{0.4}$   \\
$[-0.8,-0.6]$ &0.2&$^{0.3}_{0.2}$ &--0.1&$^{0.5}_{0.0}$ &1.0&$^{0.7}_{0.5}$
              &0.0&$^{0.4}_{0.1}$  
              &0.3&$^{1.0}_{0.5}$ &0.7&$^{0.6}_{0.4}$
              &0.8&$^{0.5}_{0.3}$   \\
$[-0.6,-0.4]$ &0.9&$^{0.4}_{0.3}$ &0.9&$^{1.3}_{0.7}$ &0.7&$^{0.6}_{0.4}$
              &0.5&$^{0.6}_{0.4}$  
              &0.6&$^{1.0}_{0.6}$ &1.1&$^{0.7}_{0.5}$
              &0.7&$^{0.4}_{0.3}$   \\
$[-0.4,-0.2]$ &0.8&$^{0.4}_{0.3}$ &0.4&$^{1.1}_{0.6}$ &0.9&$^{0.7}_{0.5}$
              &0.8&$^{0.6}_{0.5}$  
              &1.2&$^{1.2}_{0.8}$ &0.4&$^{0.6}_{0.4}$
              &0.5&$^{0.4}_{0.3}$   \\
$[-0.2,\;\;\;0.0]$ &0.8&$^{0.4}_{0.3}$ &0.5&$^{1.1}_{0.5}$ &0.8&$^{0.6}_{0.5}$
              &0.7&$^{0.6}_{0.4}$  
                   &0.5&$^{1.0}_{0.6}$ &0.4&$^{0.6}_{0.4}$
              &1.2&$^{0.5}_{0.4}$   \\
$[\;\;\;0.0,\;\;\;0.2]$ &1.6&0.3 &1.7&$^{1.4}_{0.9}$ &1.1&$^{0.7}_{0.5}$
              &1.5&$^{0.8}_{0.6}$  
                        &0.9&$^{1.1}_{0.6}$ &1.2&$^{0.7}_{0.5}$
              &1.2&0.3              \\
$[\;\;\;0.2,\;\;\;0.4]$ &2.0&0.3 &1.8&$^{1.5}_{1.0}$ &1.3&$^{0.7}_{0.5}$
              &2.1&0.5  
                        &1.2&$^{1.1}_{0.7}$ &1.6&$^{0.7}_{0.6}$
              &1.4&0.3  \\
$[\;\;\;0.4,\;\;\;0.6]$ &2.7&0.4 &1.9&$^{1.4}_{1.0}$ &2.0&$^{0.8}_{0.6}$
              &2.4&0.5  
                        &2.9&$^{1.4}_{1.1}$ &1.9&$^{0.7}_{0.6}$
              &2.6&0.4   \\
$[\;\;\;0.6,\;\;\;0.8]$ &3.1&0.5 &3.7&$^{1.8}_{1.4}$ &1.9&0.5
              &3.4&0.7  
                        &2.9&$^{1.4}_{1.1}$ &2.5&0.6
              &2.0&0.4   \\
$[\;\;\;0.8,\;\;\;1.0]$ &4.7&0.8 &6.4&$^{4.3}_{3.1}$ &4.1&$^{2.0}_{1.6}$
              &3.8&$^{2.0}_{1.6}$  
                        &3.0&$^{3.0}_{2.0}$ &3.7&$^{2.0}_{1.5}$
              &2.5&$^{1.2}_{1.0}$  \\
\hline
\hline%-----------------------------------------------------------------
\end{tabular}     
\caption{
Differential cross-sections for \tautau\ production. 
The values are for $s'/s > 0.7225$ and are corrected to no interference 
between initial- and final-state radiation. Errors are statistical only; 
systematic errors are given in Table~\ref{tab:sys_all}.
}
\label{tab:tau_angdis}
\end{center}  
\end{minipage}
\end{sideways}  
\end{table}%------------------------------------------------------------

\begin{table}[htbp]%-------------------------------------------------------
\begin{sideways}
\begin{minipage}[b]{\textheight}
\begin{center}
\vspace*{1cm}
\begin{tabular} {|c|r@{$\pm$}l|r@{$\pm$}l|r@{$\pm$}l|r@{$\pm$}l|r@{$\pm$}l|
                    r@{$\pm$}l|r@{$\pm$}l|}
\hline%-----------------------------------------------------------------
\hline%-----------------------------------------------------------------
\multicolumn{15}{|c|}{\bf \boldmath $\epem$} \\
\hline
$\ct$  &\multicolumn{14}{c|}{$\dsdcc$ / pb} \\
\hline
&\multicolumn{2}{c|}{189~GeV}  &\multicolumn{2}{c|}{192~GeV} 
&\multicolumn{2}{c|}{196~GeV}  &\multicolumn{2}{c|}{200~GeV}   
&\multicolumn{2}{c|}{202~GeV}  &\multicolumn{2}{c|}{205~GeV} 
&\multicolumn{2}{c|}{207~GeV}  \\
%\hline%-----------------------------------------------------------------
%$[-0.9,-0.7]$            & 1.3&0.2 & 1.8&$^{0.7}_{0.5}$ & 1.4&0.3 & 1.5&0.3 
%                         & 2.0&$^{0.7}_{0.5}$ & 0.9&$^{0.3}_{0.2}$
%                         & 1.5&0.2                        \\
%$[-0.7,-0.5]$            & 2.1&0.2 & 1.9&$^{0.8}_{0.6}$ & 1.5&0.3 & 2.0&0.4 
%                         & 1.7&$^{0.6}_{0.5}$ & 1.7&0.3 & 1.5&0.2 \\
%$[-0.5,-0.3]$            & 2.4&0.3 & 2.0&$^{0.8}_{0.6}$ & 2.8&0.4 & 2.0&0.4 
%                         & 3.0&0.6            & 2.4&0.4 & 2.1&0.3 \\
%$[-0.3,-0.1]$            & 3.1&0.3 & 2.0&$^{0.8}_{0.6}$ & 2.6&0.4 & 2.9&0.4 
%                         & 2.8&0.6            & 3.0&0.4 & 2.6&0.3 \\
%$[-0.1,\;\;\;0.1]$       & 4.3&0.4 & 5.2&1.0            & 4.2&0.5 & 4.9&0.6  
%                         & 4.0&0.7            & 4.8&0.6 & 3.8&0.4 \\
%$[\;\;\;0.1,\;\;\;0.3]$  & 8.3&0.5 & 8.7&1.3            & 7.9&0.7 & 7.5&0.7  
%                         & 9.3&1.1            & 7.9&0.7 & 7.6&0.5 \\
%$[\;\;\;0.3,\;\;\;0.5]$  &19.9&0.8 &19.4&1.9            &17.3&1.1 &17.0&1.1  
%                         &14.2&1.4            &15.4&1.0 &17.0&0.8 \\
%$[\;\;\;0.5,\;\;\;0.7]$  &61.2&1.3 &58.8&3.2            &56.7&1.9 &56.1&1.9  
%                         &54.1&2.7            &51.6&1.8 &50.2&1.4 \\
%$[\;\;\;0.7,\;\;\;0.9]$  &417 &4   &402 &9              &390 &5   &371 &5    
%                        &366 &7              &348 &5   &341 &4   \\
%\hline%-----------------------------------------------------------------
\hline%-----------------------------------------------------------------
$[-0.90,-0.72]$            & 1.3&0.2 & 1.4&$^{0.7}_{0.5}$ & 1.3&$^{0.4}_{0.3}$
                           & 1.5&0.3 
                           & 2.2&$^{0.7}_{0.6}$ & 0.8&$^{0.3}_{0.2}$
                           & 1.4&0.2                        \\
$[-0.72,-0.54]$            & 2.1&0.3 & 2.5&$^{0.9}_{0.7}$ & 1.5&0.3 & 1.8&0.4 
                           & 1.3&$^{0.6}_{0.4}$ & 1.7&0.4 & 1.7&0.3 \\
$[-0.54,-0.36]$            & 2.4&0.3 & 1.5&$^{0.8}_{0.5}$ & 2.2&0.4 & 2.0&0.4 
                           & 2.8&$^{0.8}_{0.7}$ & 2.3&0.4 & 1.7&0.3 \\
$[-0.36,-0.18]$            & 2.5&0.3 & 2.4&$^{0.9}_{0.7}$ & 2.9&0.5 & 2.5&0.4 
                           & 3.8&0.8            & 2.3&0.4 & 2.8&0.3 \\
$[-0.18,\;\;\;0.00]$       & 3.8&0.4 & 2.9&$^{1.0}_{0.8}$ & 3.5&0.5 & 3.9&0.5  
                           & 2.2&$^{0.8}_{0.6}$ & 3.8&0.5 & 2.8&0.4 \\
$[\;\;\;0.00,\;\;\;0.09]$  & 4.5&0.5 & 6.5&$^{2.0}_{1.6}$ & 5.1&0.9 & 5.1&0.9  
                           & 5.0&$^{1.6}_{1.2}$ & 5.7&0.9 & 3.9&0.6 \\
$[\;\;\;0.09,\;\;\;0.18]$  & 6.3&0.6 & 6.6&$^{2.1}_{1.6}$ & 5.6&0.9 & 6.2&1.0  
                           & 8.2&1.6            & 6.9&1.0 & 6.4&0.7 \\
$[\;\;\;0.18,\;\;\;0.27]$  & 9.2&0.8 & 9.2&1.9            & 8.4&1.1 & 9.7&1.2  
                           & 8.9&1.6            & 7.2&1.0 & 7.4&0.8 \\
$[\;\;\;0.27,\;\;\;0.36]$  &12.3&0.9 &13.7&2.3            &12.7&1.4 &10.0&1.2  
                           &10.8&1.8            &11.2&1.3 &11.9&1.0 \\
$[\;\;\;0.36,\;\;\;0.45]$  &19.7&1.1 &21.2&2.9            &14.9&1.5 &15.6&1.5  
                           &14.6&2.1            &14.2&1.4 &17.1&1.2 \\
$[\;\;\;0.45,\;\;\;0.54]$  &31.9&1.4 &31.0&3.5            &26.7&2.0 &28.8&2.1  
                           &24.5&2.7            &27.2&2.0 &25.1&1.5 \\
$[\;\;\;0.54,\;\;\;0.63]$  &51.6&1.8 &45.2&4.3            &50.1&2.8 &48.4&2.7  
                           &45.4&3.8            &43.8&2.5 &41.9&1.9 \\
$[\;\;\;0.63,\;\;\;0.72]$  &94.9&2.4 &91.6&5.9            &90.5&3.6 &87.1&3.5  
                           &84.4&5.0            &79.0&3.3 &77.3&2.5 \\
$[\;\;\;0.72,\;\;\;0.81]$  &215 &4   &206 &9              &199 &6   &194 &5    
                           &184 &8              &173 &5   &174 &4   \\
$[\;\;\;0.81,\;\;\;0.90]$  &684 &7   &663 &16             &640 &10  &606 &10   
                           &606 &14             &578 &9   &565 &7   \\
\hline%-----------------------------------------------------------------
\hline
\end{tabular}     
\caption{
Differential cross-sections for $\epem$  production for
$\thacol < 10\degree$. Errors are statistical only; systematic 
errors are given in Table~\ref{tab:sys_all}.
}
\label{tab:ee_angdis}
\end{center}  
\end{minipage}
\end{sideways}  
\end{table}%------------------------------------------------------------

\begin{table}[htbp]%-------------------------------------------------------
\centering
\begin{tabular}{|ll|c|c|c|c|c|c|c|}
\hline%-----------------------------------------------------------------
\hline%-----------------------------------------------------------------
          &$\sqrt{s}$ / GeV &189   & 192 & 196 & 200 & 202 & 205 & 207 \\ 
\hline
\qqbar     &$s'/s>0.01$     &0.90  &0.92 &0.97 &1.00 &1.00 &1.00 &1.04 \\
           &$s'/s>0.7225$   &0.57  &0.54 &0.53 &0.52 &0.52 &0.51 &0.54 \\
\hline
\Pgmp\Pgmm &$s'/s>0.01$     &1.12  &1.13 &1.13 &1.14 &1.19 &1.17 &1.16 \\
           &$s'/s>0.7225$   &1.08  &1.08 &1.07 &1.07 &1.08 &1.08 &1.08 \\
\hline
\Pgtp\Pgtm &$s'/s>0.01$     &2.6\phz &2.6\phz &2.6\phz &2.6\phz &2.5\phz 
                                                       &2.6\phz &2.6\phz \\
           &$s'/s>0.7225$   &2.5\phz &2.5\phz &2.5\phz &2.5\phz &2.5\phz
                                                       &2.5\phz &2.5\phz \\
\hline
\Pep\Pem   &A: $\absctepem<0.9$, $\thacol<170\degree$ 
                            &0.51  &0.46 &0.46 &0.46 &0.46 &0.59 &0.59 \\
           &B: $\absctem<0.7$, $\thacol<10\degree$ 
                            &0.31  &0.36 &0.36 &0.36 &0.36 &0.39 &0.39 \\
           &C: $\absctepem<0.96$, $\thacol<10\degree$ 
                            &0.44  &0.43 &0.43 &0.43 &0.43 &0.44 &0.44 \\
\hline
\Pep\Pem   &D: $-0.9<\ctem<-0.7$, $\thacol<10\degree$ 
                            &1.10  &1.16 &1.16 &1.16 &1.16 &1.18 &1.18 \\
           &D: $\absctem<0.7$, $\thacol<10\degree$ 
                            &0.47  &0.50 &0.50 &0.50 &0.50 &0.52 &0.52 \\
           &D: $0.7<\ctem<0.9$, $\thacol<10\degree$ 
                            &0.77  &0.86 &0.86 &0.86 &0.86 &0.89 &0.89 \\
\hline%-----------------------------------------------------------------
\hline%-----------------------------------------------------------------
\end{tabular}
\caption[]{Total systematic errors, in \%, excluding those on the luminosity 
           measurement, for each channel at each nominal
           centre-of-mass energy. For the $\epem$ final state, the
           first three rows refer to the total cross-section measurements,
           while the last three refer to the differential cross-section,
           for which extra cuts are applied. For the hadronic, $\mumu$
           and $\tautau$ final states, the values given for $s'/s > 0.7225$
           apply to both the total cross-section and the differential
           cross-section.
}
\label{tab:sys_all}
\end{table}%------------------------------------------------------------

\begin{table}[htbp]%-------------------------------------------------------
\begin{center}
\begin{tabular}{|l|c|c|}
\hline
\hline
\multicolumn{3}{|c|}{\bf \boldmath $\qqbar$} \\
\hline
                  &\boldmath $s'/s>0.01$  &\boldmath $s'/s>0.7225$  \\
\hline
MC statistics (efficiency) &0.03             &0.06 \\
MC statistics (background) &0.03             &0.04 \\
ISR modelling              &0.12             &0.02 \\
Fragmentation modelling    &0.37             &0.26 \\
Detector effects           &0.19             &0.15 \\
$s'$ determination         &0.03             &0.19 \\
\WW\ rejection cuts        &0.07             &0.13 \\
\WW\ background            &0.21             &0.26 \\
Other background           &0.87             &0.19 \\
Interference               &0.04             &0.13 \\
\hline
Total                      &1.00             &0.52 \\
\hline
\hline
\end{tabular}
\end{center}
\caption{Systematic errors, in \%, on the hadronic cross-section 
         measurements at 200~GeV. Values at other energies are
         very similar, the total errors are given in Table~\ref{tab:sys_all}.
         Errors on the luminosity measurement are given in 
         Table~\ref{tab:lumi_errors}.
        }
\label{tab:mh_syserr}
\end{table}%------------------------------------------------------------

\begin{table}%------------------------------------------------------------
\begin{center}
\begin{tabular}{|l|c|c|}
\hline
\hline
\multicolumn{3}{|c|}{\bf \boldmath $\mumu$} \\
\hline
                  &\boldmath $s'/s>0.01$  &\boldmath $s'/s>0.7225$  \\
\hline
MC statistics (efficiency)  &0.08         &0.08 \\
MC statistics (background)  &0.10         &0.08 \\
MC statistics (feedthrough) &--           &0.03 \\
Efficiency                  &1.00         &1.00 \\
ISR modelling               &0.20         &0.07 \\
Feedthrough                 &--           &0.16 \\
Cosmic background           &0.20         &0.20 \\
Other background            &0.46         &0.20 \\
Interference                &0.01         &0.12 \\
\hline
Total                       &1.14         &1.07 \\
\hline
\hline
\end{tabular}
\end{center}
\caption{Systematic errors, in \%, on the $\mumu$ cross-section 
         measurements at 200 GeV. The errors at other energies are
         very similar, the totals at each energy are given in 
         Table~\ref{tab:sys_all}. Errors on the luminosity measurement 
         are given in Table~\ref{tab:lumi_errors}.
%         For the differental cross-section measurements the systematic
%         errors are assumed to be the same as those given for the total
%         cross-section with $s'/s > 0.7225$.
        }
\label{tab:mu_syserr}
\end{table}%------------------------------------------------------------

\begin{table}%------------------------------------------------------------
\begin{center}
\begin{tabular}{|l|c|c|}
\hline
\hline
\multicolumn{3}{|c|}{\bf \boldmath $\tautau$} \\
\hline
                  &\boldmath $s'/s>0.01$  &\boldmath $s'/s>0.7225$  \\
\hline
MC statistics (efficiency)  &0.19         &0.21 \\
MC statistics (background)  &0.22         &0.20 \\
MC statistics (feedthrough) &--           &0.05 \\
Efficiency                  &2.34         &2.34 \\
ISR                         &0.20         &0.04 \\
Feedthrough                 &--           &0.15 \\
Background                  &0.99         &0.69 \\
Interference                &0.04         &0.01 \\
\hline
Total                       &2.6\phz      &2.5\phz \\
\hline
\hline
\end{tabular}
\end{center}
\caption{Systematic errors, in \%, on the $\tautau$ cross-section 
         measurements at 200 GeV. The errors at other energies are
         very similar, the totals at each energy are given in 
         Table~\ref{tab:sys_all}. Errors on the luminosity measurement 
         are given in Table~\ref{tab:lumi_errors}.
%         For the differental cross-section measurements the systematic
%         errors are assumed to be the same as those given for the total
%         cross-section with $s'/s > 0.7225$.
        }
\label{tab:tau_syserr}
\end{table}%------------------------------------------------------------

\begin{table}[htbp]
\begin{center}
\begin{tabular}{|l|c|c|c|}
\hline
\hline
\multicolumn{4}{|c|}{\bf \boldmath $\epem$} \\
\hline
    &{\bf A}  &{\bf B}  &{\bf C}  \\
    &\footnotesize{$\absctepem < 0.9$}  &\footnotesize{$\absctem < 0.7$}  
    &\footnotesize{$\absctepem < 0.96$} \\
    &\footnotesize{$\thacol < 170^{\circ}$} 
    &\footnotesize{$\thacol < 10^{\circ}$} 
    &\footnotesize{$\thacol < 10^{\circ}$} \\
\hline
MC statistics                        &0.02 &0.04    &0.02 \\
4-fermion correction                 &0.06 &--      &--   \\
Multiplicity cuts                    &0.09 &0.04    &0.03 \\
Calorimeter energy scale/resolution  &0.01 &$<0.01$ &0.08 \\
Two track requirement                &0.32 &0.30    &--   \\
Acceptance                           &0.19 &0.14    &0.39 \\
Background                           &0.25 &0.13    &0.17 \\
\hline
Total                                &0.46 &0.36    &0.43 \\
\hline
\hline
&{\bf D}  &{\bf D}  &{\bf D} \\
    &\footnotesize{$-0.9 <\ctem < -0.7$}
    &\footnotesize $\absctem < 0.7$     &\footnotesize{$+0.7 <\ctem <+ 0.9$} \\
    &\footnotesize{$\thacol < 10^{\circ}$} 
    &\footnotesize{$\thacol < 10^{\circ}$}
    &\footnotesize{$\thacol < 10^{\circ}$} \\
\hline
Multiplicity cuts                    &0.05 &0.04 &0.05 \\
Calorimeter energy scale/resolution  &0.01 &0.01 &0.01 \\
Two track requirement                &0.48 &0.30 &0.48 \\
Opposite charge requirement          &0.64 &0.37 &0.64 \\
Charge misassignment                 &0.50 &--   &--   \\
Acceptance                           &0.30 &0.10 &0.30 \\
Background                           &0.60 &0.11 &0.02 \\
\hline
Total                                &1.16 &0.50 &0.86 \\
\hline
\hline
\end{tabular}
\end{center}
\caption[]{Systematic errors, in \%, on the $\epem$ cross-section 
           and angular distribution measurements at 200~GeV. Values
           at other energies are very similar, the totals at each
           energy are given in Table~\ref{tab:sys_all}. 
           Errors on the luminosity measurement are given in 
           Table~\ref{tab:lumi_errors}.
           In the case of the angular distribution, acceptance D, the 
           errors arising from Monte Carlo statistics are included in 
           the statistical errors given in Table~\ref{tab:ee_angdis}.
         }
\label{tab:ee_syserr}
\end{table}%------------------------------------------------------------

\clearpage

\begin{table}[htbp]%-------------------------------------------------------
\centering
\begin{tabular}{|ll|c||r@{$\pm$}l|c|}
\hline%-----------------------------------------------------------------
\hline%-----------------------------------------------------------------
 & &Standard Model   &\multicolumn{3}{c|}{Data / Standard Model} \\
 & &$\chi^2$ / d.o.f &\multicolumn{2}{c|}{Mean} &$\chi^2$ / d.o.f \\ 
\hline
$\sigma(\qqbar)$       &$s'/s>0.01$     &5.9/7   &0.997&0.010 &5.8/6   \\
                       &$s'/s>0.7225$   &5.0/7   &0.990&0.011 &4.2/6   \\
\hline
$\sigma(\Pgmp\Pgmm)$   &$s'/s>0.01$     &9.8/7   &1.012&0.021 &9.5/6   \\
                       &$s'/s>0.7225$   &4.7/7   &0.994&0.028 &4.6/6   \\
$\AFB(\Pgmp\Pgmm)$     &$s'/s>0.01$     &11.6/7  &0.975&0.065 &11.5/6  \\
                       &$s'/s>0.7225$   &12.3/7  &0.999&0.040 &12.3/6  \\
\hline
$\sigma(\Pgtp\Pgtm)$   &$s'/s>0.01$     &2.6/7   &1.045&0.039 &1.3/6   \\
                       &$s'/s>0.7225$   &2.4/7   &1.052&0.044 &1.1/6   \\
$\AFB(\Pgtp\Pgtm)$     &$s'/s>0.01$     &6.3/7   &1.015&0.096 &6.2/6   \\
                       &$s'/s>0.7225$   &13.8/7  &1.033&0.057 &13.5/6  \\
\hline
$\sigma(\ell^+\ell^-)$ &$s'/s>0.01$     &12.4/14 &1.020&0.019 &11.3/13 \\
                       &$s'/s>0.7225$   &7.1/14  &1.010&0.024 &6.9/13  \\
$\AFB(\ell^+\ell^-)$   &$s'/s>0.01$     &17.9/14 &0.988&0.054 &17.9/13 \\
                       &$s'/s>0.7225$   &26.2/14 &1.010&0.033 &26.1/13 \\
\hline%-----------------------------------------------------------------
\hline%-----------------------------------------------------------------
\end{tabular}
\caption[]{Comparison of measurements with Standard Model predictions.
           The first column gives the $\chi^2$ value of the measured
           cross-sections or asymmetry values at 189--207~GeV presented 
           here with respect to the Standard Model predictions. The second 
           and third columns give the results of fits to the mean ratios of 
           data to Standard Model predictions. Values for $\ell^+\ell^-$
           are for $\mumu$ and $\tautau$ together.
}
\label{tab:sm_fit}
\end{table}%------------------------------------------------------------

\begin{table}[htbp]%----------------------------------------------------
\begin{sideways}
\begin{minipage}[b]{\textheight}
\begin{center}
\vspace*{3cm}
\begin{tabular}{|c||r@{$\pm$}l|r@{/}l||c|r@{/}l||r@{$\pm$}l|r@{/}l||c|r@{/}l|}
\hline%-----------------------------------------------------------------
\hline%-----------------------------------------------------------------
      &\multicolumn{7}{c||}{Using cross-sections}
      &\multicolumn{7}{c|}{Using ratios} \\
      &\multicolumn{4}{c||}{Fit} &\multicolumn{3}{c||}{Standard Model} 
      &\multicolumn{4}{c||}{Fit} &\multicolumn{3}{c|}{Standard Model} \\
\hline%-----------------------------------------------------------------
$\sqrt{s}$ / GeV &\multicolumn{2}{c|}{$1/\alphaem$} 
                 &\multicolumn{2}{c||}{$\chi^2$/d.o.f.} 
                 &$1/\alphaem$                       
                 &\multicolumn{2}{c||}{$\chi^2$/d.o.f.} 
                 &\multicolumn{2}{c|}{$1/\alphaem$} 
                 &\multicolumn{2}{c||}{$\chi^2$/d.o.f.} 
                 &$1/\alphaem$                       
                 &\multicolumn{2}{c|}{$\chi^2$/d.o.f.} \\
\hline%-----------------------------------------------------------------
188.6   &127.1&$^{ 3.8}_{ 3.3}$  & 2.0&4  & 127.9  & 2.0&5 
        &126.2&$^{ 4.2}_{ 3.8}$  & 1.7&3  & 127.9  & 1.9&4 \\
191.6   &134.9&$^{16.5}_{10.7}$  &10.0&4  & 127.9  &10.4&5 
        &138.1&$^{14.3}_{10.6}$  & 9.1&3  & 127.9  &10.0&4 \\
195.5   &131.2&$^{ 6.9}_{ 5.7}$  & 8.4&4  & 127.9  & 8.7&5 
        &129.9&$^{ 7.6}_{ 6.4}$  & 8.1&3  & 127.9  & 8.2&4 \\
199.5   &130.3&$^{ 7.2}_{ 5.8}$  & 6.4&4  & 127.8  & 6.6&5 
        &130.3&$^{ 8.0}_{ 6.7}$  & 6.4&3  & 127.8  & 6.5&4 \\
201.6   &134.1&$^{11.5}_{ 8.5}$  & 2.8&4  & 127.8  & 3.3&5 
        &134.0&$^{12.2}_{ 9.4}$  & 2.8&3  & 127.8  & 3.1&4 \\
204.9   &122.1&$^{ 5.5}_{ 4.4}$  & 0.7&4  & 127.8  & 1.8&5 
        &123.7&$^{ 6.8}_{ 5.8}$  & 0.6&3  & 127.8  & 0.9&4 \\
206.6   &123.4&$^{ 4.0}_{ 3.4}$  & 4.6&4  & 127.8  & 5.7&5 
        &117.9&$^{ 4.7}_{ 4.2}$  & 0.6&3  & 127.8  & 4.5&4 \\
\hline%-----------------------------------------------------------------
193.2   &127.4&$^{ 2.1}_{ 2.0}$  &59.1&59 & 127.9  &59.1&60 
        &126.7&$^{ 2.4}_{ 2.3}$  &50.4&47 & 127.9  &50.7&48 \\
\hline%-----------------------------------------------------------------
\hline%-----------------------------------------------------------------
\end{tabular}
\caption[]{Results of fits for $\alphaem$. The first seven rows show
 the fits to data at each energy, the last row the combined fit to these 
 data and measurements at 130--183~GeV~\cite{bib:OPAL-SM183,bib:OPAL-SM172}.
 The Standard Model values of $1/\alphaem$, and the $\chi^2$ between the 
 measurements and the Standard Model predictions are also given for comparison.
 Results are shown for the fits using cross-sections, and also for the
 fits to cross-section ratios, as discussed in the text.
}
\label{tab:alphaem1}
\end{center}  
\end{minipage}
\end{sideways}  
\end{table}%------------------------------------------------------------

\begin{table}
\begin{center}
\renewcommand{\arraystretch}{1.5}
\begin{tabular}{|c|r@{$\pm$}l|r@{$\pm$}l|r@{}l|}
\hline
\hline
  Parameter &
  \multicolumn{2}{c|}{ Without lepton } & 
  \multicolumn{2}{c|}{ With lepton } & 
  \multicolumn{2}{c|}{ Standard Model } \\[-1ex] 
  & \multicolumn{2}{c|}{ universality } & 
  \multicolumn{2}{c|}{ universality } & 
  \multicolumn{2}{c|}{ prediction } \\ \hline
$\mPZ$ / GeV                   &  91.1882 &   0.0033 &  91.1872 &   0.0033 & \multicolumn{2}{c|}{$-$} \\
$\Gamma_{\rm Z}$ / GeV         &   2.4945 &   0.0041 &   2.4943 &   0.0041 &   2.4960 & ${ +0.0016\atop  -0.0029 }$\\
 \hline
$r^{\rm tot}_{\rm had}$        &    2.963 &    0.009 &    2.963 &    0.009 &   2.9650 & ${ +0.0037\atop  -0.0066 }$\\
$j^{\rm tot}_{\rm had}$        &    0.131 &    0.078 &    0.144 &    0.078 &   0.2213 & ${ +0.0027\atop  -0.0059 }$\\
 \hline
$r^{\rm tot}_{\rm e}$          &  0.14134 &  0.00069 &  \multicolumn{2}{c|}{} & \multicolumn{2}{c|}{} \\ 
$r^{\rm tot}_{\mu}$            &  0.14215 &  0.00056 &  \multicolumn{2}{c|}{} & \multicolumn{2}{c|}{} \\ 
$r^{\rm tot}_{\tau}$           &  0.14228 &  0.00074 &  \multicolumn{2}{c|}{} & \multicolumn{2}{c|}{} \\ 
$r^{\rm tot}_{\ell}$           & \multicolumn{2}{c|}{} &  0.14199 &  0.00050 &  0.14270 & ${+0.00016\atop -0.00027}$\\
 \hline
$j^{\rm tot}_{\rm e}$          & $-$0.080 &    0.044 &  \multicolumn{2}{c|}{} & \multicolumn{2}{c|}{} \\ 
$j^{\rm tot}_{\mu}$            & $-$0.008 &    0.019 &  \multicolumn{2}{c|}{} & \multicolumn{2}{c|}{} \\ 
$j^{\rm tot}_{\tau}$           & $-$0.004 &    0.025 &  \multicolumn{2}{c|}{} & \multicolumn{2}{c|}{} \\ 
$j^{\rm tot}_{\ell}$           & \multicolumn{2}{c|}{} & $-$0.014 &    0.015 &  0.00439 & ${+0.00010\atop -0.00022}$\\
 \hline
$r^{\rm fb}_{\rm e}$           &  0.00138 &  0.00084 &  \multicolumn{2}{c|}{} & \multicolumn{2}{c|}{} \\ 
$r^{\rm fb}_{\mu}$             &  0.00270 &  0.00043 &  \multicolumn{2}{c|}{} & \multicolumn{2}{c|}{} \\ 
$r^{\rm fb}_{\tau}$            &  0.00248 &  0.00057 &  \multicolumn{2}{c|}{} & \multicolumn{2}{c|}{} \\ 
$r^{\rm fb}_{\ell}$            & \multicolumn{2}{c|}{} &  0.00243 &  0.00032 &  0.00280 & ${+0.00007\atop -0.00016}$\\
 \hline
$j^{\rm fb}_{\rm e}$           &    0.763 &    0.070 &  \multicolumn{2}{c|}{} & \multicolumn{2}{c|}{} \\ 
$j^{\rm fb}_{\mu}$             &    0.758 &    0.024 &  \multicolumn{2}{c|}{} & \multicolumn{2}{c|}{} \\ 
$j^{\rm fb}_{\tau}$            &    0.788 &    0.030 &  \multicolumn{2}{c|}{} & \multicolumn{2}{c|}{} \\ 
$j^{\rm fb}_{\ell}$            & \multicolumn{2}{c|}{} &    0.767 &    0.018 &   0.7987 & ${ +0.0005\atop  -0.0006}$\\
 \hline
$\chi^2/\mbox{d.o.f.}$         & \multicolumn{2}{c|}{ 207.0 / 247 } &
\multicolumn{2}{c|}{ 213.3 / 255 } & \multicolumn{2}{c|}{} \\
\hline
\hline
\end{tabular}
\end{center}
\caption[]{
Results from the fit to all \LEPone\ and \LEPtwo\ data for the S-matrix 
parameters with and without the assumption of lepton universality. 
The last column gives the Standard Model predictions.
An S-matrix fit to only \LEPone\ data without lepton universality 
gives a $\chi^2$ of 146.6 with 187 d.o.f.~\cite{bib:PR328}. 
}
\label{tab:smat}
\end{table}

\begin{table}
\begin{sideways}
\begin{minipage}[b]{\textheight}
\begin{center}
\renewcommand{\arraystretch}{1.3}
\setlength{\tabcolsep}{0.3em}
  \begin{tabular}{|rl|rr|rr|rrr|rrr|rrr|rrr|} 
  \hline
  \hline
 \multicolumn{2}{|c|}{Parameter}  &    1 &    2 &    3 &    4 &    5 &    6 &    7 &    8 &    9 &   10 &   11 &   12 &   13 &   14 &   15 &   16 \\
  \hline
   1 & $\mPZ$
 &   1.00 &   0.04 &   0.06 &$-$0.40 &$-$0.08 &   0.04 &   0.03 &$-$0.11 &$-$0.12 &$-$0.11 &$-$0.06 &   0.07 &   0.06 &   0.01 &$-$0.04 &$-$0.03 \\
   2 & $\Gamma_{\rm Z}$
 &   0.04 &   1.00 &   0.92 &$-$0.08 &   0.57 &   0.71 &   0.54 &$-$0.03 &   0.01 &   0.02 &   0.00 &   0.02 &   0.01 &   0.00 &   0.04 &   0.04 \\
  \hline
   3 & $r^{\rm tot}_{\rm had}$       
 &   0.06 &   0.92 &   1.00 &$-$0.09 &   0.57 &   0.71 &   0.54 &$-$0.04 &   0.00 &   0.01 &   0.01 &   0.02 &   0.02 &   0.00 &   0.04 &   0.04 \\
   4 & $j^{\rm tot}_{\rm had}$       
 &$-$0.40 &$-$0.08 &$-$0.09 &   1.00 &   0.00 &$-$0.07 &$-$0.05 &   0.08 &   0.09 &   0.08 &   0.03 &$-$0.04 &$-$0.03 &   0.00 &   0.03 &   0.02 \\
  \hline
   5 & $r^{\rm tot}_{\rm e}$         
 &$-$0.08 &   0.57 &   0.57 &   0.00 &   1.00 &   0.45 &   0.33 &   0.03 &   0.02 &   0.02 &   0.17 &$-$0.01 &$-$0.01 &$-$0.02 &   0.03 &   0.03 \\
   6 & $r^{\rm tot}_{\mu}$           
 &   0.04 &   0.71 &   0.71 &$-$0.07 &   0.45 &   1.00 &   0.41 &$-$0.03 &   0.08 &   0.01 &   0.01 &   0.03 &   0.01 &   0.00 &   0.09 &   0.03 \\
   7 & $r^{\rm tot}_{\tau}$          
 &   0.03 &   0.54 &   0.54 &$-$0.05 &   0.33 &   0.41 &   1.00 &$-$0.02 &   0.00 &   0.09 &   0.00 &   0.01 &   0.03 &   0.00 &   0.03 &   0.10 \\
  \hline
   8 & $j^{\rm tot}_{\rm e}$         
 &$-$0.11 &$-$0.03 &$-$0.04 &   0.08 &   0.03 &$-$0.03 &$-$0.02 &   1.00 &   0.02 &   0.02 &   0.00 &$-$0.01 &$-$0.01 &   0.24 &   0.01 &   0.00 \\
   9 & $j^{\rm tot}_{\mu}$           
 &$-$0.12 &   0.01 &   0.00 &   0.09 &   0.02 &   0.08 &   0.00 &   0.02 &   1.00 &   0.03 &   0.01 &   0.06 &$-$0.01 &   0.00 &   0.33 &   0.01 \\
  10 & $j^{\rm tot}_{\tau}$          
 &$-$0.11 &   0.02 &   0.01 &   0.08 &   0.02 &   0.01 &   0.09 &   0.02 &   0.03 &   1.00 &   0.01 &$-$0.01 &   0.06 &   0.00 &   0.01 &   0.29 \\
  \hline
  11 & $r^{\rm fb}_{\rm e}$          
 &$-$0.06 &   0.00 &   0.01 &   0.03 &   0.17 &   0.01 &   0.00 &   0.00 &   0.01 &   0.01 &   1.00 &$-$0.01 &$-$0.01 &   0.06 &   0.00 &   0.00 \\
  12 & $r^{\rm fb}_{\mu}$            
 &   0.07 &   0.02 &   0.02 &$-$0.04 &$-$0.01 &   0.03 &   0.01 &$-$0.01 &   0.06 &$-$0.01 &$-$0.01 &   1.00 &   0.02 &   0.00 &   0.12 &   0.00 \\
  13 & $r^{\rm fb}_{\tau}$           
 &   0.06 &   0.01 &   0.02 &$-$0.03 &$-$0.01 &   0.01 &   0.03 &$-$0.01 &$-$0.01 &   0.06 &$-$0.01 &   0.02 &   1.00 &   0.00 &   0.00 &   0.12 \\
  \hline
  14 & $j^{\rm fb}_{\rm e}$          
 &   0.01 &   0.00 &   0.00 &   0.00 &$-$0.02 &   0.00 &   0.00 &   0.24 &   0.00 &   0.00 &   0.06 &   0.00 &   0.00 &   1.00 &   0.00 &   0.00 \\
  15 & $j^{\rm fb}_{\mu}$            
 &$-$0.04 &   0.04 &   0.04 &   0.03 &   0.03 &   0.09 &   0.03 &   0.01 &   0.33 &   0.01 &   0.00 &   0.12 &   0.00 &   0.00 &   1.00 &   0.02 \\
  16 & $j^{\rm fb}_{\tau}$           
 &$-$0.03 &   0.04 &   0.04 &   0.02 &   0.03 &   0.03 &   0.10 &   0.00 &   0.01 &   0.29 &   0.00 &   0.00 &   0.12 &   0.00 &   0.02 &   1.00 \\
  \hline
  \hline
  \end{tabular}
\end{center}
\caption[]{
 Error correlation matrix for the S-matrix fit without lepton 
 universality.
}
\label{tab:smat_cor16}
\end{minipage}
\end{sideways}
\end{table}

\begin{table}
\begin{center}
\renewcommand{\arraystretch}{1.3}
  \begin{tabular}{|rl|rrrrrrrr|} 
  \hline
  \hline
 \multicolumn{2}{|c|}{Parameter}  &    1 &    2 &    3 &    4 &    5 &    6 &    7 &    8 \\
  \hline
   1 & $\mPZ$ 
 &   1.00 &   0.04 &   0.06 &$-$0.39 &   0.01 &$-$0.19 &   0.07 &$-$0.06 \\
   2 & $\Gamma_{\rm Z}$
 &   0.04 &   1.00 &   0.92 &$-$0.08 &   0.80 &   0.00 &   0.02 &   0.06 \\
   3 & $r^{\rm tot}_{\rm had}$       
 &   0.06 &   0.92 &   1.00 &$-$0.09 &   0.80 &$-$0.01 &   0.03 &   0.05 \\
   4 & $j^{\rm tot}_{\rm had}$       
 &$-$0.39 &$-$0.08 &$-$0.09 &   1.00 &$-$0.06 &   0.14 &$-$0.03 &   0.04 \\
   5 & $r^{\rm tot}_{\ell}$          
 &   0.01 &   0.80 &   0.80 &$-$0.06 &   1.00 &   0.06 &   0.05 &   0.09 \\
   6 & $j^{\rm tot}_{\ell}$          
 &$-$0.19 &   0.00 &$-$0.01 &   0.14 &   0.06 &   1.00 &   0.05 &   0.31 \\
   7 & $r^{\rm fb}_{\ell}$           
 &   0.07 &   0.02 &   0.03 &$-$0.03 &   0.05 &   0.05 &   1.00 &   0.11 \\
   8 & $j^{\rm fb}_{\ell}$           
 &$-$0.06 &   0.06 &   0.05 &   0.04 &   0.09 &   0.31 &   0.11 &   1.00 \\
  \hline
  \hline
  \end{tabular}
\end{center}
\caption[]{
 Error correlation matrix for the S-matrix fit assuming lepton 
 universality.
}
\label{tab:smat_cor8}
\end{table}

\begin{table}[htbp]%----------------------------------------------------------
\renewcommand{\tabcolsep}{0.15cm}
%\scalebox{0.95}{
\begin{sideways}
\begin{minipage}[b]{\textheight}{\footnotesize
\begin{center}\begin{tabular}{|cc|c|c|c|c|c|c|c|c|c|}
\hline
\hline
Channel &      &  LL  &  RR  &  LR  &  RL  &  VV  &  AA  &  
               LL+RR  &LR+RL & $\overline{\cal O}_{\mathrm{DB}}$ \\
        &      &  \scriptsize{ $[\pm1,0,0,0]$}  & 
                  \scriptsize{ $[0,\pm1,0,0]$}  & 
                  \scriptsize{ $[0,0,\pm1,0]$}  &
                  \scriptsize{ $[0,0,0,\pm1]$}  & 
                  \scriptsize{ $[\pm1,\pm1,\pm1,\pm1]$} &
                  \scriptsize{ $[\pm1,\pm1,\mp1,\mp1]$} &
                  \scriptsize{ $[\pm1,\pm1,0,0]$} & 
                  \scriptsize{ $[0,0,\pm1,\pm1]$} &
       \scriptsize{ $[\pm\frac{1}{4},\pm1,\pm\frac{1}{2},\pm\frac{1}{2}]$} \\

 \hline
\epem    &$\epsz$& $ 0.009_{-0.017}^{+0.018}$ & $ 0.009_{-0.017}^{+0.019}$ & 
                   $-0.009_{-0.009}^{+0.010}$ & $-0.009_{-0.009}^{+0.010}$ & 
                   $-0.002_{-0.004}^{+0.004}$ & $ 0.006_{-0.005}^{+0.005}$ & 
                   $ 0.004_{-0.008}^{+0.009}$ & $-0.005_{-0.005}^{+0.005}$ & 
                   $-0.003_{-0.007}^{+0.007}$ \\
         &$\lamp$& 4.7 & 4.7 & 8.1 & 8.1 &12.6 & 8.1 & 6.8 &11.8 & 9.2 \\
         &$\lamm$& 6.1 & 6.0 & 6.2 & 6.2 &10.6 &11.9 & 8.5 & 8.6 & 7.9 \\
 \hline
\mumu    &$\epsz$& $-0.002_{-0.009}^{+0.009}$ & $-0.002_{-0.009}^{+0.009}$ & 
                   $ 0.002_{-0.013}^{+0.012}$ & $ 0.002_{-0.013}^{+0.012}$ & 
                   $ 0.000_{-0.003}^{+0.003}$ & $-0.001_{-0.004}^{+0.004}$ & 
                   $-0.001_{-0.004}^{+0.004}$ & $ 0.001_{-0.006}^{+0.006}$ & 
                   $ 0.000_{-0.006}^{+0.006}$ \\
         &$\lamp$& 8.1 & 7.7 & 6.3 & 6.3 &12.7 &11.3 &11.1 & 8.7 & 9.6 \\
         &$\lamm$& 7.3 & 7.0 & 6.3 & 6.3 &12.4 &10.2 &10.1 & 9.3 & 9.2 \\
 \hline
\tautau  &$\epsz$& $ 0.015_{-0.014}^{+0.013}$ & $ 0.017_{-0.015}^{+0.014}$ & 
                   $-0.002_{-0.020}^{+0.018}$ & $-0.002_{-0.020}^{+0.018}$ & 
                   $ 0.004_{-0.005}^{+0.005}$ & $ 0.008_{-0.007}^{+0.007}$ & 
                   $ 0.008_{-0.007}^{+0.007}$ & $-0.001_{-0.010}^{+0.009}$ & 
                   $ 0.007_{-0.009}^{+0.009}$ \\
         &$\lamp$& 4.9 & 4.7 & 5.7 & 5.7 & 8.6 & 6.7 & 6.8 & 7.8 & 6.4 \\
         &$\lamm$& 7.2 & 6.9 & 4.6 & 4.6 &11.1 &10.7 &10.0 & 7.0 & 8.4 \\
 \hline
$ \lept$ &$\epsz$& $ 0.004_{-0.007}^{+0.007}$ & $ 0.004_{-0.007}^{+0.007}$ & 
                   $-0.005_{-0.007}^{+0.007}$ & $-0.005_{-0.007}^{+0.007}$ & 
                   $ 0.000_{-0.002}^{+0.002}$ & $ 0.003_{-0.003}^{+0.003}$ & 
                   $ 0.002_{-0.004}^{+0.004}$ & $-0.002_{-0.004}^{+0.004}$ & 
                   $ 0.000_{-0.004}^{+0.004}$ \\
         &$\lamp$& 7.7 & 7.4 & 9.3 & 9.3 &15.2 &10.5 &10.6 &13.2 &11.2 \\
         &$\lamm$& 9.5 & 9.2 & 7.3 & 7.3 &15.1 &15.4 &13.3 &10.3 &11.4 \\
 \hline
\qqbar   &$\epsz$& $-0.021_{-0.038}^{+0.021}$ & $ 0.018_{-0.022}^{+0.023}$ & 
                   $ 0.007_{-0.020}^{+0.020}$ & $ 0.011_{-0.010}^{+0.013}$ & 
                   $ 0.011_{-0.012}^{+0.012}$ & $-0.009_{-0.025}^{+0.009}$ & 
                   $-0.007_{-0.014}^{+0.014}$ & $ 0.010_{-0.009}^{+0.045}$ & 
                   $ 0.011_{-0.010}^{+0.052}$ \\
         &$\lamp$& 8.2 & 4.3 & 4.9 & 3.1 & 5.7 &12.0 & 7.5 & 3.9 & 3.7 \\
         &$\lamm$& 3.7 & 7.0 & 6.1 & 9.3 &10.4 & 5.0 & 5.7 &10.3 & 9.7 \\
 \hline
combined &$\epsz$& $ 0.001_{-0.006}^{+0.006}$ & $ 0.006_{-0.007}^{+0.007}$ & 
                   $-0.003_{-0.007}^{+0.007}$ & $ 0.001_{-0.005}^{+0.006}$ & 
                   $ 0.000_{-0.002}^{+0.002}$ & $ 0.001_{-0.003}^{+0.003}$ & 
                   $ 0.002_{-0.003}^{+0.003}$ & $-0.001_{-0.003}^{+0.003}$ & 
                   $ 0.002_{-0.004}^{+0.004}$ \\
         &$\lamp$& 9.2 & 7.2 & 9.4 & 9.0 &14.7 &12.6 &11.1 &12.9 &10.6 \\
         &$\lamm$& 9.4 &10.1 & 7.8 &10.1 &16.2 &14.9 &13.2 &12.3 &13.0 \\
 \hline
\uubar   &$\epsz$& $ 0.009_{-0.009}^{+0.009}$ & $ 0.014_{-0.013}^{+0.015}$ & 
                   $ 0.048_{-0.047}^{+0.095}$ & $ 0.018_{-0.044}^{+0.045}$ & 
                   $ 0.005_{-0.004}^{+0.005}$ & $ 0.007_{-0.007}^{+0.008}$ & 
                   $ 0.006_{-0.005}^{+0.005}$ & $ 0.038_{-0.038}^{+0.052}$ & 
                   $ 0.009_{-0.009}^{+0.009}$ \\
         &$\lamp$& 5.9 & 4.8 & 2.4 & 3.3 & 8.5 & 6.7 & 7.8 & 3.0 & 6.0 \\
         &$\lamm$& 9.1 & 7.7 & 5.4 & 4.1 &13.0 &10.7 &11.8 & 6.2 & 9.3 \\
 \hline
\ddbar   &$\epsz$& $-0.011_{-0.011}^{+0.011}$ & $-0.033_{-0.140}^{+0.032}$ & 
                   $-0.036_{-0.049}^{+0.047}$ & $ 0.045_{-0.043}^{+0.092}$ & 
                   $-0.009_{-0.010}^{+0.009}$ & $-0.007_{-0.008}^{+0.007}$ & 
                   $-0.008_{-0.008}^{+0.008}$ & $ 0.019_{-0.032}^{+0.032}$ & 
                   $-0.034_{-0.085}^{+0.033}$ \\
         &$\lamp$& 8.6 & 6.0 & 4.6 & 2.4 & 9.9 &10.9 &10.2 & 3.7 & 6.3 \\
         &$\lamm$& 5.5 & 2.2 & 2.9 & 5.4 & 5.6 & 6.7 & 6.5 & 5.2 & 2.7 \\
 \hline
$ \ud$   &$\epsz$& $ 0.018_{-0.032}^{+0.032}$ & $ 0.040_{-0.039}^{+0.050}$ & 
                   $ 0.018_{-0.032}^{+0.032}$ & $ 0.039_{-0.039}^{+0.051}$ & 
                   $ 0.011_{-0.011}^{+0.050}$ & $ 0.000_{-0.016}^{+0.015}$ & 
                   $ 0.028_{-0.027}^{+0.035}$ & $ 0.027_{-0.027}^{+0.035}$ & 
                   $ 0.016_{-0.015}^{+0.025}$ \\
         &$\lamp$& 3.7 & 3.0 & 3.7 & 3.0 & 3.8 & 6.1 & 3.6 & 3.6 & 2.8 \\
         &$\lamm$& 5.1 & 6.2 & 5.2 & 6.2 & 9.7 & 6.1 & 7.3 & 7.3 & 7.9 \\
 \hline
 \hline
\end{tabular}\end{center}}
\caption[foo]{\label{tab:ccres}
  Results of the contact interaction fits to the non-radiative hadron
  and lepton-pair data. The numbers in square brackets are the values of
  [$\eta_{\mathrm{LL}}$,$\eta_{\mathrm{RR}}$,$\eta_{\mathrm{LR}}$,
  $\!\eta_{\mathrm{RL}}$] which define the models.
  $\epsz$ is the fitted value of $\varepsilon = 1/\Lambda^{2}$,
  $\Lambda_{\pm}$ are the 95\% confidence level limits; the values
  for $\Lambda_+$ and $\Lambda_-$ correspond to the upper and lower signs,
  respectively, of the $\eta_{ij}$ values. 
  The units of $\Lambda$ are TeV, those of $\epsz$ are $\mathrm{TeV}^{-2}$.
}
\end{minipage}
\end{sideways}
%}
\end{table}%------------------------------------------------------------

\begin{table}[htbp]
\begin{center}
\renewcommand{\arraystretch}{1.3}
\begin{tabular}{|c|ccccc|} \hline \hline
Model: & $\chi$ & $\psi$ & $\eta$ & LR &SSM\\\hline
$m_{\rm Z'}^{\rm low}$ / GeV  &  781   &  366   &  515   &  518  &  1018 \\
$\tm^{\rm up}$ / mrad         &  1.94  &  2.58  &  3.31  &  1.90 &  0.91 \\
$\tm^{\rm low}$ / mrad        &--0.99  &--1.29  &--4.47  &--0.98 &--4.22 \\
\hline
\hline
\end{tabular}
\end{center}
\caption[ ]{One-dimensional limits at 95\% confidence level on the \zp\ mass,
$m_{\rm Z'}^{\rm low}$, and the mixing angle, $\tm^{\rm up}$ and 
$\tm^{\rm low}$, for various \zp\ models. 
The Z mass is free during the fit and the other three
Standard Model parameters ($\alphas$, $\mtop$ and $\mHiggs$) are 
fixed at their default values.
}
\label{tab:zptab}
\end{table}

%-----------------------------------------------------------------------
%       Figures
%-----------------------------------------------------------------------
\clearpage
%%%%%%%%%%%%%%%%%%%%%%%%%%%%%%%%%%%%%%%%%%%%%%%%%%%%%%%%%%
%
%%%%%%%%%%%%%%%%%%%%%%%%%%%%%%%%%%%%%%%%%%%%%%%%%%%%%%%%%%
%
%%%%%%%%%%%%%%%%%%%%%%%%%%%%%%%%%%%%%%%%%%%%%%%%%%%%%%%%%%
%
%%%%%%%%%%%%%%%%%%%%%%%%%%%%%%%%%%%%%%%%%%%%%%%%%%%%%%%%%%
%
%%%%%%%%%%%%%%%%%%%%%%%%%%%%%%%%%%%%%%%%%%%%%%%%%%%%%%%%%%
%
%%%%%%%%%%%%%%%%%%%%%%%%%%%%%%%%%%%%%%%%%%%%%%%%%%%%%%%%%%
%
%%%%%%%%%%%%%%%%%%%%%%%%%%%%%%%%%%%%%%%%%%%%%%%%%%%%%%%%%%
%
%%%%%%%%%%%%%%%%%%%%%%%%%%%%%%%%%%%%%%%%%%%%%%%%%%%%%%%%%%
%
%%%%%%%%%%%%%%%%%%%%%%%%%%%%%%%%%%%%%%%%%%%%%%%%%%%%%%%%%%
%%%%%%%%%%%%%%%%%%%%%%%%%%%%%%%%%%%%%%%%%%%%%%%%%%%%%%%%%%
%
%%%%%%%%%%%%%%%%%%%%%%%%%%%%%%%%%%%%%%%%%%%%%%%%%%%%%%%%%%
%
%%%%%%%%%%%%%%%%%%%%%%%%%%%%%%%%%%%%%%%%%%%%%%%%%%%%%%%%%%
%
%%%%%%%%%%%%%%%%%%%%%%%%%%%%%%%%%%%%%%%%%%%%%%%%%%%%%%%%%%
%
%%%%%%%%%%%%%%%%%%%%%%%%%%%%%%%%%%%%%%%%%%%%%%%%%%%%%%%%%%
%
%%%%%%%%%%%%%%%%%%%%%%%%%%%%%%%%%%%%%%%%%%%%%%%%%%%%%%%%%%
%
%%%%%%%%%%%%%%%%%%%%%%%%%%%%%%%%%%%%%%%%%%%%%%%%%%%%%%%%%%
%
\clearpage
%%%%%%%%%%%%%%%%%%%%%%%%%%%%%%%%%%%%%%%%%%%%%%%%%%%%%%%%%%
%
%%%%%%%%%%%%%%%%%%%%%%%%%%%%%%%%%%%%%%%%%%%%%%%%%%%%%%%%%%
%
%%%%%%%%%%%%%%%%%%%%%%%%%%%%%%%%%%%%%%%%%%%%%%%%%%%%%%%%%%
%
%%%%%%%%%%%%%%%%%%%%%%%%%%%%%%%%%%%%%%%%%%%%%%%%%%%%%%%%%%
%%%%%%%%%%%%%%%%%%%%%%%%%%%%%%%%%%%%%%%%%%%%%%%%%%%%%%%%%%
%
%%%%%%%%%%%%%%%%%%%%%%%%%%%%%%%%%%%%%%%%%%%%%%%%%%%%%%%%%%
%
%%%%%%%%%%%%%%%%%%%%%%%%%%%%%%%%%%%%%%%%%%%%%%%%%%%%%%%%%%


\begin{thebibliography}{99}

\bibitem{bib:OPAL-SM189}
  OPAL Collab., G.~Abbiendi et~al., 
  \EPJ\ {\bf C13} (2000) 553.

\bibitem{bib:OPAL-SM183}
  OPAL Collab., G.~Abbiendi et~al., 
  \EPJ\ {\bf C6} (1999) 1.

\bibitem{bib:OPAL-SM172}
  OPAL Collab., K.~Ackerstaff et~al., 
  \EPJ\ {\bf C2} (1998) 441.
  % DOI 10.1007/s100529800851.

\bibitem{bib:ADL-SM}
  \ALEPHColl, D.~Buskulic et~al., \PhysLett\ {\bf B378} (1996) 373; \\
  \ALEPHColl, R.~Barate et~al., \EPJ\ {\bf C12} (2000) 183; \\
  \DELPHIColl, P.~Abreu et~al., \EPJ\ {\bf C11} (1999) 383; \\
  \DELPHIColl, P.~Abreu et~al., \PhysLett\ {\bf B485} (2000) 45; \\
  \LthreeColl, M.~Acciarri et~al., \PhysLett\ {\bf B370} (1996) 195; \\
  \LthreeColl, M.~Acciarri et~al., \PhysLett\ {\bf B407} (1997) 361; \\
  %\bibitem{bib:L3-rpvsnu}
  \LthreeColl, M.~Acciarri et al., \PhysLett\ {\bf B414} (1997) 373; \\
  %\bibitem{bib:L3-newphys}
  \LthreeColl, M.~Acciarri et al., \PhysLett\ {\bf B433} (1998) 163; \\
  %\bibitem{bib:L3-LSG} LSG
  \LthreeColl, M.~Acciarri et al., \PhysLett\ {\bf B464} (1999) 135; \\
  \LthreeColl, M.~Acciarri et al., \PhysLett\ {\bf B470} (1999) 281; \\
  \LthreeColl, M.~Acciarri et al., \PhysLett\ {\bf B479} (2000) 101; \\
  \LthreeColl, M.~Acciarri et al., \PhysLett\ {\bf B489} (2000) 81.

\bibitem{bib:smatrix}
  A.~Leike, T.~Riemann and T.~Rose, \PhysLett\ {\bf B273} (1991) 513; \\
  T.~Riemann, \PhysLett\ {\bf B293} (1992) 451.

\bibitem{bib:zfitter}
  D.~Bardin et~al., \CPC\ {\bf 133} (2001) 229; \\
  D.~Bardin et~al., \PhysLett\ {\bf B255} (1991) 290; \\
  D.~Bardin et~al., \NPhys\ {\bf B351} (1991) 1; \\
  D.~Bardin et~al., \ZPhys\ {\bf C44} (1989) 493.\\
  We use \ZFITTER\ version 6.30 with default parameters, except 
  {\tt INTF}=0, {\tt FINR}=0, {\tt BOXD}=2, {\tt CONV}=2 and
  {\tt ALEM}=2,
  and with the following input parameters: $\mPZ$=91.1852~GeV,
  $\mtop$=174.3~GeV, $\mHiggs$=115~GeV,
  $\Delta\alpha_{\rm had}^{(5)}$=0.02761, 
  $\alphas(\mPZ)$=0.1185.
  
\bibitem{bib:LEPMCWS}
  {\sl Report of the Working Groups on Precision Calculations for LEP2
  Physics}, CERN 2000-009, ed. S.~Jadach, G.~Passarino and R.~Pittau, 
  pages 269--378.

\bibitem{bib:KK2f}
  S.~Jadach, B.F.L.~Ward and Z.~W\c{a}s, \PhysLett\ {\bf B449} (1999) 97; \\
  S.~Jadach et al., \CPC\ {\bf 130} (2000) 260.

\bibitem{bib:grc4f}
 %Four-fermion: grc4f \\
  J.~Fujimoto et~al., \CPC\ {\bf 100} (1997) 128.

\bibitem{bib:koralw}
  S.~Jadach et al., \CPC\ {\bf 119} (1999) 272.
  
%\bibitem{bib:ADD}
% N.~Arkani-Hamed, S.~Dimopoulos and G.~Dvali,
% \PhysLett\  {\bf B429} (1998) 263; \\
% I.~Antoniadis, N.~Arkani-Hamed, S.~Dimopoulos and G.~Dvali,
% \PhysLett\  {\bf B436} (1998) 257; \\
% N.~Arkani-Hamed, S.~Dimopoulos and G.~Dvali, \PhysRev\ {\bf D59}
% (1999) 86004.

%\bibitem{bib:Giudice}
% G.F.~Giudice, R.~Rattazzi and J.D.~Wells, \NPhys\ {\bf B544} (1999) 3.

\bibitem{bib:OPAL-detector}
  \OPALColl, K.~Ahmet et~al., \NIM\ {\bf A305} (1991) 275.

\bibitem{bib:OPAL-SI}
  S.~Anderson et~al., \NIM\ {\bf A403} (1998) 326.

\bibitem{bib:OPAL-TR}
  M.~Arignon et~al., \NIM\ {\bf A313} (1992) 103; \\
  M.~Arignon et~al., \NIM\ {\bf A333} (1993) 330.

\bibitem{bib:OPAL-DAQ}
  J.T.~Baines et~al., \NIM\ {\bf A325} (1993) 271; \\
  D.G.~Charlton, F.~Meijers, T.J.~Smith and P.S.~Wells, \NIM\ {\bf A325} 
  (1993) 129.

\bibitem{bib:OPAL-lumi}
  OPAL Collab., G.~Abbiendi et~al., 
  \EPJ\ {\bf C14} (2000) 373.

\bibitem{bib:ELEP}
  %LEP Energy Working Group, private communication; \\
  LEP Energy Working Group, {\sl Evaluation of the LEP centre-of-mass
  energy for data taken in 2000}, LEP Energy Working Group 01/01 
  (March 2001); \\
  LEP Energy Working Group, {\sl Evaluation of the LEP centre-of-mass
  energy for data taken in 1999}, LEP Energy Working Group 00/01 
  (June 2000); \\
  LEP Energy Working Group, {\sl Evaluation of the LEP centre-of-mass
  energy for data taken in 1998}, LEP Energy Working Group 99/01 
  (March 1998); \\
  See {\tt http://lepecal.web.cern.ch/LEPECAL/}; \\
  LEP Energy Working Group, A~Blondel et al., \EPJ\ {\bf C11} (1999) 573.

\bibitem{bib:pythia}
  %Multihadrons: PYTHIA \\
  T.~Sj\"ostrand et al., \CPC\ {\bf 135} (2001) 238.

\bibitem{bib:herwig}
 G.~Marchesini et~al., \CPC\ {\bf 67} (1992) 465; \\
 % Herwig 6.5
 G.~Corcella et al., JHEP {\bf 0101} (2001) 101.

\bibitem{bib:ariadne}
  L.~L{\"o}nnblad, \CPC\ {\bf 71} (1992) 15.

\bibitem{bib:OPAL-tune}
 OPAL Collab., G.~Alexander et al., \ZPhys\ {\bf C69} (1996) 543; \\
 OPAL Collab., G.~Abbiendi et al., CERN-EP-2003-031, submitted to \EPJ\
 {\bf C}.

\bibitem{bib:bhwide}
 %\epem: BHWIDE \\
  S.~Jadach, W.~Placzek and B.F.L.~Ward, \PhysLett\ {\bf B390} (1997) 298.

\bibitem{bib:koralz}
S.~Jadach, B.F.L.~Ward and Z.~W\c{a}s, Comput. Phys. Commun. {\bf 79} 
(1994) 503.

\bibitem{bib:phojet}
 %Two-photon: PHOJET \\
  R.~Engel and J.~Ranft, \PhysRev\ {\bf D54} (1996) 4244.

\bibitem{bib:twogen}
 %Two-photon: TWOGEN \\
  A.~Buijs et al., \CPC\ {\bf 79} (1994) 523.

\bibitem{bib:OPAL-f2gam}
 \OPALColl\, K.~Ackerstaff et~al., \ZPhys\ {\bf C74} (1997) 33.

%\bibitem{bib:excalibur}
% %Four-fermion: EXCALIBUR \\
%  F.A.~Berends, R.~Pittau and R.~Kleiss, \CPC\ {\bf 85} (1995) 437.

\bibitem{bib:bdk}
 F.A.~Berends, P.H.~Daverveldt and R.~Kleiss, \NPhys\ {\bf B253} (1985) 421; \\
 F.A.~Berends, P.H.~Daverveldt and R.~Kleiss, \CPC\ {\bf 40} (1986) 271, 285
 and 309.

\bibitem{bib:vermaseren}
J.A.M.~Vermaseren, Nucl.~Phys. {\bf B229} (1983) 347.

\bibitem{bib:radcor}
F.A.~Berends and R.~Kleiss, Nucl. Phys. {\bf B186} (1981) 22.
 
\bibitem{bib:teegg}
D.~Karlen, Nucl.\ Phys.\ {\bf B289} (1987) 23.

\bibitem{bib:gopal}
  J.~Allison et~al., \NIM\ {\bf A317} (1992) 47.

\bibitem{bib:bhlumi}
  S.~Jadach et~al., \CPC\ {\bf 102} (1997) 229.

\bibitem{bib:bhlumi_err}
  %W.~Placzek, S.~Jadach, M.~Melles, B.F.L.~Ward and S.A.~Jost,
  W.~Placzek et~al., hep-ph/9903381.

%\bibitem{bib:Chris}
%  C.G.~Ainsley, {\it Studies of $Z/\gamma \rightarrow \qqbar$ events with
%  the OPAL detector at LEPII}, Ph.D. Thesis, University of Cambridge,
%  RAL-TH-2003-001, January 2003.

\bibitem{bib:OPAL-WW189} 
  OPAL Collab., G.~Abbiendi et~al., \PhysLett\ {\bf B493} (2000) 249.

\bibitem{bib:durham}
  S.~Catani et al., \PhysLett\ {\bf B269} (1991) 432.

\bibitem{bib:MT}
 OPAL Collab., K.~Ackerstaff et al.,
 \EPJ\ {\bf C2} (1998) 213.

\bibitem{bib:OPAL-SM130} %PR154
  \OPALColl, G.~Alexander et~al., \PhysLett\ {\bf B376} (1996) 232.
 
\bibitem{bib:kandy}
  S.~Jadach et~al., \CPC\ {\bf 140} (2001) 475.

\bibitem{bib:w420}
  R.K.~Ellis et al., \NPhys\ {\bf B178} (1981) 421.

%\bibitem{bib:George}
% G.~Anagnostou, {\it Analysis of Tau and Muon Pairs at Collision Energies
% 192-209 GeV with the OPAL Detector at LEP}, Ph.D. Thesis, University of
% Birmingham, RAL-TH-2002-011, ISSN 1362-0215, November 2002.

\bibitem{bib:PR328}
  OPAL Collab., G.~Abbiendi et~al., \EPJ\ {\bf C19} (2001) 587.
 %{\sl Precise Determination of the Z Resonance Parameters at LEP: 
 %    Zedometry},CERN-EP-2000-148, 

\bibitem{bib:OPAL-gg}
  OPAL Collab., G.~Abbiendi et~al., \EPJ\ {\bf C26} (2003) 331. 
  %{\sl Multi-photon production in \epem\ collisions at 
  % $\sqrt{s}$ =181--209~GeV},

\bibitem{bib:pdg2000}
 Particle Data Group, D.E.~Groom et al., \EPJ\ {\bf C15} (2000) 1.

\bibitem{bib:dal5h}
 H.~Burkhardt and B.~Pietrzyk, \PhysLett\ {\bf B513} (2001) 46.

\bibitem{bib:alrun}
 TOPAZ Collab., I.~Levine et~al., \PRL\ {\bf 78} (1997) 424. 

\bibitem{bib:MK_alphaem}
 M.~Kobel, {\sl Direct Measurements of the Electromagnetic Coupling
 Constant at Large $q^2$}, FREIBURG-EHEP 97-13,
 Contributed paper to the XVIII International Symposium on Lepton
 Photon Interactions, Hamburg, July 1997.

\bibitem{bib:Mohr}
 P.J.~Mohr and B.N.~Taylor, Rev.~Mod.~Phys. {\bf 72} (2000) 351.

\bibitem{bib:bhlumi_dal5h}
 H.~Burkhardt and B.~Pietrzyk, \PhysLett\ {\bf B356} (1995) 398.

\bibitem{bib:smatasy} 
 S.~Kirsch and T.~Riemann, \CPC\ {\bf 88} (1995) 89. 

\bibitem{bib:L3-smat} 
 L3 Collab., M. Acciarri et al., \PhysLett\ {\bf B489} (2000) 93.

\bibitem{bib:delphi-smat} 
 DELPHI Collab., P. Abreu et al., \EPJ\ {\bf C11} (1999) 383.

\bibitem{bib:alibaba} 
 W.~Beenakker, F.A.~Berends and S.C.~van~der~Marck, 
 Nucl. Phys. {\bf B349} (1991) 323.

\bibitem{bib:Eichten}
  E.~Eichten, K.~Lane and M.~Peskin, \PRL\ {\bf 50} (1983) 811.

\bibitem{bib:contacttable}
 {See for example {\sl Contact interactions and 
  new heavy bosons at HERA: a model independent analysis}, 
  P. Haberl, F. Schrempp and H.U. Martyn, 
  in {Proceedings, Physics at HERA}, {\bf vol. 2}, (1991) 1133. }

\bibitem{bib:OPAL-CI} %PR168
  \OPALColl, G.~Alexander et~al., \PhysLett\ {\bf B387} (1996) 432.

%\bibitem{bib:CInew2}
%  G.J.~Gounaris, D.T.~Papadamou, F.M.~Renard, 
%  \PhysRev\ {\bf D56} (1997) 3970.

\bibitem{bib:CI-ep}
  H1 Collab., C.~Adloff et al., \PhysLett\ {\bf B479} (2000) 358; \\
  ZEUS Collab., J.~Breitweg et al., \EPJ\ {\bf C14} (2000) 239.

\bibitem{bib:CI-pp}
  D0 Collab., B.~Abbott et al., \PhysRev\ D Rapid.~Comm. {\bf 62} (2000)
  031101; \\
  CDF Collab., F.~Abe et al., \PRL\ {\bf 79} (1997) 2198.

\bibitem{bib:CI-atomic}
  A.~Deandrea, \PhysLett\ {\bf B409} (1997) 277.

\bibitem{bib:altarelli}
 G.~Altarelli et al., Phys. Lett. {\bf B318} (1993) 139.

\bibitem{bib:zpGUT}
  G.~B\'elanger and S. Godfrey, Phys. Rev. {\bf D35} (1987) 378.

\bibitem{bib:zpLR}
  G.~Senjanovic, Nucl. Phys. {\bf B153} (1979) 334.

\bibitem{bib:zefit}
   A.~Leike, S.~Riemann and T.~Riemann, hep-ph/9808374 (1991);  \\
   A.~Leike, S.~Riemann and T.~Riemann, Phys. Lett. {\bf B291} (1992)
   187.

\bibitem{bib:zp_indep}
 A.~Leike, \ZPhys\ {\bf C62} (1994) 265.

%\bibitem{bib:LEP}
% LEP Electroweak working group,
% {\it A combination of preliminary electroweak measurements and constraints
% on the Standard Model}, CERN-EP/99-15 (1999).

%\bibitem{bib:L3-zp}
% L3 Collaboration, 
% {\it Limits on an additional heavy gauge boson Z' from the L3 experiment}, 
% L3 Note 2282. {\it Will have to go unless now published.}
%\bibitem{bib:delphi-zp}
% DELPHI Collaboration, {\it Results on fermion-pair production at LEP 
% energies up to 183 GeV}, \\ DELPHI Note 98-139 (1998). {\it Ditto}

%\bibitem{bib:rpvsup}   
%  J.~Wess,  J.~Bagger,    {\it Supersymmetry  and
%    Supergravity}  (Princeton  University Press, 1983);  \\  
%  H.P.~Nilles, \PhysRep\ {\bf 110}   (1984) 1; \\ 
%  H.E.~Haber,  G.E.~Kane,  \PhysRep\ {\bf 117} (1985) 75;  \\
%  R.~Barbieri, Riv.~Nuovo~Cim. {\bf 11} (1988) 1; \\
%  P.~West, {\it Introduction to  Supersymmetry and Supergravity} (World
%  Scientific, 1986). 

%\bibitem{bib:rpvsnu}
%  J.~Kalinowski, R.~R\"{u}ckl, H.~Spiesberger, P.M.~Zerwas,
%    \PhysLett\ {\bf B406} (1997) 314.

%\bibitem{bib:ALEPH-rpv}
% \ALEPHColl, R.~Barate et al., \EPJ\ {\bf C4} (1998) 433; \\
% \OPALColl, G.~Abbiendi et al., {\sl Search for R-parity Violating Decays
%  of Scalar Fermions at LEP}, CERN-EP/99-043, submitted to \EPJ\ C.

%\bibitem{bib:Hewett}
% J.L.~Hewett, \PRL\ {\bf 82} (1999) 4765.

%\bibitem{bib:Peskin}
% E.A.~Mirabelli, M.~Perelstein, M.E.~Peskin, \PRL\ {\bf 82} (1999) 2236.

%\bibitem{bib:Rizzo}
% T.G.~Rizzo, \PhysRev\ {\bf D59} (1999) 115010.

%\bibitem{bib:BK}
% F.A.~Berends and R.~Kleiss, Nucl. Phys. {\bf B186} (1981) 22.

\end{thebibliography}
\end{document}